\documentclass[preprint,aps,preprintnumbers,amsmath,amssymb,superscriptaddress,floatfix,longbibliography]{revtex4-2}
\usepackage{acro}
\usepackage{bm}
\usepackage{booktabs}
\usepackage{color}
\usepackage{derivative}
\usepackage{graphicx}
\usepackage[colorlinks,linkcolor=blue,urlcolor=blue,citecolor=blue,bookmarks=false]{hyperref}
\usepackage{cleveref}
\usepackage[utf8]{inputenc}
\usepackage[version=4]{mhchem}
\usepackage{physics2}
\usephysicsmodule{ab,ab.braket}
\usepackage{siunitx}
\usepackage{soul}
\usepackage[colorinlistoftodos]{todonotes}
\usepackage{orcidlink}

\newcommand{\simlt}
      {\ifmmode       { \raisebox{-.8em}{$<$}\atop\sim}
         \else        {$\raisebox{-.8em}{$<$}\atop\sim$}
      \fi}

\graphicspath{{images/}}

\DeclareAcronym{STM}{short=STM, long=scanning tunneling microscope}
\DeclareAcronym{ARPES}{short=ARPES, long=angle resolved photoemission spectroscopy}
\DeclareAcronym{QPI}{short=QPI, long=quasiparticle interference}
\DeclareAcronym{LDOS}{short=LDOS, long=local density of states}
\DeclareAcronym{JDOS}{short=JDOS, long=joint density of states}
\DeclareAcronym{DFT}{short=DFT, long=density functional theory}
\DeclareAcronym{BdG}{short=BdG, long=Bogoliubov-de Gennes}
\DeclareAcronym{BQPI}{short=BQPI, long=Bogoliubov quasiparticle interference}
\DeclareAcronym{FT}{short=FT, long=Fourier transform}
\DeclareAcronym{dLDOS}{short=dLDOS, long=discrete lattice local density of states}
\DeclareAcronym{cLDOS}{short=cLDOS, long=continuum local density of states}

\crefname{equation}{Eq.}{Eqs.}
\Crefname{equation}{Equation}{Equations}
\crefname{figure}{Fig.}{Figs.}
\Crefname{figure}{Figure}{Figures}
\crefname{section}{Sec.}{Secs.}
\Crefname{section}{Section}{Sections}
\crefrangelabelformat{section}{#3#1#4--#5#2#6}

\begin{document}


\title{Quasiparticle interference as a tool to study quantum materials}
\author{Luke C. Rhodes \orcidlink{0000-0003-2468-4059}}
\affiliation{SUPA, School of Physics and Astronomy, University of St Andrews, North Haugh, St Andrews, KY16 9SS, United Kingdom}
\author{Yuhki Kohsaka \orcidlink{0000-0002-6459-2661}}
\affiliation{Department of Physics, Kyoto University, Kyoto 606-8502, Japan}
\author{Tetsuo Hanaguri \orcidlink{0000-0003-2896-0081}}
\affiliation{RIKEN Center for Emergent Matter Science, 2-1 Hirosawa, Wako, Saitama 351-0198, Japan}
\author{Carolina de Almeida Marques \orcidlink{0000-0002-3804-096X}}
\affiliation{SUPA, School of Physics and Astronomy, University of St Andrews, North Haugh, St Andrews, KY16 9SS, United Kingdom}
\author{Peter Wahl \orcidlink{0000-0002-8635-1519}}
\affiliation{SUPA, School of Physics and Astronomy, University of St Andrews, North Haugh, St Andrews, KY16 9SS, United Kingdom}
\affiliation{Physikalisches Institut, Universität Bonn, Nussallee 12, 53115 Bonn, Germany}

\date{\today}

\begin{abstract}
To understand the properties of quantum materials a detailed knowledge of the material's low energy electronic structure is key. Details of the electronic structure drive the ground state through electronic instabilities, electronic correlation effects, new electronic orders or just the absence of electronic states near the Fermi energy -- making a realistic and detailed understanding crucial to be able to control and design properties of quantum materials. The past 25 years have seen a significant improvement in experimental techniques to observe the true electronic structure, in particular in techniques such as Angle resolved photoemission spectroscopy (ARPES) where energy resolutions of \qty{2}{\milli\eV} are routinely achievable now, which however is limited to zero magnetic field and only provides information about the occupied states. Scanning tunneling microscopy (STM) achieves a significantly better energy resolution (\qty{<100}{\micro\eV}) and can operate at temperatures well below \qty{50}{\milli\K} and in magnetic fields. While per se a real-space technique, by imaging quasiparticle interference (QPI) STM can also provide information about the electronic structure. This technique has been used over the past decades to study a wide range of quantum materials to understand correlated electron behaviour. Recent theoretical progress now enables routine modelling of QPI, a key requirement to interpret the complex data. Here, we review the principles of QPI, its origin, experimental detection, and the physical insight gained from the study of QPI and possible future directions for this technique.
\end{abstract}


\maketitle
\tableofcontents
\section{Introduction}
A detailed understanding of the low energy electronic structure of materials is a prerequisite to understand their properties and be able to exploit them for technological applications. Key techniques to establish the electronic structure of materials include quantum oscillations, a technique that can determine the geometry of the Fermi surface, and \ac{ARPES}, a spectroscopic technique enabling imaging of the electronic structure at the surface in the occupied states. \Ac{STM}, a real space imaging technique, is not traditionally considered a method to determine the electronic structure. It does, however, via imaging of how defects affect the electronic states in an otherwise perfect crystal lattice, allow determining the electronic dispersion relation. It allows inherently to do so in the occupied and unoccupied states of a material, with an energy resolution determined primarily by the temperature of the experiment and in magnetic field.

Different from \ac{ARPES} and quantum oscillations, \ac{QPI} imaging requires defects. While single crystals of many quantum materials are grown optimizing for the highest possible purities, defect are ubiquitous, and often can have significant influence on the properties of a materials, e.g. due to many body interactions through the Kondo effect or as dopants as in high temperature cuprate superconductors.

\ac{STM} enables to directly measure the influence defects have on the electronic states in a material. This has led to the detection of what is now called \ac{QPI}~\cite{crommie_imaging_1993,hasegawa_direct_1993}, where quasiparticles scattering off of defects produce observable modulations to the materials \ac{LDOS}, and whose \ac{FT}, as we will discuss in this review, reveals information about the materials underlying electronic structure with extremely high energy resolution. Since the first observation of \ac{QPI}, it has developed into a technique that can be used to map the electronic structure with very high precision. Key insights that can be gained from such measurements and which are not easily accessible otherwise are the low energy electronic structure in the occupied and unoccupied states, information about the symmetry of the superconducting order parameter, influence of magnetic field on the electronic structure, the spin-orbital texture of the electronic states, and properties of the scatterers. Unravelling this information often requires modelling to be able to evaluate the importance of different contributions to the \ac{QPI} signal.


\section{History}
\begin{figure}
    \centering
    \includegraphics[width=0.5\textwidth]{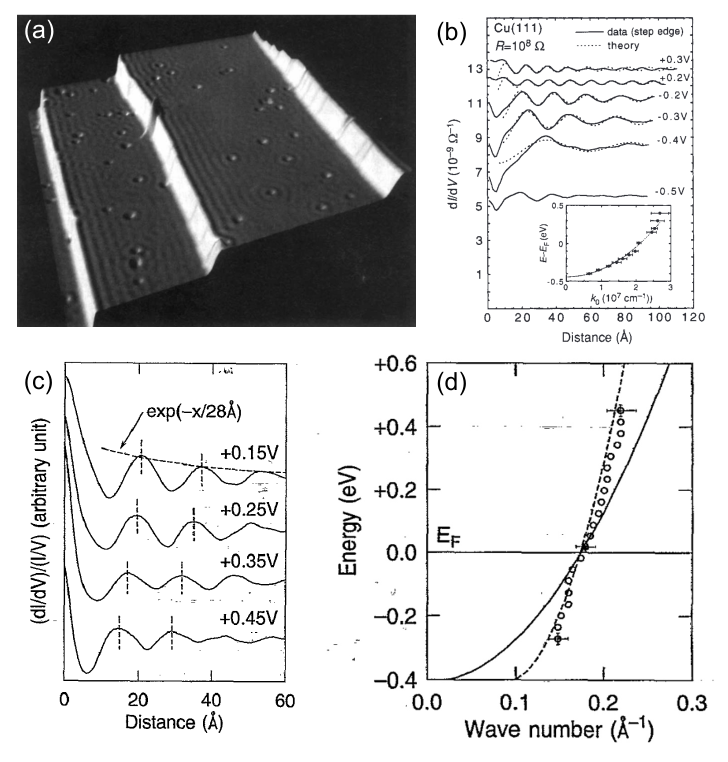}
    \caption{
        First experiments on \ac{QPI} in noble metal (111) surface states.
        (a) Topographic \ac{STM} image showing standing wave patterns next to a step edge on Cu(111).
        (b) shows the dispersion relation extracted from analyzing the wave length of the standing wave patterns in the differential conductance as a function of energy.
        (c, d) shows a measurement of the wave length and resulting wave vector for the case of Au(111).
        Panels a and b reprinted with permission from ref.~\onlinecite{crommie_imaging_1993}, \copyright (1993) by Springer Nature; panels c and d with permission from \cite{hasegawa_direct_1993}, \copyright (1993) American Physical Society.
    }
    \label{fig:111qpi}
\end{figure}

\Ac{QPI} was first reported in two papers in 1993 by Mike Crommie~\cite{crommie_imaging_1993} and Yukio Hasegawa~\cite{hasegawa_direct_1993}, showing standing wave patterns due to quantum interference in the surface states of Cu(111) and Au(111), respectively (\cref{fig:111qpi}). The noble metal (111) surfaces exhibit surface states that are quasi-free two-dimensional electronic states living in a directional bulk band gap. Subsequently, these studies were extended to resonator structures constructed by atomic manipulation~\cite{crommie_confinement_1993}. The interference patterns were described and well captured by scattering theory~\cite{heller_scattering_1994}. Application of \ac{QPI} to strongly correlated electron materials was pioneered by J. C. Séamus Davis and J. E. Hoffman who used \ac{QPI} to map out the structure of the superconducting gap in high-$T_\mathrm{c}$ superconducting cuprates~\cite{Hoffman_Imaging_2002}. Since then, \ac{QPI} has been applied to many complex materials (often called ``quantum materials''), from heavy fermion materials via high temperature superconductors and iron-based superconductors to graphene and topological insulators~\cite{hoffman_spectroscopic_2011,avraham_quasiparticle_2018}.

\section{Experimental detection}
\subsection{What \ac{STM} measures - theory of elastic tunneling}
\label{sec:elastictunneling}

\begin{figure}
    \centering
    \includegraphics[width=0.8\textwidth]{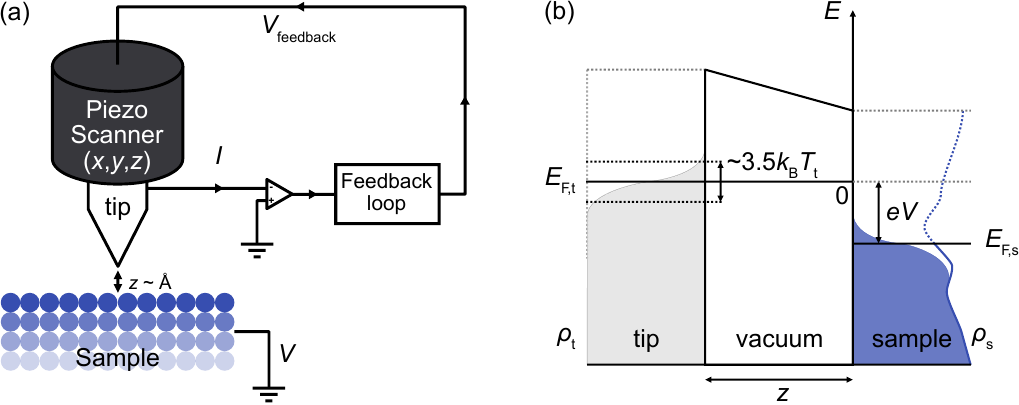}
    \caption{
        Illustration of the working principle of \ac{STM}.
        (a) Sketch of the measurement geometry of an \ac{STM}, where the tip mount on a scan piezo is at a distance $z$ of typically $\sim\qty{5}{\AA}$ from the sample. By applying a bias voltage $V_0$ between the tip and the sample, a tunneling current $I$ is created that can be measured. The current $I$ is compared to a setpoint current $I_0$ by a feedback loop that feeds a voltage $V_\mathrm{feedback}$ to the piezo scanner to adjust $z$ to keep the current $I$ constant and equal to $I_0$.
        (b) Sketch of the tunneling process between a tip with constant DOS, $\rho_\mathrm t$, and a sample with a DOS with characteristic features, $\rho_\mathrm s(E)$, separated by a distance $z$. By applying a bias $V$ to the sample, keeping the tip at ground (\qty{0}{V}), the Fermi level of the sample, $E_{\mathrm{F,s}}$, is shifted by an energy $eV$ with respect to the Fermi level of the tip, $E_{\mathrm{F,r}}$. The current will be dominated by the tunneling of electrons from the occupied states of the tip (gray shaded area) through the vacuum potential barrier to the unoccupied states of the sample. The derivative of the Fermi-Dirac function in \cref{eq:g_finiteT}, due to the temperature of the tip $T_{\mathrm{t}}$, results in an energy broadening of $\sim 3.5k_{\mathrm{B}}T_{\mathrm{t}}$ of the tunneling spectrum.
        Figure adapted from \cite{de_almeida_marques_imaging_2022}.
    }
    \label{fig:stm-sketch}
\end{figure}

In an \ac{STM}, the surface is probed using a metal tip, illustrated in \cref{fig:stm-sketch}(a).  When the tip is brought sufficiently close to the sample surface, typically to a tip-sample distance of about $\sim\qty{5}{\AA}$, electrons can tunnel between the tip and the surface. Applying a voltage $V$ between the two then results in a tunneling current $I$.  This tunneling current is highly sensitive to the tip-sample distance $z$.  For a vaccum gap, the distance dependence is governed by the wave function overlap between tip and sample, resulting in an exponential dependence.  Tunneling spectroscopy was used already in the 1960s to study the electronic density of states of superconductors in planar tunneling junctions~\cite{giaever_energy_1960}, where it revealed the superconducting gap, a discovery which confirmed BCS theory of superconductivity and was later recognized by the Nobel Prize.  It is in this context that the theory of tunneling was developed by John Bardeen~\cite{bardeen_tunnelling_1961}.  Following the development of the \ac{STM} by Binnig and Rohrer~\cite{binnig_scanning_1982}, the theory was later extended for \ac{STM} by Tersoff and Hamann~\cite{tersoff_theory_1983} and Lang~\cite{lang_spectroscopy_1986}. From the theory of tunneling, one can show that
\begin{align}
    I(V,z)
    \propto
    e^{-\kappa z}
    \int_{-\infty}^\infty
    \rho_\mathrm{t}(E)
    \rho_\mathrm{s}(E+eV)
    M(E, E+eV)
    \ab[
        f(E,T_\mathrm{t}) - f(E+eV,T_\mathrm{s})
    ]
    \odif{E}.
    \label{eq:current}
\end{align}
Here, $f(E,T)=\ab[\ab(\exp(E/k_\mathrm{B}T)+1)]^{-1}$ is the Fermi-Dirac function at temperature $T$, where $k_\mathrm{B}$ is the Boltzmann constant.  The subscripts t and s refer to the tip and sample, respectively; $T_{\mathrm{t},\mathrm{s}}$ are their temperatures and $\rho_{\mathrm{t},\mathrm{s}}$ their densities of states.  The quantity $e$ is the elementary charge, so that $eV$ represents the energy shift associated with the applied bias voltage $V$. The tunneling process is illustrated in \cref{fig:stm-sketch}(b). Besides the tip-sample distance $z$, the tunneling current depends on the tip and sample density of states, $\rho_\mathrm t$ and $\rho_\mathrm s$, and on the tunneling matrix element $M$.  The factor $e^{-\kappa z}$ represents the exponential dependence of the tunneling probability on the tip-sample distance $z$, where $\kappa$ describes the potential barrier between the tip and sample and can be estimated from the work functions of the two. For small bias voltages $V$, $\kappa$ can be approximated as independent of the applied voltage.

Suppose that $\rho_\mathrm{t}$ and $M$ are constant in the energy range of interest.  Then, the differential conductance $g(V)=\odv{I}/{V}$ is given by
\begin{align}
    g(V)
    \equiv
    \odv{I}{V}
    \propto
    e^{-\kappa z}
    \int_{-\infty}^\infty
    \rho_\mathrm{s}(E)
    \pdv{f(E-eV,T_\mathrm{t})}{V}
    \odif{E}.
    \label{eq:g_finiteT}
\end{align}
Thus, $g(V)$ is proportional to the convolution of $\rho_\mathrm{s}(E)$ and the derivative of the Fermi-Dirac function.  The full width at half maximum of this thermal broadening function is approximately $3.5k_\mathrm{B}T_\mathrm{t}$, which sets the thermal contribution to the energy resolution and typically limits the energy resolution of tunneling spectroscopy~\cite{ast_sensing_2016}.   If a lock-in amplifier is used to measure $g(V)$, the modulation amplitude introduces additional broadening~\cite{singh_construction_2013}.

In the low temperature limit, the derivative of the Fermi-Dirac function approaches a delta function, and the differential conductance becomes proportional to the sample density of states at the energy selected by the bias voltage,
\begin{align}
    g(V) \propto \rho_\mathrm{s}(eV).
    \label{eq:g_zeroT}
\end{align}
This relation holds under the assumptions described above, namely that the tip density of states and the tunneling matrix element are energy independent, and that the tunneling current is dominated by elastic tunneling.  In general, tunneling spectroscopy measures a convolution of $\rho_\mathrm{t}(E)$ and $\rho_\mathrm{s}(E)$.  Therefore, great care must be taken to ensure that features in the tunneling spectrum genuinely originate from the sample density of states.

Apart from the improved mechanical  stability of the \ac{STM} tip and sample at low temperatures, because of the thermal broadening introduced in \cref{eq:g_finiteT}, \ac{QPI} is most powerful as a low-temperature technique, where sharp spectral features can be resolved.  The development of \ac{STM}s that can routinely perform measurements at dilution refridgerator temperatures~\cite{song_invited_2010,singh_construction_2013,assig_10_2013,machida_dilution_2018,fernandez-lomana_millikelvin_2021} has significantly expanded the range of physics questions that can be addressed by \ac{QPI}, both by providing access to the correlated ground states in quantum materials and by improving the attainable energy resolution.

\subsection{Inelastic tunneling}
\label{sec:inelastictunneling}
Apart from the contribution from elastic tunneling introduced in \cref{sec:elastictunneling}, also inelastic processes contribute to the tunneling current. These processes are due to tunneling electrons exchanging energy with a bosonic mode, e.g. a spin or vibrational excitation. Similar to a Raman process, in principle this energy exchange can result in the tunneling electron proceeding with a higher or lower energy, however in practice, at the temperatures relevant for tunneling spectroscopy, the bosonic modes are not populated, so only the energy loss process is relevant. In this case, once the energy of the tunneling electrons $eV$ exceeds that of a bosonic mode $\hbar\Omega$, the electron can excite that mode, losing its energy. For an individual mode, such as excitation of a single vibrational mode or a single spin flip, such an excitation is delta-function like and results in step-like features in the tunneling spectrum~\cite{lauhon_single-molecule_1999,heinrich_single-atom_2004}. In the case of collective excitations, such as magnon or phonon excitations, the excitation will be an energy continuum, which is expected to result in a background of inelastic excitations that increases with the modulus of the bias voltage $|V|$~\cite{kirtley_inelastic-tunneling_1990}. While the impact of coupling to vibrational modes is typically small, in specific cases it can become a large and dominant contribution due to phonon-assisted tunneling processes~\cite{zhang_giant_2008}. The coupling to electronic excitations such as spin excitations is often significantly larger than for vibrational excitations~\cite{loth_spin-polarized_2010}. Inelastic processes can play an important role where the tunneling process is not through a vacuum tunneling gap~\cite{gupta_strongly_2005,trainer_probing_2021}, highlighting the importance of a vacuum environment for the interpretation of the tunneling spectra and \ac{QPI}.

\subsection{\Ac{QPI} imaging}
\label{sec:qpiimaging}
The simplest way to measure \ac{QPI} in an \ac{STM} is by recording topographic images at a small bias voltage $V$: because the current depends on the density of states of the sample, the local variation in the density of states due to \ac{QPI} results in variations in the tunneling current that are compensated by the feedback loop~\cite{Petersen1998}. This method is however limited to very small voltages and effectively only recording the scattering on the Fermi surface because the measured signal is due to the integrated density of states between the Fermi energy and the applied bias, so at larger bias voltages, the contributions from a wider energy range will be included, resulting in significant broadening of the signal. Furthermore, to extract dispersion relations, it is usually desirable to measure the \ac{QPI} signal as a function of energy.

\begin{figure}
    \centering
    \includegraphics[width=0.95\textwidth]{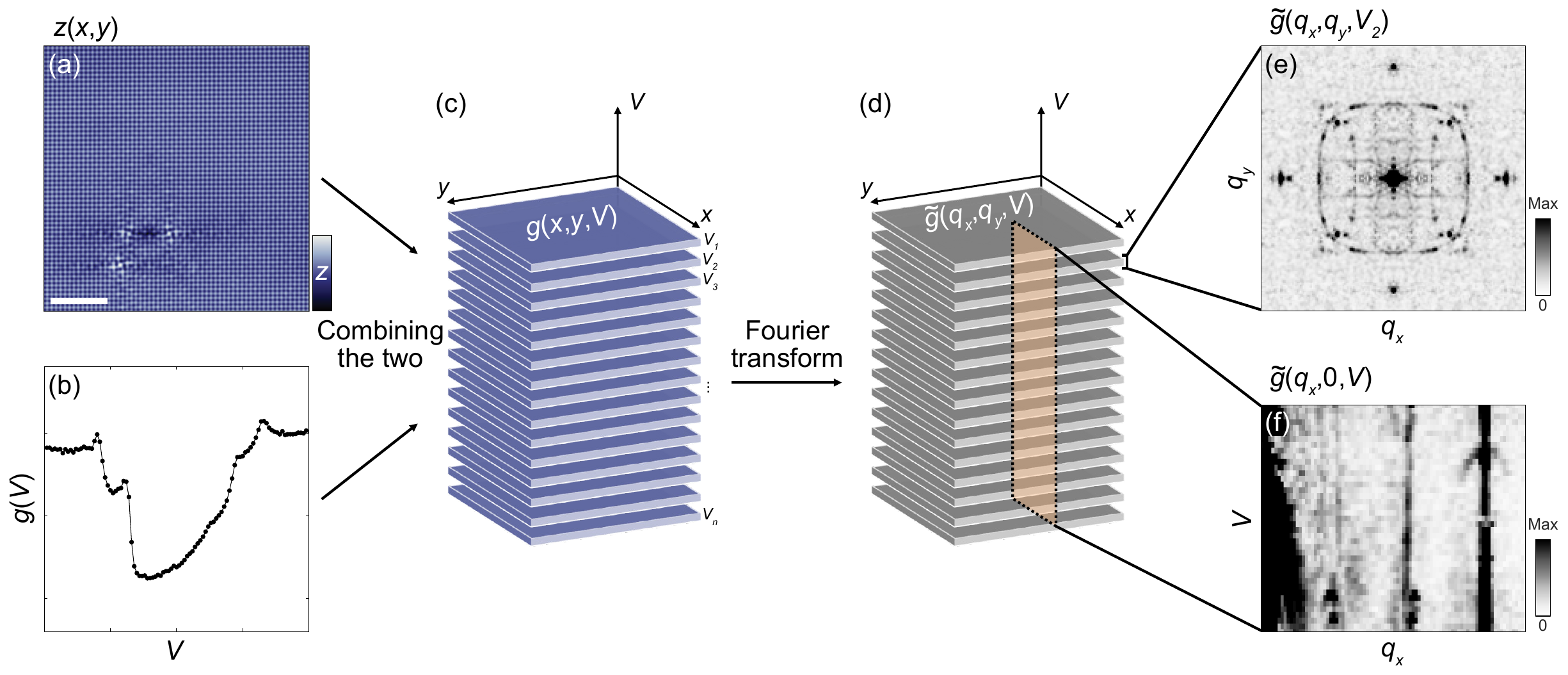}
    \caption{
        Measurement of Quasi-particle interference.
        (a) Topographic \ac{STM} image $z(x,y)$, taken in constant-current mode. The topographic image contains information about the sample morphology, but with contributions from the density of states through the tunneling current.
        (b) Differential conductance spectrum $g(\bm{r},V)$ taken at a single pixel, which often can be interpreted in terms of the local density of states $\rho(\bm{r},eV)$.
        (c) A spectrum $g(\bm{r},V)$ is taken at each pixel of a topography $z(\bm{r}=(x,y))$, resulting in a three-dimensional data set $g(x,y,V)$.
        (d) To analyze the dominant scattering vectors and their energy dispersion in reciprocal space, the \ac{FT} of the slices at constant bias voltage $V_n$ is taken, creating a new datacube  $|\tilde{g}(q_x,q_y,V)|$.
        (e) \ac{FT} of the image at a single bias $V_2$, $|\tilde{g}(q_x,q_y,V_2)|$,  showing the atomic peaks and clear \ac{QPI} patterns.
        (f) Cross sectional cut of the datacube $|\tilde{g}(q_x,0,V)|$ after taking the \ac{FT} of each individual energy layer, showing the energy dispersion of the scattering vectors.
    }
    \label{fig:qpi-dataset}
\end{figure}

With an \ac{STM}, the differential conductance can be recorded as a function of position, $g(\bm{r},V)$.  In a typical \ac{QPI} experiment, spectra are acquired on a spatial grid, \cref{fig:qpi-dataset}(a,b), resulting in a three-dimensional data set as a function of position and bias voltage, \cref{fig:qpi-dataset}(c).  Assuming that the tunneling is dominated by elastic tunneling, the differential conductance at each point is related to the \ac{LDOS} as $g(\bm{r},V)=C(\bm{r})\rho_\mathrm{s}(\bm{r},V)$, where $C(\bm{r})$ contains factors such as the tip-sample distance, the tip density of states, and the tunneling matrix element.

In spectroscopic maps, the feedback loop is usually used to stabilize the tip-sample distance at the tunneling setpoint $(V_0, I_0)$ at each position and is then turned off while the spectrum is acquired.  Since the setpoint determines the local tip height, the prefactor $C(\bm{r})$ can vary from point to point.  This gives rise to the setpoint effect, which introduces an additional position-dependent weighting into $g(\bm{r},V)$, as described in \cref{sec:setpoint_effect}.  As a result, the measured conductance modulations do not simply reflect the spatial variations of the \ac{LDOS} at the energy $eV$, an effect that must be taken into account when interpreting the data.

One way to avoid this problem is to record a map in constant-height mode, where the tip height $z$ is kept fixed during data acquisition.  For a sufficiently flat surface, this gives $g(\bm{r},V)=C\rho_\mathrm{s}(\bm{r},V)$, with a position-independent prefactor $C$.  Constant-height maps therefore do not suffer from the setpoint effect.  In practice, however, the acquisition time of such maps is limited by piezo creep, thermal drift, and the risk of tip crashes due to the surface morphology.  For this reason, long spectroscopic maps are usually acquired stabilizing the tip-sample distance at each point using constant-current mode, despite the resulting setpoint effect.

It is also possible to record \ac{QPI} in constant current mode without turning off the feedback loop. In this case, the lock in amplifier used to record the differential conductance is operated at a frequency $f$ higher than the characteristic frequency of the feedback loop, so that the modulation in the current at that frequency does not result in a movement of the tip while still keeping it below the cut-off frequency of the current amplifier. The differential conductance can then simply be recorded while scanning a topographic image, however care must be taken that the scanning is sufficiently slow to avoid the lock-in signal to leak into adjacent pixels.  This procedure yields $g(\bm{r},V_0)$, which is proportional to the Feenstra function at $V_0$ because of the setpoint effect (see \cref{eq:setpoint_effect,eq:feenstra} in \cref{sec:setpoint_effect}).

\subsection{Data processing and improving signal to noise}

The acquisition of spectroscopic maps $g(\bm{r},V)$ by \ac{STM} involves raster-scanning the tip across the surface. The measurement process can introduce artifacts due to the nature of the scanning process as well as details of the tip shape. The scanning is controlled using piezoelectric components, which introduces inherent distortions due to the hysteresis and creep which depend on the history of applied voltages. In the commonly used raster scanning, the scan process itself exhibits fast and slow scan directions that are inequivalent and so will induce uncertainties in the position of the \ac{STM} tip. These uncertainties can be as small as a few picometres but also orders of magnitude larger, on the order of nanometers, and will be different along the fast and slow scan directions and hence anisotropic. These displacements in the tip position, $\delta \bm{u}(\bm{r})$, result in distortions in topographic images and \ac{QPI} maps so that the imaged atomic lattice deviates from an ideal one, typically resulting in a certain amount of symmetry breaking and artifacts in the \ac{FT}. These distortions are important to account for in the interpretation of \ac{QPI} data and to be aware of where a quantitative analysis of lattice distortions or anisotropy is the goal.
An example of these distortions is shown in the \ac{FT} of a topography of \ce{Sr2RuO4}, \cref{fig:linear_lf_transform} (a, b). Despite the tetragonal symmetry of the material, the two independent sets of atomic peaks along the $x$ and $y$ directions do not show-up as delta-function-like maxima and have an intrinsic width due to variations of the periodicity across the image due to piezo creep. They also do not have the same magnitude $|\bm{q}_\mathrm{at}|$ and are not at a \qty{90}{\degree} angle from each other, as would be expected for an ideal square lattice. This is clearly seen in the inset of \cref{fig:linear_lf_transform}(b): when taking the position of $\bm{q}_{t1}$ and rotating it by \qty{90}{\degree} ($\bm{q}_{i2}$, red circle), it does not fall on top of $\bm{q}_{t2}$ (blue asterisk).

\begin{figure}
    \includegraphics[width=\textwidth]{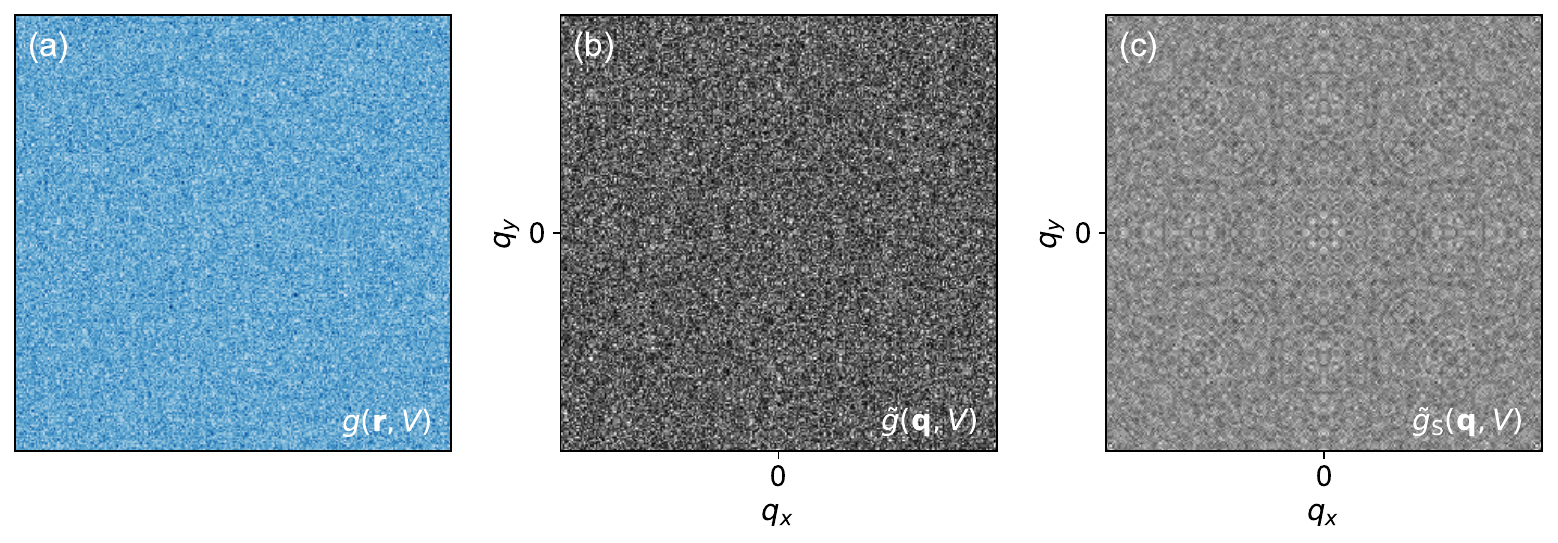}
    \caption{Signal from noise. (a) synthetic conductance map $g(\bm{r},V)$ consisting of random noise. (b) Modulus of \ac{FT} $|\tilde{g}(\bm{q},V)|$ of (a), showing also only random noise. (c) Modulus of the \ac{FT} as in (b) following symmetrization along the horizontal, vertical and diagonal as might be applied for material with tetragonal symmetry. The symmetrized map $|\tilde{g}_\mathrm{S}(\bm{q},V)|$ exhibits patterns that could be mistaken for \ac{QPI} signal.}
    \label{symmetrizednoise}
\end{figure}

Removing such artifacts from the measured images allows one to identify true anisotropies in the electronic states of the sample and/or use of symmetrisation of the \ac{FT} to increase the signal-to-noise ratio and remove anisotropies in the intensities introduced by the tip shape. Two main strategies are commonly used to correct for drift effects: (1) drift correction by linear transformation that maps the atomic peaks in reciprocal space to positions that obey the symmetry of the crystal lattice and (2) mapping the atomic positions in real space to those of an ideal lattice prior to taking the \ac{FT}, often done using the Lawler-Fujita algorithm. The former works well if the drift is predominantly linear and can even be applied when the atomic peaks are close to or beyond the Nyquist limit (e.g. following suitable unfolding, see \cref{sec:anti-aliasing}), while the latter can also compensate for non-linear drift effects, but requires a higher sampling of atomic resolution if a real-space image is to be reconstructed on grid points. Following the drift correction, \ac{QPI} maps can be symmetrized. Care must be taken that the symmetrization does not suppress real physical features breaking the assumed symmetry, at the same time, it is important to be aware that symmetrization of noise can exhibit features that look like real signal, in particular along the symmetry axes, where the symmetrization does not result in increased signal-to-noise. An example is shown in \cref{symmetrizednoise}, where an image of random white noise (\cref{symmetrizednoise}(a)) is first Fourier transformed (\cref{symmetrizednoise}(b)), still showing only white noise, but where, following symmetrization (\cref{symmetrizednoise}(c)), seemingly well-defined features appear out of the noise. Experimental strategies to safeguard against interpretation of such artifacts include ensuring that features exist also in the unsymmetrized \ac{QPI} data, their consistency across multiple datasets and tracing the dispersion of \ac{QPI} features across multiple energies.


\subsubsection{Drift correction by linear transformation}
\label{sec:linear-transform}

\begin{figure}
    \centering
    \includegraphics[width=1\textwidth]{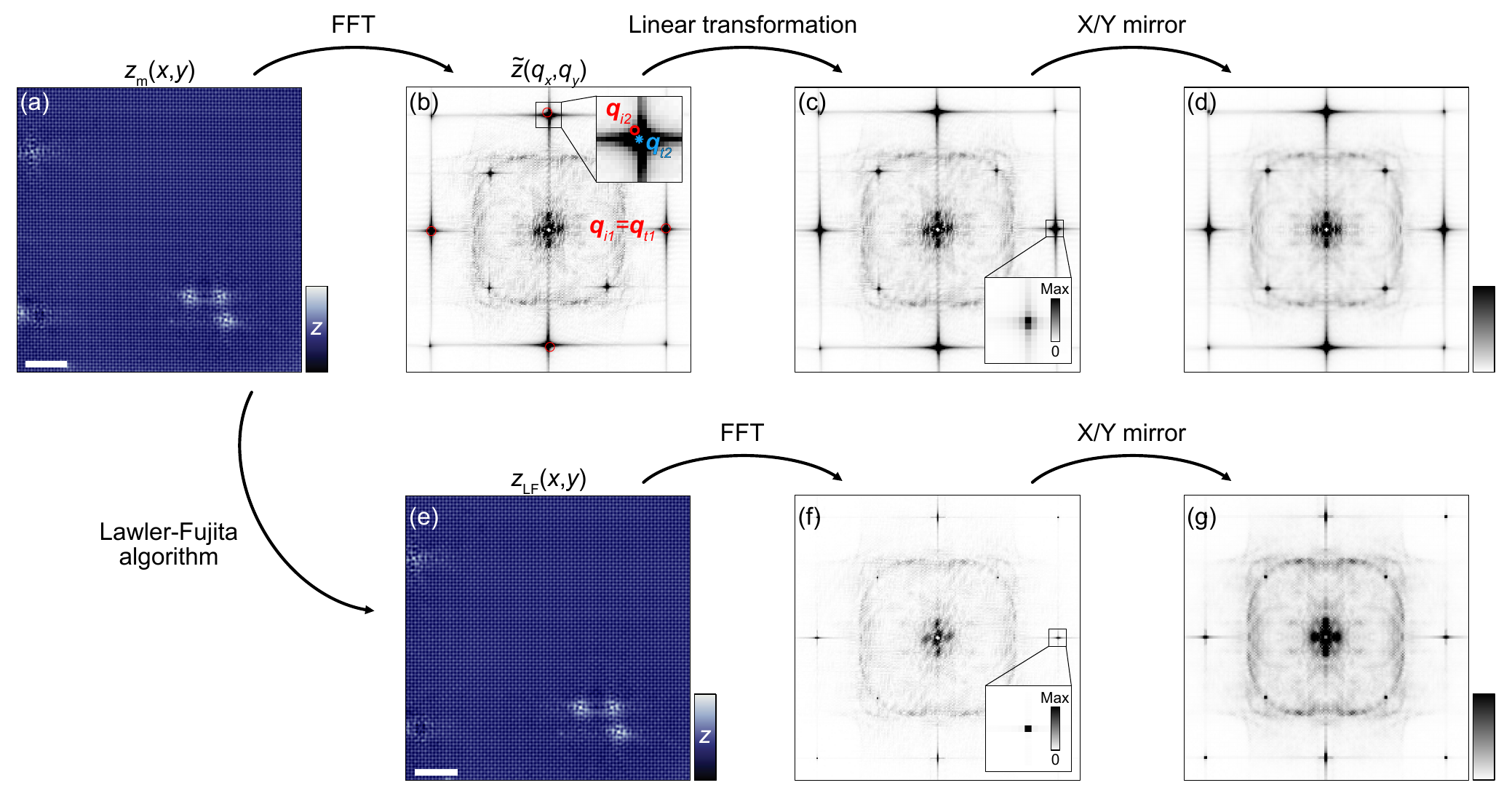}
    \caption{
        Linear transformation and Lawler-Fujita algorithm.
        (a) Constant-current topography, as measured on a sample of \ce{Sr2RuO4}. The scale bar indicates \qty{5}{nm}.
        (b) Absolute value of \ac{FT} of (a). The atomic peaks along the vertical and horizontal directions are not at the same distance from the center of the image ($\bm{q}=(0,0)$), and the angle between them is not \qty{90}{\degree}. The blue stars indicate the position of the atomic peaks, while the red circles indicate the positions where they will be mapped after the linear transformation at exactly $(1,0)$ and $(0,1)$. The inset shows a zoom-in at the atomic peak at $\bm{q}_\mathrm{at}=(0,1)$, showing how it deviates from the position for an ideal square lattice.
        (c) \ac{FT} after a linear transformation $\cal{A}$ that maps the position of the atomic peaks (blue stars in (b)) to the ideal positions at the red circles in (b). The inset shows a zoom-in of one of the atomic peaks, showing that the peak intensity remains broad.
        (d) Shows (b) after symmetrising along the horizontal or vertical directions to increase signal-to-noise ratio.
        The broadening of the atomic peaks is avoided when using the Lawler-Fujita algorithm for drift correction: (e) same topography as in (a), but after applying the Lawler-Fujita algorithm (scale bar: \qty{5}{nm}).
        (f) Absolute value of the \ac{FT} of (e). The atomic peaks show no distortion, and the inset shows a zoom-in of the atomic peak at $\bm{q}=(1,0)$, showing that the intensity is concentrated on a single pixel, as expected for an ideal lattice.
        (g) shows the \ac{FT} in (f) after symmetrizing along the horizontal or vertical directions.
        Fig. adapted from ref.~\cite{marques2022atomic}. 
        }
    \label{fig:linear_lf_transform}
\end{figure}

To correct for this anisotropy due to piezo creep and drift during acquisition, one can define a linear transformation ${\cal A}$  that maps the atomic peaks $\bm{q}_{t1}$, $\bm{q}_{t2}$ in the reciprocal space onto the atomic peaks of an ideal lattice $\bm{q}_{i1}$, $\bm{q}_{i2}$.

The equations
\begin{equation}
\bm{q}^\mathrm{i}_{1,2}={\cal A} \bm{q}^\mathrm{t}_{1,2}.
\end{equation}
uniquely define the transformation ${\cal A}$ which can then be used to map each point of the Fourier transform to its corrected location. The resulting image is shown in \cref{fig:linear_lf_transform}(c). Now, the atomic peaks occur in the positions expected for a square lattice, with one pixel precision. To improve signal-to-noise ratio, the image can now be symmetrised, by rotating and/or mirroring the image according to the symmetry group of the sample. Here, as an example, symmetrisation along the $x$ or $y$ direction is applied, producing the image in \cref{fig:linear_lf_transform}(d).

\subsubsection{Drift correction by Lawler-Fujita algorithm}
\label{sec:lawler-fujita}

The linear transformation described above acts on the absolute value of the Fourier transform (FT) and does not allow reconstructing a real-space image free of drift effects, since changes to the periodicity of the lattice due to  non-linear drift effects are encoded in the width and phase at the atomic peaks $\bm{q}_{t1}$, $\bm{q}_{t2}$. It also can only correct for linear drift, whereas in reality the drift normally also contains non-linear contributions.

To correct for such non-linear drift in an \ac{STM} image, one can use the Lawler-Fujita algorithm~\cite{Lawler2010}. It defines a transformation based on the topography that maps the distorted atomic lattice onto an ideal lattice in real space. This transformation is determined by comparing the measured topographic image $z_{\mathrm{m}}(x,y)$ (\cref{fig:linear_lf_transform}(a)) with the expected ideal lattice, allowing the reconstruction of a corrected $z_{\mathrm{LF}}(x,y)$ image. The transformation is obtained using a phase sensitive lock-in detection comparing to a reference wave to obtain the displacement map $\delta\bm{u}(\bm{r})$ required to locally map the image onto an ideal lattice.

For the specific case shown in \cref{fig:linear_lf_transform}, an ideal square lattice can be described by
\begin{equation}
    T(\bm{r})=\cos(\bm{q}_{i1}\cdot\bm{r}+\phi_1)+\cos(\bm{q}_{i2}\cdot\bm{r}+\phi_2),
\end{equation}
where $\bm{q}_{i1}$ and $\bm{q}_{i2}$ are the $\bm{q}_{i1,2}$-vectors correspond to the atomic peaks of the ideal lattice as before, and $\phi_1$ and $\phi_2$ are phases which can be chosen to be zero. To obtain the local phases $\Theta_n(\bm{r})$ (with $n=1, 2$), which encode the correction that is required to map the topography onto the ideal lattice, a phase-sensitive detection for each of the two independent spatial directions is performed by calculating
\begin{eqnarray*}
    X_n(\bm{r})&=&{\cal L}( z_{\mathrm{m}}(\bm{r})\cos(\bm{q}_{in}\cdot\bm{r}),\Lambda)\\
    Y_n(\bm{r})&=&{\cal L}( -z_{\mathrm{m}}(\bm{r})\sin(\bm{q}_{in}\cdot\bm{r}),\Lambda),\\
\end{eqnarray*}
with $n=1, 2$, designating the direction of each atomic peak of the ideal lattice.
To enable the phase-sensitive detection and extract the slow variations of the phase due to drift, a low-pass filter $\cal L$ with width $\Lambda \gg \frac{\pi}{|\bm{q}_{in}|}$ is applied. From the corresponding filtered maps $X_n(\bm{r})$ and $Y_n(\bm{r})$ for each of the directions of the atomic peaks the phases $\Theta_n(\bm{r})$ associated with the small displacements $\delta u_n(\bm{r})$ are obtained from

\begin{equation}
\label{eq:phase_li}
    \Theta_n(\bm{r})=\mathrm{arctan2}(Y_n(\bm{r}),X_n(\bm{r})).
\end{equation}

These phase maps will exhibit phase slips of $\sim 2\pi$ each time the displacement $\delta u_n(\bm{r})$ matches a full lattice constant, which have to be eliminated before converting the phase maps $\Theta_n(\bm{r})$ to the displacement field $\delta\bm{u}(\bm{r})$.

The last step is to obtain the displacement field $\delta \bm{u}(\bm{r})$ from the phase maps $\Theta_n(\bm{r})$. The phase $\Theta_n(\bm{r})$ is equal to
\begin{equation}
    \Theta_n(\bm{r})=\bm{q_{in}}\cdot\delta\bm{u}(\bm{r})
\end{equation}
and thus, the displacement field $\delta\bm{u}(\bm{r})$ is obtained by inverting this linear equation and combining it with \cref{eq:phase_li}. Knowing the displacement field $\delta\bm{u}(\bm{r})$, the distorted topographic image in $z_\mathrm{m}(x,y)$ can be mapped onto a ideal lattice in $z_{\mathrm{LF}}(x,y)$, \cref{fig:linear_lf_transform}(e). One of the advantages of this method is that now the position of the atoms in real space is known and it can be used for further analysis.

After applying the Lawler-Fujita algorithm, the absolute value of the FT of $z_{\mathrm{LF}}(x,y)$, \cref{fig:linear_lf_transform}(f), shows the atomic peaks with the width of a single pixel, reflecting the uniform periodicity of the atomic lattice in the reconstructed real-space image. Again, the drift corrected image can be used to, e.g., apply symmetrisation to the resulting \ac{QPI} (compare \cref{fig:linear_lf_transform}(g)).

\subsubsection{Extending $\bm{q}$-space beyond the Nyquist limit and correcting aliasing}
\label{sec:anti-aliasing}

For some measurements, for example where \ac{QPI} is dominated by small $\bm{q}$ vectors, the area of interest in reciprocal space can be significantly smaller than the area enclosed by the atomic peaks. In this case, acquiring spectroscopic maps with the pixel count required by the Nyquist theorem can become too time consuming. As a way to optimize measurement time and allow mapping over larger areas, data can be acquired in real space with lower pixel density than required by the Nyquist theorem for the atomic lattice. This results in $g(\bm{r},V)$ maps which are undersampled and the atomic peaks being folded to smaller wave vectors due to aliasing. This folding can be corrected for to reconstruct the reciprocal space beyond the Nyquist theorem using an unfolding algorithm.

\begin{figure}
    \centering
    \includegraphics[width=1\textwidth]{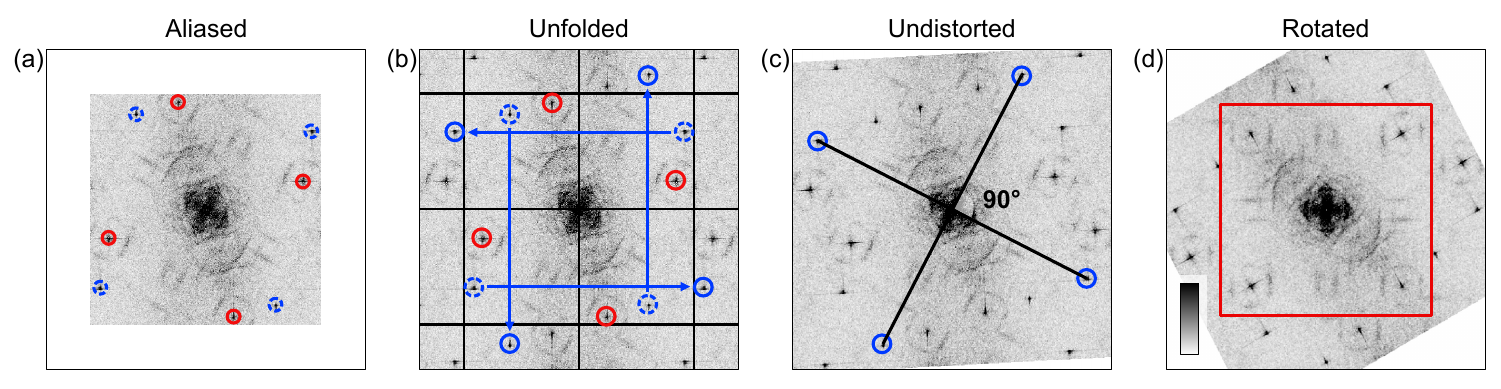}
    \caption{
        Unfolding procedure.
        (a) \Ac{FT} of a map layer of a differential conductance map $g_\mathrm{m}(\bm{r},V=\qty{1.5}{\mV})$ (here for \ce{Sr3Ru2O7}, with aliased atomic peaks indicated by the dotted blue circles. The peaks associated with the orthorhombic unit cell due to the octahedral rotations are indicated by red circles.
        (b) Unfolded version of \textbf{a}, where the atomic peaks (blue dotted circles) are mapped to their correct positions.
        (c) Differential conductance map $g(\bm{r},V=\qty{1.5}{\mV})$ after drift correction by using a geometrical transformation to map the atomic peaks onto an ideal lattice.
        (d) Rotated $g(\bm{r},V)$ map to align the atomic peaks with the horizontal and vertical axes.
        Adapted from Ref.~\cite{marques2022atomic}.
    }
    \label{fig:antialiasing}
\end{figure}

An example for the \ac{FT} of an image acquired with undersampling is shown in \cref{fig:antialiasing}(a), where the atomic peaks appear folded to lower $q$-value (blue circles). The unfolding algorithm maps the folded areas to their original positions in $\bm{q}$-space (see \cref{fig:antialiasing}(b)). After applying the unfolding algorithm, a linear transformation as discussed in \cref{sec:linear-transform} can be applied to map the atomic peaks onto an ideal lattice, \cref{fig:antialiasing}(c), and rotated so that they align with the $x$ and $y$ directions, \cref{fig:antialiasing}(d).

Because the aliasing superimposes features from outside the area covered by the Nyquist frequency into it, for the acquisition of such maps it is worth ensuring that features of interest in the \ac{QPI} do not become superimposed by, e.g., atomic peaks or other \ac{QPI} features. This can often be achieved by acquiring the map at an angle with the atomic lattice, which avoids folding the atomic peaks towards $\bm{q}=(0,0)$.

\subsection{Setpoint effect}
\label{sec:setpoint_effect}
As discussed in \cref{sec:qpiimaging}, under the assumptions of low temperature, an energy-independent tunneling matrix element, and a constant tip density of states, the tunneling spectrum is proportional to the sample density of states (\cref{eq:g_zeroT}).  An individual spectrum can therefore often be interpreted in terms of the \ac{LDOS}.  Its spatial dependence, however, requires more care, because the proportionality factor between $g(\bm{r},V)$ and the \ac{LDOS} can vary with position.  This position-dependent normalization is known as the setpoint effect. It can modify the apparent intensity of features in \ac{QPI} maps~\cite{li_local_1997,macdonald_dispersing_2016}, although its influence can be reduced by an appropriate choice of measurement parameters.  Using the same approximations as for the differential conductance, the tunneling current $I$ can be written as
\begin{align}
    I(\bm{r},V)
    =
    C(\bm{r})
    \int_0^{eV}
    \rho_\mathrm{s}(\bm{r},E)\odif{E}.
\end{align}
In constant-current operation, the feedback loop of the \ac{STM} regulates the tip-sample distance $z$ such that the current at the setpoint bias voltage $V_0$ is fixed to the setpoint current $I_0$.  Therefore,
\begin{align}
    C(\bm{r})
    =
    \dfrac{I_0}{\displaystyle\int_0^{eV_0}\rho_\mathrm{s}(\bm{r},E)\odif{E}}.
\end{align}
The tunneling current and the differential conductance measured after stabilizing the setpoint are then
\begin{align}
    I(\bm{r},V; V_0, I_0)
    &=
    \frac{
        I_0\displaystyle\int_0^{eV}\rho_\mathrm{s}(\bm{r},E)\odif{E}
    }{
        \displaystyle\int_0^{eV_0}\rho_\mathrm{s}(\bm{r},E)\odif{E}
    },
    \\
    g(\bm{r},V; V_0, I_0)
    &=
    \frac{
        I_0e\rho_\mathrm{s}(\bm{r},eV)
    }{
        \displaystyle\int_0^{eV_0}\rho_\mathrm{s}(\bm{r},E)\odif{E}
    }.
    \label{eq:setpoint_effect}
\end{align}
Throughout this section, we explicitly indicate the dependence of $I$ and $g$ on the setpoint parameters $V_0$ and $I_0$.  When no confusion can arise, these arguments are omitted.  The important point is the denominator: spatial variations of the \ac{LDOS} integrated up to the setpoint energy $eV_0$ introduces a position-dependent normalization of the measured differential conductance.  This is the setpoint effect. As a result, modulations in $g(\bm{r},V)$ do not simply reflect the \ac{LDOS} at the energy $eV$, but are also affected by the \ac{LDOS} integrated over the setpoint bias window, i.e. between the Fermi energy and the energy of the setpoint bias $eV_0$.

\begin{figure}
    \includegraphics[width=0.75\textwidth]{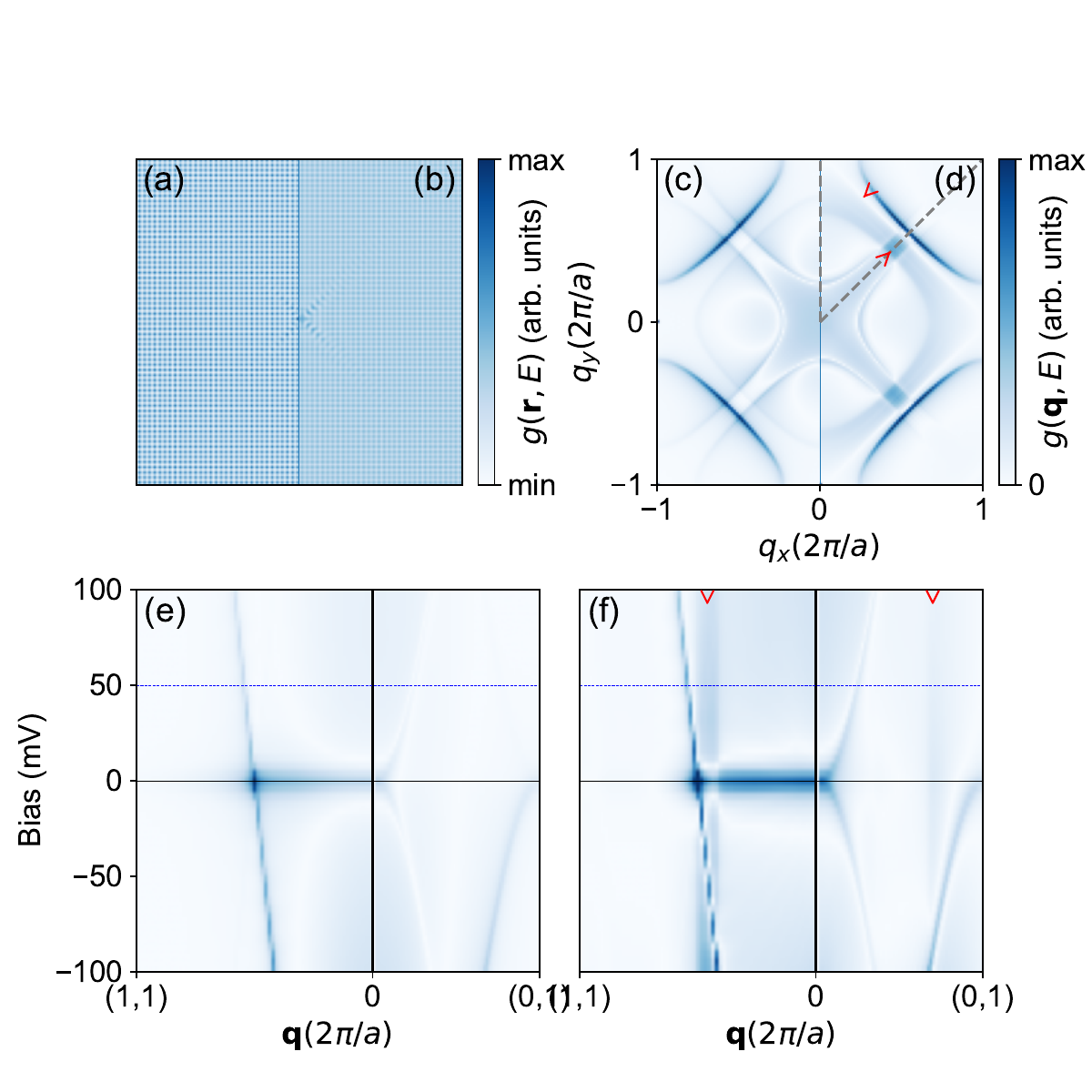}
   \caption{(a, b) Simulated real space image of the LDOS (a) and differential conductance (b) for a nearest neighbour tight-binding model on a square lattice. For the differential conductance a setpoint of $V=\qty{-100}{mV}$ has been assumed. The suppression of the atomic corrugation in (b) compared to (a) is clearly visible, showing the QPI more prominently than in (a). (c, d) Fourier transforms of (a) and (b), respectively, showing how the setpoint effect affects the QPI signal. Broad features appear in addition to the QPI signal due to the setpoint effect (two of the features are indicated with red arrowheads). (e, f) QPI dispersion in the LDOS (e) and the differential conductance (f) along the cut shown as grey dashed line in (d). In the differential conductance, non-dispersive artifacts due to the setpoint effect (highlighted by red arrowheads) can be seen. The bias voltage of (a-d) is shown as blue dashed line.}
    \label{fig:qpisetpointeffect}
\end{figure}

This effect is particularly transparent in the \ac{FT} of the differential conductance, $\tilde{g}(\bm{q},V; V_0, I_0)$.  Since the denominator in \cref{eq:setpoint_effect} is independent of $V$, the setpoint effect can introduce features whose wavevectors do not disperse with energy, as demonstrated in \cref{fig:qpisetpointeffect}.  The setpoint effect also imposes a simple constraint on the finite-$\bm{q}$ components of the conductance map.  Since the integrated conductance up to the setpoint bias is fixed by the setpoint current, one obtains
\begin{align}
    \int_0^{V_0} \tilde{g}(\bm{q},V; V_0, I_0) \odif{V} = 0 \quad
    (\bm{q}\neq 0)
\end{align}
Thus, any finite-$\bm{q}$ modulation appearing at one energy must be compensated by modulations at other energies within the setpoint bias window.  These compensating modulations can appear with opposite phase and may give rise to apparent features in the differential conductance map.  For this reason, phase inversions between $V=0$ and $V=V_0$ should be interpreted with care.

Likewise, this implies that energy-independent spatial modulations in the LDOS then to be suppressed in $g(\bm{r}, V )$. This is most notably seen for the atomic contrast, which typically appears weaker in the differential conductance, but is often dominant in a calculated LDOS map\cite{kreisel_quasi-particle_2021}. Fig.~\ref{fig:qpisetpointeffect} shows simulated LDOS and differential conductance maps to show the effects of the setpoint effect on QPI measurements. Fig.~\ref{fig:qpisetpointeffect}(a, b) show the reduced atomic corrugation due to the setpoint effect. 

To assess whether features in a $\odv{I}/{V}$ map originate from the setpoint effect, one can either recalculate the map for different setpoint parameters $V_1$ and $I_1$ using \cref{eq:setpoint_effect},
\begin{align}
    g(\bm{r}, V; V_1, I_1)
    =
    g(\bm{r}, V;V_0, I_0)
    \dfrac{I_1}{I(\bm{r}, V_1; V_0, I_0)},
\end{align}
or analyze quantities that are free of the setpoint effect.

Particularly in \ac{QPI} studies of superconductors, one way to mitigate the setpoint effect is to consider ratios of differential conductances at opposite bias voltages, such as
\begin{align}
    Z(\bm{r},V)
    =
    \dfrac{g(\bm{r},+V; V_0, I_0)}{g(\bm{r},-V; V_0, I_0)}
    =
    \dfrac{\rho_\mathrm{s}(\bm{r},+eV)}{\rho_\mathrm{s}(\bm{r},-eV)}.
\end{align}
In this ratio, the position-dependent normalization factor largely cancels.  This procedure successfully enhances \ac{QPI} signals in superconductors, where the relevant quasiparticle states at positive and negative bias are related by particle-hole symmetry~\cite{Hanaguri2007NatPhys,Kohsaka2008Nature}.  A limitation of this approach is that it directly compares the conductance maps at two different energies, $+eV$ and $-eV$.  It is therefore most naturally applied to phenomena that are approximately symmetric about the Fermi energy, but its interpretation can be less straightforward in systems with strong particle-hole asymmetry.  Conversely, a modulation that is strongly suppressed by this normalization satisfies a necessary condition for being a setpoint-induced mirage feature. This observation is relevant to the checkerboard electronic crystal state reported in Ref.~\onlinecite{Hanaguri2004}, whose contrast is largely removed by the conductance ratio~\cite{Hanaguri2007NatPhys}.

A closely related normalization is the Feenstra function~\cite{Feenstra_1994},
\begin{align}
    L(\bm{r},V)
    =
    \dfrac{g(\bm{r},V; V_0, I_0)V}{I(\bm{r},V; V_0, I_0)}
    =
    \dfrac{eV\rho_\mathrm{s}(\bm{r},eV)}{\displaystyle\int_0^{eV}\rho_\mathrm{s}(\bm{r},E)\odif{E}}.
    \label{eq:feenstra}
\end{align}
The Feenstra function normalizes the local spectral weight at energy $eV$ by the spectral weight integrated over the energy window defined by the bias voltage. A quantity proportional to the Feenstra function can be directly measured by recording the differential conductance in constant
current mode (where $V_0=V$ in \cref{eq:feenstra}, compare \cref{sec:qpiimaging}). In contrast to the conductance ratio $Z(\bm{r},V)$, it does not compare positive and negative bias voltages, and therefore does not rely on an approximate symmetry about the Fermi energy.  A practical limitation when calculating \cref{eq:feenstra}, however, is that the normalization involves division by the tunneling current.  The Feenstra function can therefore become sensitive to offsets and noise when the current is small, for example at low bias or inside an energy gap. Also, the physical interpretation of the Feenstra function is less straightforward than of the bare density of states.

Thus, the raw differential conductance, the conductance ratio, and the Feenstra function should be regarded as complementary representations of the same spectroscopic data.  They differ in how the conductance is normalized: the raw conductance retains the setpoint normalization, the conductance ratio compares opposite bias polarities, and the Feenstra function normalizes the conductance by the current at the same bias.  Comparing these representations is useful for assessing whether a feature is robust or strongly affected by the setpoint normalization.

\subsection{Resolution of \ac{QPI} maps}
While the energy resolution of \ac{QPI} is primarily given by the temperature and possibly the amplitude of the lock-in modulation used, the situation is a bit more involved for the $\bm{q}$-space resolution.
\Ac{QPI} is typically analyzed from the \ac{FT} of the differential conductance map $g(\bm{r},V)$. The resolution in $\bm{q}$-space is then determined by the parameters of the map -- the lateral size $x$ of the map sets the smallest spacing $\delta q_x=1/x$ that can be detected and, together with the number of measurement points $n_x$ the maximum scattering vector $q_\textrm{max}=n_x/(2x)$ given by the Nyquist limit. Fourier components with a larger frequency (shorter wave length) will be aliased, an effect that can be corrected for as long as the aliased features do not overlap with features of interest (see \cref{sec:anti-aliasing}). Typically, $n_x$ is set such that for a given size $x$ of the area the atomic speaks remain below the Nyquist limit.

While in principle this would suggest that larger maps per se result in larger resolution, in reality the maximum useful size of a map is given by the decay length of the \ac{QPI} patterns, which is determined, for a point scatterer, by a geometric $1/r$ decay (where $r$ is the distance from a scattering centre) and the spatial decay due to the quasiparticle lifetime. Usually, the intrinsic $1/r$ decay dominates, and the useful size of \ac{QPI} maps is then given by the length scale over which the scattering signal becomes smaller than the signal-to-noise ratio of the differential conductance. Notably, a larger density of measurement points (i.e. larger $n_x$) for a fixed image size results in an increase in the signal to noise ratio in the \ac{QPI} signal in the \ac{FT}, whereas increasing the lateral size at same pixel number decreases it~\cite{chi_quasiparticle_2025}.

\section{Theoretical description}
\label{sec:theory}
\ac{QPI} is surprisingly well described by scattering theory, where the defects are considered point-like and electrons propagate otherwise in an unperturbed material. In the following, we will introduce the scattering theory used to describe \ac{QPI} and the different levels of modelling that have been used to understand \ac{QPI}.

\subsection{Impurity scattering in periodic systems}
We begin with a brief overview of the scattering theory of defects in periodic lattices which forms the basis for understanding \ac{QPI} as measured by \ac{STM}.  For a comprehensive introduction into the scattering theory, we refer the reader to the textbook by Economou~\cite{economou_greens_2006}.
The Hamiltonian that describes such a defect interacting with electrons states of the host system can be written as
\begin{align}
    \hat{H} = \hat{H}_0 + \hat{H}_\mathrm{imp}
\end{align}
where $\hat{H}_0$ is the Hamiltonian of the clean periodic system, and $\hat{H}_\mathrm{imp}$ is the perturbation due to the impurity.  In reality, the impurity Hamiltonian can take many forms depending on the shape and type of impurity under consideration.  For example, a surface can be regarded as an extended impurity~\cite{Pinon_Surface_2020}, as can corrals of defects arranged in specific geometries~\cite{Crommie_QuantumCorrals_1995} and step edges.  However, for the purpose of modelling \ac{QPI} and probing the electronic structure of a material, the most instructive model is often a single point defect.  Such defects may correspond to a missing atom or an atom of the host material substituted by a different species.  In a simple case, $\hat{H}_\mathrm{imp}=\hat{V}\ab(\bm{r}-\bm{r}_\mathrm{imp})$ where the impurity is located at $\bm{r}=\bm{r}_\mathrm{imp}$, and the strength and character of the impurity potential are defined by the matrix $\hat{V}$.

\subsection{Green's function and $T$-matrix formalism}
\label{sec:tmatrix}
To understand how a localized impurity influences the \ac{LDOS}, it is useful to formulate the problem in terms of Green's functions.  We define the Green's function of the unperturbed system $\hat{H}_0$ as
\begin{align}
    \hat{G}_0(\omega) = \ab[\omega-\hat{H}_0]^{-1},
\end{align}
and the Green's function in the presence of the impurity as
\begin{align}
    \hat{G}(\omega) = \ab[\omega-(\hat{H}_0 + \hat{H}_\mathrm{imp})]^{-1}.
\end{align}
These two Green's functions are related by the Dyson equation
\begin{align}
    \hat{G} = \hat{G}_0 + \hat{G}_0\hat{H}_\mathrm{imp}\hat{G}.
\end{align}
Iterating this equation gives
\begin{align}
    \hat{G}
    =
    \hat{G}_0
    + \hat{G}_0\hat{H}_\mathrm{imp}\hat{G}_0
    + \hat{G}_0\hat{H}_\mathrm{imp}\hat{G}_0\hat{H}_\mathrm{imp}\hat{G}_0
    + \cdots.
\end{align}
This expansion describes propagation in the clean system, interrupted by repeated scattering events from the impurity.  The repeated impurity scattering processes can be collected into the $T$-matrix,
\begin{align}
    \hat{T}
    =
    \hat{H}_\mathrm{imp}
    + \hat{H}_\mathrm{imp}\hat{G}_0\hat{H}_\mathrm{imp}
    + \hat{H}_\mathrm{imp}\hat{G}_0\hat{H}_\mathrm{imp}\hat{G}_0\hat{H}_\mathrm{imp}
    + \cdots.
\end{align}
With this definition, the full Green's function can be written compactly as
\begin{align}
    \hat{G} = \hat{G}_0 + \hat{G}_0\hat{T}\hat{G}_0.
\end{align}
When the series for the $T$-matrix converges, it can be summed as
\begin{align}
    \hat{T}
    &=
    \hat{H}_\mathrm{imp}
    \ab[
        \hat{1}
        + \hat{G}_0\hat{H}_\mathrm{imp}
        + (\hat{G}_0\hat{H}_\mathrm{imp})^2
        + \cdots
    ]
    \\
    &=
    \hat{H}_\mathrm{imp}
    \ab[\hat{1} - \hat{G}_0 \hat{H}_\mathrm{imp}]^{-1}.
\end{align}
To express these results in real space, we define
\begin{align}
    \braket< \bm{r} | \hat{G}(\omega) | \bm{r}^\prime >
    &=
    \hat{G}\ab(\bm{r},\bm{r}^\prime,\omega),\\
    \braket< \bm{r} | \hat{H}_\mathrm{imp} | \bm{r}^\prime >
    &=
    \hat{V}\ab(\bm{r},\bm{r}^\prime),
\end{align}
and similarly for $\hat{G}_0$ and $\hat{T}$.
The full Green's function then becomes
\begin{align}
    \hat{G}(\bm{r},\bm{r}^\prime,\omega)
    =
    \hat{G}_0(\bm{r},\bm{r}^\prime,\omega)
    +
    \int\odif{\bm{x}}\odif{\bm{x}^\prime}\,
    \hat{G}_0(\bm{r},\bm{x},\omega)
    \hat{T}(\bm{x},\bm{x}^\prime,\omega)
    \hat{G}_0(\bm{x}^\prime,\bm{r}^\prime,\omega).
\end{align}
For the translationally invariant unperturbed system,
\begin{align}
    \hat{G}_0(\bm{r},\bm{r}^\prime,\omega)
    =
    \hat{G}_0(\bm{r}-\bm{r}^\prime,\omega),
\end{align}
so that
\begin{align}
    \hat{G}(\bm{r},\bm{r}^\prime,\omega)
    &=
    \hat{G}_0(\bm{r}-\bm{r}^\prime,\omega)
    +
    \int\odif{\bm{x}}\odif{\bm{x}^\prime}\,
    \hat{G}_0(\bm{r}-\bm{x},\omega)
    \hat{T}(\bm{x},\bm{x}^\prime,\omega)
    \hat{G}_0(\bm{x}^\prime-\bm{r}^\prime,\omega),
    \label{eq:GF_realspace}\\
    \hat{T}(\bm{r},\bm{r}^\prime,\omega)
    &= \hat{V}(\bm{r},\bm{r}^\prime)
    +
    \int\odif{\bm{x}}\odif{\bm{x}^\prime}\,
    \hat{V}(\bm{r},\bm{x})
    \hat{G}_0(\bm{x}-\bm{x}^\prime,\omega)
    \hat{T}(\bm{x}^\prime,\bm{r}^\prime,\omega).
\end{align}
When the impurity potential is sufficiently weak, the higher-order terms in the $T$-matrix expansion can be neglected.  Keeping only the leading term gives
\begin{align}
    \hat{T}(\bm{r},\bm{r}^\prime,\omega)
    \approx
    \hat{V}(\bm{r},\bm{r}^\prime),
\end{align}
which corresponds to the Born approximation.  Substituting this into \cref{eq:GF_realspace}, the full Green's function becomes
\begin{align}
    \hat{G}(\bm{r},\bm{r}^\prime,\omega)
    \approx
    \hat{G}_0(\bm{r}-\bm{r}^\prime,\omega)
    +
    \int\odif{\bm{x}}\odif{\bm{x}^\prime}\,
    \hat{G}_0(\bm{r}-\bm{x},\omega)
    \hat{V}(\bm{x},\bm{x}^\prime)
    \hat{G}_0(\bm{x}^\prime-\bm{r}^\prime,\omega).
\end{align}

\subsection{\Ac{LDOS} and \ac{QPI} response}
\noindent The \ac{LDOS} is obtained from the real-space Green's function as
\begin{align}
    \rho(\bm{r},\omega)
    =
    -\dfrac{1}{\pi}
    \operatorname{Im}
    \operatorname{Tr}
    \hat{G}(\bm{r},\bm{r},\omega)
    \label{eq:LDOS_rspace}
\end{align}
Taking the \ac{FT} of the \ac{LDOS} gives
\begin{align}
    \rho(\bm{q},\omega)
    &=
    -\dfrac{1}{2\pi i}
    \int\dfrac{\odif{\bm{k}}}{(2\pi)^d}
    \operatorname{Tr}
    \ab[
        \hat{G}(\bm{k},\bm{k}-\bm{q},\omega)
        -
        \hat{G}^*(\bm{k},\bm{k}+\bm{q},\omega)
    ],
    \label{eq:LDOS_qspace}
    \\
    \hat{G}(\bm{k},\bm{k}^\prime,\omega)
    &=
    \hat{G}_0(\bm{k},\omega)\delta_{\bm{k},\bm{k}^\prime}
    +
    \hat{G}_0(\bm{k},\omega)
    \hat{T}(\bm{k},\bm{k}^\prime,\omega)
    \hat{G}_0(\bm{k}^\prime,\omega),
    \label{eq:GF_kspace}
\end{align}
where $d$ is the spatial dimension of the system.  The first term in \cref{eq:GF_kspace} gives the $\bm{q}=0$ component of the \ac{LDOS}, which corresponds to the spatially homogeneous density of states of the unperturbed system.  The second term gives the $\bm{q}\neq 0$ components, which correspond to modulations of the \ac{LDOS} due to scattering from impurities.  These modulations are \ac{QPI} patterns.  The $\bm{q}\neq 0$ component of \cref{eq:LDOS_qspace} gives the \ac{QPI} response,
\begin{align}
    \delta\rho(\bm{q},\omega)
    &=
    -\dfrac{1}{2\pi i}
    \ab[
        \Lambda(\bm{q},\omega)
        -
        \Lambda^*(-\bm{q},\omega)
    ],
    \label{eq:LDOS_QPI}
    \\
    \Lambda(\bm{q},\omega)
    &=
    \int\dfrac{\odif{\bm{k}}}{(2\pi)^d}
    \operatorname{Tr}
    \ab[
        \hat{G}_0(\bm{k},\omega)
        \hat{T}(\bm{k},\bm{k}-\bm{q},\omega)
        \hat{G}_0(\bm{k}-\bm{q},\omega)
    ].
    \label{eq:LDOS_QPI_component}
\end{align}
This expression shows that the Fourier-transformed \ac{LDOS} modulation is generated by scattering processes that connect electronic states with momenta $\bm{k}$ and $\bm{k}-\bm{q}$.  The contribution is therefore enhanced when both Green's functions have substantial spectral weight at the same energy.  Thus, the \ac{QPI} signal encodes information about the electronic structure through scattering vectors $\bm{q}$ connecting states on a constant-energy contour.

In the Born approximation, the $T$-matrix is replaced by the impurity potential.  To illustrate the role of the impurity form factor, let us consider a scalar impurity potential, $\hat{V}=V(\bm{r}-\bm{r}_\mathrm{imp})\hat{1}$.  The corresponding momentum-space matrix element is $e^{-i\bm{q}\cdot\bm{r}_{\mathrm{imp}}} V(\bm{q})$, where $V(\bm{q}) = \int\odif{\bm{r}}\, e^{-i\bm{q}\cdot\bm{r}} V(\bm{r})$ is the \ac{FT} of the spatial profile of the localized impurity potential.  Within the Born approximation, \cref{eq:LDOS_QPI} therefore becomes
\begin{align}
    \delta\rho(\bm{q},\omega)
    \approx
    -\dfrac{1}{\pi}
    e^{-i\bm{q}\cdot\bm{r}_{\mathrm{imp}}}
    V(\bm{q})
    \int\dfrac{\odif{\bm{k}}}{(2\pi)^d}
    \operatorname{Im}
    \operatorname{Tr}
    \ab[
        \hat{G}_0(\bm{k},\omega)
        \hat{G}_0(\bm{k}-\bm{q},\omega)
    ].
    \label{eq:LDOS_QPI_Born_scalar}
\end{align}
The phase factor $e^{-i\bm{q}\cdot\bm{r}_{\mathrm{imp}}}$ reflects the position of the impurity and, in principle, should be summed over all impurities in the sample.  In practice, however, this phase factor can often be neglected in numerical calculations (see also \cref{sec:impurity_density}).  $V(\bm{q})$ acts as a form factor determined by the spatial profile of the impurity potential.  Thus, the \ac{QPI} signal is not determined solely by the electronic structure, but is also weighted by the momentum dependence of the impurity potential.  In general, a short-ranged impurity gives a broad $V(\bm{q})$ in momentum space, whereas a slowly varying or charged impurity suppresses large-$q$ scattering.  For instance, in three dimensions, a screened Coulomb-like potential, $V(r)\propto e^{-\kappa r}/r$, gives $V(q)\propto 1/(q^2+\kappa^2)$.

\subsection{Selection rules}
\label{sec:selection_rules}
The $T$-matrix formalism also provides a useful way to understand how the internal structure of electronic states and impurity potentials gives rise to selection rules in \ac{QPI}. Expanding the unperturbed Green's function in terms of band eigenstates $\ket|\psi_n(\bm{k})>$ with eigenenergies $\epsilon_n(\bm{k})$ gives
\begin{align}
    \hat{G}_0(\bm{k},\omega)
    &=
    \sum_n
    \dfrac{\ket|\psi_n(\bm{k})>\bra<\psi_n(\bm{k})|}{D_n(\bm{k},\omega)},
    \\
    D_m(\bm{k},\omega)
    &=
    \omega - \epsilon_m(\bm{k}) + i\eta
\end{align}
Here, $\eta$ is a positive infinitesimal for the retarded Green's function, or a phenomenological broadening when taken to be finite. Substituting this expression into \cref{eq:LDOS_QPI_component} gives terms of the form
\begin{align}
    \Lambda(\bm{q},\omega)
    &=
    \int\dfrac{\odif{\bm{k}}}{(2\pi)^d}
    \sum_{mn}
    \dfrac{M_{mn}(\bm{k},\bm{k}-\bm{q},\omega)}{
        D_m(\bm{k},\omega)
        D_n(\bm{k}-\bm{q},\omega)
    },
\end{align}
where the matrix element factor is
\begin{align}
    M_{mn}(\bm{k},\bm{k}-\bm{q},\omega)
    &=
    \braket< \psi_m(\bm{k}) | \hat{T}(\bm{k},\bm{k}-\bm{q}, \omega) |\psi_n(\bm{k}-\bm{q})>
    \braket< \psi_n(\bm{k}-\bm{q}) | \psi_m(\bm{k}) >.
\end{align}
The first factor is the scattering amplitude between the two states, while the second factor arises from their overlap in the internal Hilbert space.  If the matrix-element factor vanishes, the corresponding scattering process is suppressed in the \ac{QPI} signal.  Such selection rules may originate from the orbital or spin character of the electronic states, or from the matrix structure of the impurity potential.  In the following, let us consider the Born approximation to illustrate this point.  The $T$-matrix is replaced by the impurity potential, and
\cref{eq:LDOS_QPI} becomes
\begin{align}
    \delta\rho(\bm{q},\omega)
    &\approx
    -\dfrac{1}{\pi}
    \int\dfrac{\odif{\bm{k}}}{(2\pi)^d}
    \sum_{mn}
    \operatorname{Im}
    \ab[
        D_m^{-1}(\bm{k},\omega)
        D_n^{-1}(\bm{k}-\bm{q},\omega)
    ]
    M_{mn}(\bm{k},\bm{k}-\bm{q}),
    \label{eq:LDOS_QPI_Born}
    \\
    M_{mn}(\bm{k},\bm{k}^\prime)
    &=
    \braket< \psi_m(\bm{k}) | \hat{V}(\bm{k},\bm{k}^\prime) |\psi_n(\bm{k}^\prime) >
    \braket< \psi_n(\bm{k}^\prime) | \psi_m(\bm{k}) >.
    \label{eq:matrix_element_factor_Born}
\end{align}
Hereafter, $M_{mn}$ denotes this Born-approximation matrix-element factor unless otherwise stated.

\subsubsection{Spin selection in scalar scattering}
\label{sec:spin_selection}
For a scalar impurity potential,
\begin{align}
    \hat{V}_\mathrm{s} = V(\bm{r}-\bm{r}_\mathrm{imp})\hat{1}
\end{align}
the matrix-element factor, apart from the overall phase factor associated with the impurity position, becomes
\begin{align}
    M_{mn}^{\mathrm{s}}(\bm{k},\bm{k}^\prime)
    =
    V(\bm{q})
    \bigl| \braket< \psi_m(\bm{k}) | \psi_n(\bm{k}^\prime) > \bigr|^2,
\end{align}
where $\bm{k}^\prime = \bm{k}-\bm{q}$.  This factor encodes the overlap of the internal degrees of freedom of the two states. In the special case of a two-component spin state, the projector can be written as
\begin{align}
    \ket|\psi_n(\bm{k})>\bra<\psi_n(\bm{k})|
    =
    \dfrac{1}{2}
    \ab[\hat{1} + \bm{s}_n(\bm{k})\cdot\hat{\bm{\sigma}}]
\end{align}
where $\bm{s}_n(\bm{k})$ is the spin polarization of the state $\ket|\psi_n(\bm{k})>$.  The scalar matrix-element factor then reduces to
\begin{align}
    M_{mn}^{\mathrm{s}}(\bm{k},\bm{k}^\prime)
    &=
    V(\bm{q})
    \dfrac{1 + \bm{s}_m(\bm{k}) \cdot \bm{s}_n(\bm{k}^\prime)}{2}.
\end{align}
Thus, the \ac{QPI} intensity depends on the relative spin angle between the two states.  In particular, if the two states have opposite spin polarizations, the matrix-element factor vanishes and the corresponding scattering process is suppressed in the \ac{QPI} signal.  This type of spin selection rule is particularly relevant for topological insulators~\cite{roushan_topological_2009,Zhang2009, alpichshev_stm_2010,hanaguri_momentum-resolved_2010,Beidenkopf2011,alpichshev_stm_2012}.

\subsubsection{Time-reversal-odd scattering}
\label{sec:time_reversal_odd_selection}
\noindent We now compare the matrix-element factor \cref{eq:matrix_element_factor_Born} with its time-reversal counterpart.  Here $\hat{\Theta}$ denotes the antiunitary time-reversal operator acting on the internal degrees of freedom of the electronic states.  In a time-reversal-symmetric host system, $\hat{\Theta}$ maps an eigenstate at $\bm{k}$ to its time-reversal partner at $-\bm{k}$,
\begin{align}
    \ket|\psi_{\bar{m}}(-\bm{k})>
    =
    \hat{\Theta} \ket|\psi_m(\bm{k})>,
\end{align}
up to an arbitrary phase.  The time-reversal counterpart of $M_{mn}(\bm{k},\bm{k}^\prime)$ is then
\begin{align}
    M_{\bar{n}\bar{m}}(-\bm{k}^\prime,-\bm{k})
    =
    \braket< \psi_{\bar n}(-\bm{k}^\prime) | \hat{V}(-\bm{k}^\prime,-\bm{k}) |\psi_{\bar m}(-\bm{k}) >
    \braket< \psi_{\bar m}(-\bm{k}) | \psi_{\bar n}(-\bm{k}^\prime) >.
\end{align}
Using the antiunitary property of $\hat{\Theta}$ and the Hermiticity of the impurity potential, this can be rewritten as
\begin{align}
    M_{\bar{n}\bar{m}}(-\bm{k}^\prime,-\bm{k})
    =
    \braket<
        \psi_m(\bm{k}) |
        \hat{\Theta}\hat{V}(-\bm{k},-\bm{k}^\prime)\hat{\Theta}^{-1} |
        \psi_n(\bm{k}^\prime)
    >
    \braket< \psi_n(\bm{k}^\prime) | \psi_m(\bm{k}) >.
\end{align}
For an impurity potential that is odd under time reversal,
\begin{align}
    \hat{\Theta} \hat{V}_\mathrm{odd}(-\bm{k},-\bm{k}^\prime) \hat{\Theta}^{-1}
    =
    -\hat{V}_\mathrm{odd}(\bm{k},\bm{k}^\prime).
\end{align}
Therefore,
\begin{align}
    M_{\bar{n}\bar{m}}^\mathrm{odd}(-\bm{k}^\prime,-\bm{k})
    =
    -\braket< \psi_m(\bm{k}) | \hat{V}_\mathrm{odd}(\bm{k},\bm{k}^\prime) | \psi_n(\bm{k}^\prime) >
    =
    -M_{mn}^{\mathrm{odd}}(\bm{k},\bm{k}^\prime).
\end{align}
If the host electronic system is time-reversal symmetric, the corresponding Kramers partners have the same energy.  Therefore, the two scattering processes $(\bm{k}^\prime,n)\rightarrow(\bm{k},m)$ and $(-\bm{k},m)\rightarrow(-\bm{k}^\prime,n)$ have identical energy denominators in the \ac{QPI} response.  Their matrix-element factors, however, have opposite signs.  As a result, their contributions to \ac{QPI} response cancel in the Born approximation~\cite{kohsaka_spin-orbit_2017}.  This is consistent with experimental observations that the \ac{QPI} patterns of topological surface states are not substantially changed by magnetic impurities~\cite{Beidenkopf2011,Strozecka2011}.

\subsubsection{Spin-orbit scattering}
\label{sec:spin_orbit_selection}
A further distinct case is spin-orbit scattering from the impurity.  This can be described by an impurity term of the form
\begin{align}
    \hat{V}_\mathrm{SO}
    =
    \dfrac{\lambda_\mathrm{SO}}{\hbar}
    \hat{\bm{\sigma}}
    \cdot
    \ab[\nabla V(\bm{r}-\bm{r}_\mathrm{imp})\hat{1}\times\hat{\bm{p}}],
\end{align}
where $\lambda_\mathrm{SO}$ parameterizes the strength of the spin-orbit scattering and $\hat{\bm{p}}$ is the momentum operator.  With the convention of $\bm{q}=\bm{k}-\bm{k}^\prime$, the corresponding matrix element factor is
\begin{align}
    M_{mn}^{\mathrm{SO}}(\bm{k},\bm{k}^\prime)
    =
    i\lambda_\mathrm{SO}
    V(\bm{q})
    \braket< \psi_m(\bm{k}) | \hat{\bm{\sigma}}\cdot\ab(\bm{k}\times\bm{k}^\prime) | \psi_n(\bm{k}^\prime)>
    \braket< \psi_n(\bm{k}^\prime) | \psi_m(\bm{k}) >.
\end{align}
Thus, spin-orbit scattering introduces a spin- and momentum-dependent matrix element through the factor $\hat{\bm{\sigma}}\cdot\ab(\bm{k}\times\bm{k}^\prime)$.  The resulting \ac{QPI} signal is sensitive not only to the spin texture of the electronic states, but also to the geometry of the scattering process. This type of scattering is relevant for systems with strong spin-orbit coupling and has been quantitatively demonstrated in a giant Rashba material~\cite{kohsaka_spin-orbit_2017}.

\subsubsection{Superconducting states}
\label{sec:superconducting_selection}
We now extend the matrix-element argument to a superconducting state. For simplicity, we consider a single-band superconductor and use the reduced Nambu basis
$(\hat{c}_{\bm{k}\uparrow}, \hat{c}_{-\bm{k}\downarrow}^\dagger)^\top$, where $\hat{c}_{\bm{k}\sigma}^\dagger$ creates an electron with momentum $\bm{k}$ and spin $\sigma$.  Expanding the Green's function in terms of \ac{BdG} eigenstates gives
\begin{align}
    \hat{G}_0(\bm{k},\omega)
    &=
    \sum_{s=\pm}
    \dfrac{\ket|\psi^{(s)}(\bm{k})> \bra<\psi^{(s)}(\bm{k})|}{D^{(s)}(\bm{k},\omega)},
    \label{eq:GF_Nambu_expansion}
    \\
    D^{(s)}(\bm{k},\omega)
    &=
    \omega - sE(\bm{k}) + i\eta,
    \\
    E(\bm{k})
    &=
    \sqrt{\xi(\bm{k})^2 + \ab|\Delta(\bm{k})|^2},
\end{align}
where $s=+$ and $s=-$ denote the positive- and negative-energy \ac{BdG} branches, respectively.  Here, $E(\bm{k})$ is the positive quasiparticle energy, $\ket|\psi^{(s)}(\bm{k})>$ is the \ac{BdG} eigenstate with energy $sE(\bm{k})$, $\xi(\bm{k})$ is the normal-state electron energy relative to the Fermi level, and $\Delta(\bm{k})$ is the superconducting gap.  In a superconductor, the physical \ac{LDOS} measured by \ac{STM} is obtained from the electron component of the Nambu Green's function.  Therefore, within the Born approximation, the \ac{QPI} matrix-element factor of a superconducting state is
\begin{align}
    M^{(s)}(\bm{k},\bm{k}^\prime)
    =
    \dfrac{1}{2}
    \braket< \psi^{(s)}(\bm{k}) | \hat{V}(\bm{k},\bm{k}^\prime) | \psi^{(s)}(\bm{k}^\prime) >
    \braket< \psi^{(s)}(\bm{k}^\prime) | \ab(\hat{\tau}_0+\hat{\tau}_3) | \psi^{(s)}(\bm{k})>,
\end{align}
where $\hat{\tau}_0$ and $\hat{\tau}_3$ are the identity matrix and a Pauli matrix acting in Nambu space, respectively.  This plays the same role as $M_{mn}(\bm{k},\bm{k}^\prime)$ in the normal state (\cref{eq:matrix_element_factor_Born}), but it is now evaluated using \ac{BdG} eigenstates.

We classify the impurity potential by its parity under time reversal.  Let $V_p(\bm{k},\bm{k}^\prime)$ be the scalar impurity matrix element for the electron component $\hat{c}_{\bm{k}\uparrow}^\dagger \hat{c}_{\bm{k}^\prime\uparrow}$, where $p=\pm 1$ denotes the parity of the impurity potential under time reversal.  The time-reversed counterpart satisfies $V_p(-\bm{k}^\prime,-\bm{k}) = p V_p(\bm{k},\bm{k}^\prime)$.  The corresponding impurity potential in Nambu space is
\begin{align}
    \hat{V}_p(\bm{k},\bm{k}^\prime)
    =
    \begin{pmatrix}
        V_p(\bm{k},\bm{k}^\prime) & 0\\
        0 & -V_p(-\bm{k}^\prime,-\bm{k})
    \end{pmatrix}
    =
    \begin{cases}
        V_p(\bm{k},\bm{k}^\prime) \hat{\tau_3}, \quad p=+1, \\
        V_p(\bm{k},\bm{k}^\prime) \hat{\tau_0}, \quad p=-1.
    \end{cases}
\end{align}

We now evaluate the matrix-element factor for the \ac{BdG} eigenstates.  We write the positive-energy states as $\ket|\psi^{(+)}({\bm k})> = (u_{\bm k}, v_{\bm k})^\top$ and choose the corresponding negative-energy state as $\ket|\psi^{(-)}({\bm k})> = (-v_{\bm k}^*, u_{\bm k}^*)^\top$.  The corresponding matrix-element factors are denoted by $M_p^{(+)}$ and $M_p^{(-)}$, with $p$ defined above, and are given by
\begin{align}
    M_p^{(+)}(\bm{k},\bm{k}^\prime)
    &=
    V_p(\bm{k},\bm{k}^\prime)
    \ab(
        u_{\bm k}^*
        u_{\bm{k}^\prime}
        -p
        v_{\bm k}^*
        v_{\bm{k}^\prime}
    )
    u_{\bm{k}}
    u_{\bm{k}^\prime}^*,
    \\
    M_p^{(-)}(\bm{k},\bm{k}^\prime)
    &=
    V_p(\bm{k},\bm{k}^\prime)
    \ab(
        v_{\bm k}
        v_{\bm{k}^\prime}^*
        -p
        u_{\bm k}
        u_{\bm{k}^\prime}^*
    )
    v_{\bm{k}}^*
    v_{\bm{k}^\prime}.
\end{align}
In a superconductor, the matrix-element factor is weighted by the coherence factors~\cite{Pereg2008PRB,Maltseva2009PRB,Hirschfeld2015PRB,Yamakawa2015PRB}.  To gain further insight, we use the following relationships between the coherence factors:
\begin{align}
    u_{\bm k}u_{\bm k}^*
    &=
    \dfrac{1}{2}
    \ab[
        1 + s \dfrac{\xi(\bm{k})}{E(\bm{k})}
    ],
    \label{eq:coherence_factor_uu}
    \\
    v_{\bm k}v_{\bm k}^*
    &=
    \dfrac{1}{2}
    \ab[
        1 - s \dfrac{\xi(\bm{k})}{E(\bm{k})}
    ],
    \label{eq:coherence_factor_vv}
    \\
    u_{\bm k} v_{\bm k}^*
    &=
    s \dfrac{\Delta(\bm{k})}{2E(\bm{k})}.
    \label{eq:coherence_factor_uv}
\end{align}
With these relations, we can write
\begin{align}
    M_p^{(s)}(\bm{k},\bm{k}^\prime)
    &=
    \dfrac{V_p(\bm{k},\bm{k}^\prime)}{4}
    \ab\{
        \ab[1 + s\dfrac{\xi(\bm{k})}{E(\bm{k})}]
        \ab[1 + s\dfrac{\xi(\bm{k}^\prime)}{E(\bm{k}^\prime)}]
        -
        p \dfrac{
            \Delta(\bm{k}) \Delta^*(\bm{k}^\prime)
        }{
            E(\bm{k}) E(\bm{k}^\prime)
        }
    \}.
\end{align}

As a further simplification, we consider singlet superconductors with time-reversal symmetry, so that $\Delta(\bm{k}) = \Delta({-\bm{k}})$ and $\Delta(\bm{k})=\Delta^*(-\bm{k})$, i.e., the superconducting gap is real, $\Delta(\bm{k}) = \Delta^*(\bm{k})$.  In this case, the matrix-element factor satisfies
\begin{align}
    M_p^{(s)}(\bm{k},\bm{k}^\prime)
    &=
    p M_p^{(s)}(-\bm{k}^\prime,-\bm{k}).
\end{align}
Therefore, for time-reversal-odd impurity scattering ($p=-1$), the \ac{QPI} response is suppressed to first order in $\hat{V}$ due to the cancellation mechanism discussed in \cref{sec:time_reversal_odd_selection}.

For time-reversal-even impurity scattering ($p=+1$), the matrix-element factor is symmetric under time reversal, and the cancellation mechanism does not apply.  Near the Fermi surface, where $\xi(\bm{k})\approx\xi(\bm{k}^\prime)\approx 0$, this simplifies to
\begin{align}
    M_p^{(s)}(\bm{k},\bm{k}^\prime)
    \approx
    \dfrac{V_p(\bm{k},\bm{k}^\prime)}{4}
    \ab[1 - p \operatorname{sgn}(\Delta(\bm{k}) \Delta(\bm{k}^\prime))].
    \label{eq:matrix_element_factor_SC_nearEF_even}
\end{align}
When the dominant contributions to the \ac{QPI} response come from states near the Fermi surface, the coherence factor for time-reversal-even impurity scattering ($p=+1$) enhances the \ac{QPI} response from scattering between regions with opposite gap signs and suppresses that from scattering between regions with the same gap sign.  Since $s$ does not appear in \cref{eq:matrix_element_factor_SC_nearEF_even}, this matrix-element factor remains in phase between positive- and negative-energy contributions.

So far, we have considered $\tau_0$- and $\tau_3$-type impurity potentials.  In superconductors, the impurity potential can also have a $\tau_1$ component, which corresponds to a local modulation of the superconducting gap~\cite{Pereg2008PRB,Maltseva2009PRB,Hirschfeld2015PRB},
\begin{align}
    \hat{V}_{\tau_1}(\bm{k},\bm{k}^\prime)
    =
    V_{\tau_1}(\bm{k},\bm{k}^\prime)
    \hat{\tau}_1.
\end{align}
In this case, the matrix-element factor is
\begin{align}
    M^{(+)}_{\tau_1}(\bm{k},\bm{k}^\prime)
    &=
    V_{\tau_1}(\bm{k},\bm{k}^\prime)
    \ab(
        u_{\bm k}^*
        v_{\bm{k}^\prime}
        +
        v_{\bm k}^*
        u_{\bm{k}^\prime}
    )
    u_{\bm{k}}
    u_{\bm{k}^\prime}^*,
    \\
    M^{(-)}_{\tau_1}(\bm{k},\bm{k}^\prime)
    &=
    -V_{\tau_1}(\bm{k},\bm{k}^\prime)
    \ab(
        v_{\bm k}
        u_{\bm{k}^\prime}^*
        +
        u_{\bm k}
        v_{\bm{k}^\prime}^*
    )
    v_{\bm{k}}^*
    v_{\bm{k}^\prime}.
\end{align}
Using the relationships between the coherence factors (\cref{eq:coherence_factor_uu,eq:coherence_factor_vv,eq:coherence_factor_uv}), we can write
\begin{align}
    M^{(s)}_{\tau_1}(\bm{k},\bm{k}^\prime)
    &=
    \dfrac{sV_{\tau_1}(\bm{k},\bm{k}^\prime)}{4}
    \ab\{
        \ab[1 + s\dfrac{\xi(\bm{k})}{E(\bm{k})}]
        \dfrac{\Delta^*(\bm{k}^\prime)}{E(\bm{k}^\prime)}
        +
        \ab[1 + s\dfrac{\xi(\bm{k}^\prime)}{E(\bm{k}^\prime)}]
        \dfrac{\Delta(\bm{k})}{E(\bm{k})}
    \}.
\end{align}
For a real gap, this reduces near the Fermi surface to
\begin{align}
    M^{(s)}_{\tau_1}(\bm{k},\bm{k}^\prime)
    \approx
    \dfrac{sV_{\tau_1}(\bm{k},\bm{k}^\prime)}{4}
    \ab\{
        \operatorname{sgn}\ab[\Delta(\bm{k})]
        +
        \operatorname{sgn}\ab[\Delta(\bm{k}^\prime)]
    \}.
\end{align}
Thus, the coherence factor for $\tau_1$-type impurity scattering enhances the \ac{QPI} contribution from scattering between regions with the same gap sign and suppresses that from scattering between regions with opposite gap signs.  Because of the prefactor $s$, the matrix-element factor is out of phase between positive- and negative-energy contributions.  These behaviors are opposite to those of time-reversal-even ($\tau_3$-type) impurity scattering.

To discuss the relative phase of the \ac{QPI} response between positive and negative energies in a superconducting state, we also need to consider the contribution from the denominator in \cref{eq:GF_Nambu_expansion}.  The key term is the imaginary part of the product of the denominators, as shown in \cref{eq:LDOS_QPI_Born} for the normal state.  In the superconducting state, the corresponding term satisfies the following relation:
\begin{align}
    \operatorname{Im}
    \ab[
        \dfrac{1}{
            D^{(+)}(\bm{k},\omega)
            D^{(+)}(\bm{k}-\bm{q},\omega)
        }
    ]
    &=
    -\operatorname{Im}
    \ab[
        \dfrac{1}{
            D^{(-)}(\bm{k},-\omega)
            D^{(-)}(\bm{k}-\bm{q},-\omega)
        }
    ].
\end{align}
Therefore, in addition to the matrix-element factor, the contribution from the denominator introduces a $\pi$ phase shift between positive and negative energies.

The above results can be summarized most simply in the Fermi-surface limit, $\xi(\bm{k})\approx\xi(\bm{k}^\prime)\approx 0$, as shown in \cref{tab:QPI_SC_selection}.  In this limit, the time-reversal-even scattering ($\tau_3$) is allowed only for scattering between regions with opposite gap signs, whereas the contribution from the $\tau_1$ term is allowed only for scattering between regions with the same gap sign.  The \ac{QPI} contribution from time-reversal-odd scattering ($\tau_0$) is suppressed by the time-reversal cancellation mechanism. The signs in parentheses indicate the relative phase between positive- and negative-energy \ac{QPI} amplitudes.

The effects due to the coherence factor underlying these selection rules have been discussed in detail in Refs.~\onlinecite{Pereg2008PRB,Maltseva2009PRB,Hirschfeld2015PRB,Yamakawa2015PRB}.  Experimental results consistent with the gap-sign selection summarized here have been reported for a cuprate superconductor~\cite{hanaguri_coherence_2009,gu_directly_2019}.  A broader discussion of phase-sensitive \ac{BQPI}, including results on iron-based superconductors, is given in \cref{sec:phase_sensitive_BQPI}.

\begin{table}
    \centering
    \begin{tabular}{lc@{\hspace{1em}}c}
        \toprule
        Scattering type & \multicolumn{2}{c}{gap-sign relation} \\
        \cmidrule(lr){2-3} & same sign & opposite sign \\
        \midrule
        time-reversal-even ($\tau_3$) & $\times$ & \checkmark $(-)$ \\
        time-reversal-odd ($\tau_0$) & $\times$ & $\times$ \\
        $\tau_1$ & \checkmark $(+)$ & $\times$ \\
        \bottomrule
    \end{tabular}
    \caption{
        Selection rules for the dominant \ac{QPI} contribution in the Fermi-surface limit of a single-band spin-singlet superconductor.  The symbols $\checkmark$ and $\times$ indicate whether the contribution is allowed or suppressed for scattering between regions with the same or opposite gap signs.  The signs in parentheses denote the relative phase between positive- and negative-energy \ac{QPI} amplitudes, with $+$ and $-$ indicating the same and opposite phases, respectively.  For time-reversal-odd scattering, the suppression arises from the time-reversal cancellation mechanism.
    }
    \label{tab:QPI_SC_selection}
\end{table}

\subsubsection{Velocity selection from denominator factors}
\label{sec:velocity_selection}
In addition to the selection rules originating from the matrix-element factors, we now consider the selection imposed by the denominator.  To isolate this effect, consider
\begin{align}
    I_{mn}(\bm{q},\omega)
    =
    \int\dfrac{\odif{\bm{k}}}{(2\pi)^2}
    D_m^{-1}(\bm{k},\omega)
    D_n^{-1}(\bm{k}-\bm{q},\omega).
\end{align}
Here we discuss the two-dimensional case for simplicity.  A large contribution can arise when both denominators become small simultaneously.  We therefore consider a point $\bm{k}_0$ satisfying
\begin{align}
    \epsilon_m(\bm{k}_0) = \epsilon_n(\bm{k}_0-\bm{q}) = \omega.
\end{align}
Writing $\bm{k}=\bm{k}_0+\delta\bm{k}$, and keeping terms linear in $\delta\bm{k}$, we obtain
\begin{align}
    \epsilon_m(\bm{k})-\omega
    &\approx
    \hbar\bm{v}_m(\bm{k}_0)\cdot\delta\bm{k},
    \\
    \epsilon_n(\bm{k}-\bm{q})-\omega
    &\approx
    \hbar\bm{v}_n(\bm{k}_0-\bm{q})\cdot\delta\bm{k},
\end{align}
where $\bm{v}_l(\bm{k})=\nabla_{\bm{k}}\epsilon_l(\bm{k})/\hbar$ ($l=m,n$) is the group velocity.  Thus, near $\bm{k}_0$, the two denominators are controlled by the projections of $\delta\bm{k}$ onto the group velocities.  When the two velocities form a finite angle, the variables $\xi_1=\hbar\bm{v}_m(\bm{k}_0)\cdot\delta\bm{k}$ and $\xi_2=\hbar\bm{v}_n(\bm{k}_0-\bm{q})\cdot\delta\bm{k}$ can be used as local integration variables.  The area element is then
\begin{align}
    \odif{\ab(\delta\bm{k})}
    =
    \dfrac{
        \odif{\xi_1}\odif{\xi_2}
    }{
        \hbar^2\ab|\bm{v}_m(\bm{k}_0)\times\bm{v}_n(\bm{k}_0-\bm{q})|
    }.
\end{align}
This Jacobian factor becomes large as the two velocities approach collinear configuration.  The exactly collinear cases, however, must be treated separately because the above change of variables becomes singular.

For parallel group velocities, we take $\delta k_\parallel$ along the common velocity direction and $\delta k_\perp$ perpendicular to it.  Including the leading curvature term in the transverse direction, the two denominators have the local form
\begin{align}
    D_m(\bm{k},\omega)
    &\approx
    -\hbar \ab|\bm{v}_m(\bm{k}_0)|\delta k_\parallel
    + \alpha_m(\bm{k}_0) \ab(\delta k_\perp)^2
    + i\eta,
    \\
    D_n(\bm{k}-\bm{q},\omega)
    &\approx
    -\hbar \ab|\bm{v}_n(\bm{k}_0-\bm{q})|\delta k_\parallel
    + \alpha_n(\bm{k}_0-\bm{q}) \ab(\delta k_\perp)^2
    + i\eta.
\end{align}
For fixed $\delta k_\perp$, the two poles in the complex $\delta k_\parallel$ plane lie on the same side of the real axis. Therefore, although the velocities are collinear, this configuration does not produce the leading singular contribution.

For antiparallel group velocities, the corresponding local form is
\begin{align}
    D_m(\bm{k},\omega)
    &\approx
    -\hbar \ab|\bm{v}_m(\bm{k}_0)|\delta k_\parallel
    + \alpha_m(\bm{k}_0) \ab(\delta k_\perp)^2
    + i\eta,
    \\
    D_n(\bm{k}-\bm{q},\omega)
    &\approx
    +\hbar \ab|\bm{v}_n(\bm{k}_0-\bm{q})|\delta k_\parallel
    + \alpha_n(\bm{k}_0-\bm{q}) \ab(\delta k_\perp)^2
    + i\eta.
\end{align}
Now the two poles lie on opposite sides of the real $\delta k_\parallel$ axis.  In the small-$\eta$ limit, this gives the leading singular contribution from the denominator factor.

Thus, the denominator part of the \ac{QPI} response is maximized, in this leading singular sense, when the scattering wave vector $\bm{q}$ connects two constant-energy states with antiparallel group velocities, $\bm{v}(\bm{k}) \parallel -\bm{v}(\bm{k}-\bm{q})$.  For example, this condition is satisfied for $\bm{k}$ and $-\bm{k}$ on a circular constant-energy contour, giving rise to a circular \ac{QPI} pattern with radius $2\ab|\bm{k}|$.  Parallel group velocities are also collinear, but they do not produce the same leading contribution because their poles lie on the same side of the integration contour. This effect has been observed in bilayer graphene as the absence of \ac{QPI} signals from parallel group velocities~\cite{Jolie2018}.

\section{Numerical modelling}
Although measuring \ac{QPI} is possible and readily achieved with scanning tunneling microscopy, extracting the electronic structure from such measurements is more challenging. This is due to the fact that the \ac{QPI} only indirectly probes the electronic structure through the scattering vector $\bm{q}=\bm{k}-\bm{k}^\prime$ which does not give a direct one-to-one corrrespondence to the underlying electronic structure, with the added complexity of the aforementioned selection rules. As such, numerical modelling has been essential to compare the electronic structure obtained e.g. from a tight-binding model against the experimental data.

\subsection{Joint density of states}
A simplification often used to model \ac{QPI} is the so-called \ac{JDOS} calculation~\cite{Hoffman_Imaging_2002,mcelroy_relating_2003,markiewicz_bridging_2004}, obtained from the autocorrelation of the electronic band structure or spectral function,
\begin{align}
    \rho_\mathrm{JDOS}(\bm{q},\omega)
    =
    \sum_{\bm{k}}
    \ab[
        \operatorname{Im}
        \operatorname{Tr}
        \hat{G}_0(\bm{k},\omega)
        \cdot
        \operatorname{Im}
        \operatorname{Tr}
        \hat{G}_0(\bm{k}+\bm{q},\omega)
    ].
    \label{eq:JDOS}
\end{align}
Autocorrelation functions encode all possible vectors that can connect two points in the original function, as such this can be viewed as a zeroth order approximation to \ac{QPI}, that calculates all possible scattering vectors allowed by the electronic structure model at a constant energy.

However, while this approximation provides a conceptual picture of what to expect from a \ac{QPI} measurement, it neglects several effects that can qualitatively affect the measured signal.  In particular, \ac{JDOS} calculations keep only the product of spectral weights and neglect terms involving the real part of the Green's function, which arises from the product of two complex Green's functions in \cref{eq:LDOS_QPI_component}.  Consequently, the velocity-selection rule (\cref{sec:velocity_selection}) is absent from \ac{JDOS} calculations.  For the same reason, \ac{JDOS} calculations also neglect selection rules encoded in the matrix-element factors (\cref{sec:spin_selection,sec:time_reversal_odd_selection,sec:spin_orbit_selection,sec:superconducting_selection}).  Some of these selection rules can be introduced ad-hoc by an additional scattering matrix element $\hat{T}(\bm{q},\bm{k})$ to incorporate spin or orbital selectivity of a scattering process~\cite{roushan_topological_2009},
\begin{align}
    \rho_\mathrm{JDOS}(\bm{q},\omega)
    =
    \sum_{\bm{k}}
    \ab[
        \hat{T}(\bm{q},\bm{k})
        \cdot
        \operatorname{Im}
        \operatorname{Tr}
        \hat{G}_0(\bm{k},\omega)
        \cdot
        \operatorname{Im}
        \operatorname{Tr}
        \hat{G}_0(\bm{k}+\bm{q},\omega)
    ].
    \label{eq:tJDOS}
\end{align}
While such \ac{JDOS} calculations can provide valuable insights into the expected \ac{QPI} patterns, mathematically it is not possible to derive the \ac{JDOS} expression \cref{eq:LDOS_QPI} under any algebraic approximation, an inequality that has been discussed previously in the literature~\cite{derry_quasiparticle_2015,kohsaka_spin-orbit_2017}. Therefore, the results of \ac{JDOS} calculations should be considered with caution.

\subsection{Tight-binding formulation}
In order to apply the Green's function method to calculate \ac{QPI} patterns, it is often useful to start from a tight-binding model to describe the underlying electronic structure. The tight-binding models can be written as
\begin{align}
    H(\bm{k})
    =
    \sum_{\ab<l>}
    \hat{t}_l
    e^{i\bm{k}\cdot\bm{g}_l},
    \label{eq:tbhamiltonian}
\end{align}
where the sum is over the nearest neighbours $\ab<l>$, $\bm{g}_l$ are the corresponding lattice vectors and $\hat{t}_l$ the matrices containing the hopping terms (or, for $l=0$, the on-site term). The tight-binding model can either be directly obtained from Slater-Koster integrals~\cite{slater_simplified_1954} and then adjusted to match the spectral function as measure, e.g., by \ac{ARPES}, or obtained from {\it ab-initio} calculations. Using a lattice Green's function based on \cref{eq:tbhamiltonian} and the $T$-matrix formalism, the \ac{dLDOS} can be obtained from \cref{eq:LDOS_rspace}.

One method is to use density functional theory to obtain a tight-binding model. Some \ac{DFT} codes can directly project onto localized orbitals by themselves, alternatively Wannier90~\cite{mostofi_updated_2014,pizzi_wannier90_2020} has developed into a quasi-standard to perform this projection. From the ab-initio calculations, the real-space charge density is explicitly calculated which can then be projected onto localized wavefunctions, using, e.g., localized spherical harmonic orbitals directly. In this way, not only the tight-binding band structure is obtained, but also the underlying basis functions in real space. For 2D materials it accurately captures subtle details associated with the crystal structure, such as bond angle rotation and orbital hybridization. For such a Wannierisation to actually capture the vacuum overlap of the wave functions which is relevant for \ac{QPI} requires slab calculations, including a vacuum within the \ac{DFT} simulation itself.

\subsection{Continuum \ac{QPI}}
\label{sec:cQPI}
The equations above provide valuable insight into the general phenomena giving rise to \ac{QPI} and the information contained in the \ac{QPI} signal, however when it comes to quantitative comparison with experimental data, neither of those methods enable a 1-1 correspondence between model and experimental data. The reason is that the treatment so far has neglected how \ac{STM} actually measures \ac{QPI}. \Ac{STM} measures the \ac{LDOS} as a function of a continuous real space vector, $\bm{r}=(r_x,r_y,r_z)$ and at a height $r_z$ several angstroms above the surface of the sample. The overlap of the wavefunctions associated with the \ac{QPI} scattering and \ac{STM} tip is important to capture this.

Within the tight-binding framework, this poses a challenge.  The lattice Green's function is defined in a discrete orbital basis. Therefore, the \ac{LDOS} obtained directly from a tight-binding calculation is defined only on the discrete lattice and orbital degrees of freedom.  It does not provide a continuous real-space \ac{LDOS} and misses intra-unit-cell variations of the \ac{LDOS}. This becomes in particular an issue where the unit cell contains multiple atoms.  The calculated \ac{QPI} pattern is periodic in reciprocal space, so features outside the first Brillouin zone are folded back into it. Experimental \ac{QPI} patterns, however, can extend beyond the first Brillouin zone and show nontrivial intensity variations.  In addition, the real-space form factors of the orbitals are not included in the lattice \ac{LDOS}, but can strongly affect the intensity of the observable \ac{QPI} wave vectors.  To obtain a more realistic description of \ac{QPI}, it is therefore useful to transform the lattice Green's function into a continuum representation using suitable Wannier functions~\cite{choubey_visualization_2014,kreisel_interpretation_2015}.  The continuum Green's function is written as
\begin{align}
    \hat{G}(\bm{r},\bm{r}^\prime,\omega)
    &=
    \sum_{\bm{R},\bm{R}^\prime}
    \hat{W}_{\bm{R}}(\bm{r})
    \hat{G}(\bm{R},\bm{R}^\prime,\omega)
    \hat{W}^\dagger_{\bm{R}^\prime}(\bm{r}^\prime),
    \label{eq:Wannier_transformation}
    \\
    \hat{W}_{\bm{R}}(\bm{r})
    &=
    \operatorname{diag}\ab[
        w_1(\bm{r}-\bm{R}-\bm{\tau}_1),
        w_2(\bm{r}-\bm{R}-\bm{\tau}_2),
        \dots
    ],
    \label{eq:continuumGF}
\end{align}
where $w_i(\bm{r}-\bm{R}-\bm{\tau}_i)$ is the Wannier orbital associated with orbital $i$ at lattice site $\bm{R}$ and intra-unit-cell displacement $\bm{\tau}_i$.  This representation allows the \ac{LDOS} to be evaluated at arbitrary positions in real space through \cref{eq:LDOS_rspace}, thereby incorporating the real-space structure of the Wannier orbitals into the simulated \ac{QPI} pattern.

The choice of Wannier functions thus becomes an important factor in the accurate simulation of an \ac{STM} measurement. One solution is to use simple approximations for the spherical harmonics e.g. Gaussian- or Slater-type orbitals, that capture the radial decay of the site-centered orbital and the symmetry of the wavefunction being considered. This leaves the radius of these wavefunctions as a free parameter, however suitable values can be chosen based on the distribution of hopping parameters. For example, if only nearest neighbour hopping is considered, Most of the orbital weight will be contained within the distance of the atomic bond.
This scheme, typically using Wannier functions obtained from \ac{DFT} calculations, has successfully been applied to a number of quantum materials~\cite{choubey_visualization_2014,kreisel_interpretation_2015,kreisel_towards_2016,kreisel_quasi-particle_2021,chandrasekaran_engineering_2024,marques_spin-orbit_2024,Armitage_Electronic_2025} and there are codes publicly available to perform the continuum transformation~\cite{wahl_calcqpi_2025,wahl_codebase_2025}.

The Wannier transformation in \cref{eq:Wannier_transformation} can be Fourier transformed to give
\begin{align}
    \hat{G}(\bm{k}, \bm{k}^\prime, \omega)
    &=
    \hat{W}(\bm{k})
    \ab[
        \sum_{\bm{R},\bm{R}^\prime}
        e^{-i\bm{k}\cdot\bm{R}}
        \hat{G}(\bm{R},\bm{R}^\prime,\omega)
        e^{i\bm{k}^\prime\cdot\bm{R}^\prime}
    ]
    \hat{W}^\dagger(\bm{k}^\prime),
    \\
    \hat{W}(\bm{k})
    &=
    \operatorname{diag}\ab[
        e^{-i\bm{k}\cdot\bm{\tau}_1}\tilde{w}_1(\bm{k}),
        e^{-i\bm{k}\cdot\bm{\tau}_2}\tilde{w}_2(\bm{k}),
        \dots
    ],
    \\
    \tilde{w}_i(\bm{k})
    &=
    \int
    w_i(\bm{r})
    e^{-i\bm{k}\cdot\bm{r}}
    \odif{\bm{r}}.
\end{align}
Compared to \cref{eq:LDOS_qspace,eq:GF_kspace}, these equations have two additional terms: the \ac{FT}s of the real-space Wannier functions, $\tilde{w}_i(\bm{k})$, and a spatial phase factor, $e^{i\bm{k}\cdot(\bm{\tau}_i-\bm{\tau}_j)}$, which accounts for the distance between the start and end position of the scattering event.  These terms influence the Fourier-transformed \ac{QPI} image in two important ways.  First, the momentum-space Wannier functions act as a decay envelope, that suppresses scattering vectors at large $\bm{q}$. Depending on the symmetry of the Wannier functions, they can also suppress scattering vectors along particular directions.  Second, the spatial phase factor can enhance or suppress the intensity of particular scattering vectors depending on the relative phase of the contributing terms, which depend sensitively on the interatomic distances.  As a result, the simulation becomes independent of the choice of unit cell, enabling accurate \ac{QPI} calculations for supercell structures.

\begin{figure}
    \includegraphics[width=\textwidth]{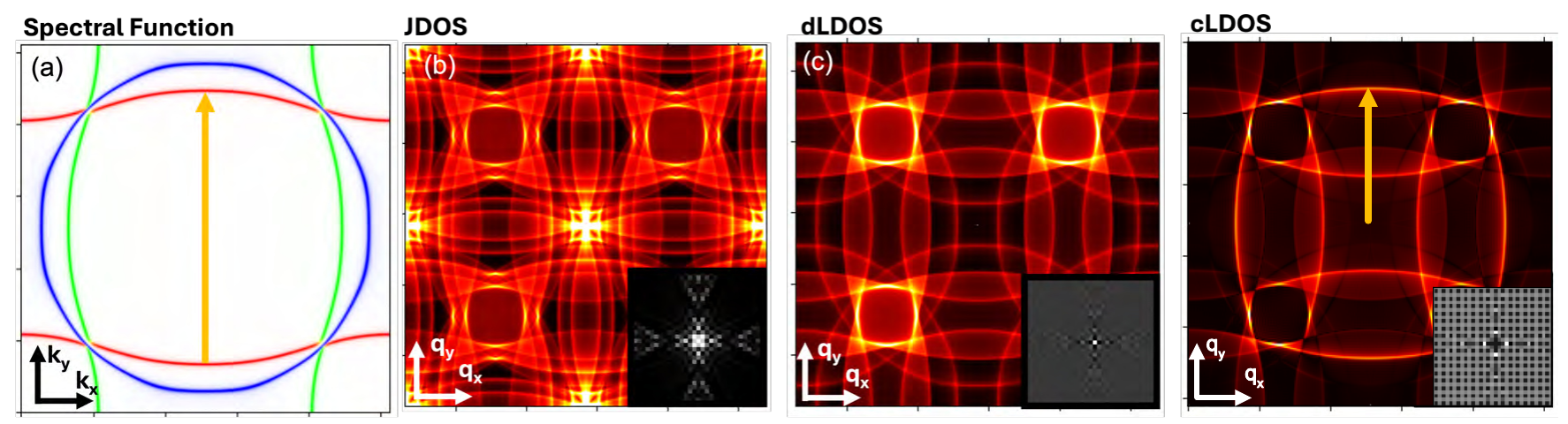}
    \caption{Comparison of the \ac{JDOS}, \ac{FT} of the \ac{LDOS} using a \ac{dLDOS} and \ac{FT} of the \ac{LDOS} using the \ac{cLDOS}. All calculations are performed at the Fermi level with the same underlying tight binding model of a \ac{DFT}-derived \ce{Sr2RuO4} monolayer. (a) Spectral function at the Fermi level for the tight binding model. Blue corresponds to dominant $d_{xy}$ orbital character, red and green correspond to dominant $d_{xz}/d_{yz}$ character, respectively.
    (b) \Ac{JDOS} calculation (\cref{eq:JDOS}) (c) an \ac{LDOS} simulation from a lattice model (discrete \ac{LDOS}, \cref{eq:LDOS_rspace}) (d) \ac{cLDOS} simulation following the continuum transformation (\cref{eq:continuumGF}) . The inset shows a close up of the calculated real-space defect. The orange arrow in (a, d) highlights the dominant scattering vector in the cLDOS simulation. (b, c, d) are all shown between $\pm \frac{2\pi}{a}$.}
    \label{fig:jdosdldoscldos}
\end{figure}

\cref{fig:jdosdldoscldos} shows a side-by-side comparison of the approaches discussed here for a tight binding model for \ce{Sr2RuO4}. The figure shows the spectral function (\cref{fig:jdosdldoscldos}(a)), \ac{JDOS} (\cref{fig:jdosdldoscldos}(b)), \ac{dLDOS} (\cref{fig:jdosdldoscldos}(c)) and \ac{cLDOS} (\cref{fig:jdosdldoscldos}(d)) for the same three-orbital tight-binding model. What is clearly noticeable is that with increased realism of the calculation, more and more scattering terms become suppressed. The \ac{JDOS} exhibits by far the most complex scattering pattern, because it ignores any selection rules for the scattering vectors, most importantly here the orbital selection rules (cf. also \cref{sec:selection_rules}). The \ac{dLDOS} obtained from the lattice Green's function formalism takes care of the orbital selection rules, significantly reducing the allowed scattering vectors, but ignores the matrix element effects associated with the overlap of the electronic wave functions in the sample with those of the tip. The \ac{cLDOS} incorporates these, here suppressing the contribution from some of the bands (notably the bands with predominant $d_{xy}$ orbital character), as well as an overall envelope from the finite spatial extent of the wave functions, resulting in a suppression of \ac{QPI} signal at large $\bm{q}$-vectors.

This route of discretising the full \ac{DFT} electronic structure to the tight binding limit and then interpolating back to the continuum is on one hand computationally much faster than solving the full continuum ab-initio calculation, but moreover has the benefit that the tight-binding model can be adjusted to more closely match the experimental single-particle electronic structure and allows for incorporation of phenomena beyond \ac{DFT}.

Assumptions and hypothesis can now be tested, e.g. by introducing the relevant phenomena at the tight binding level and comparing the consequences for experimental measurements. For examples, including spin-orbit coupling~\cite{kreisel_quasi-particle_2021}, charge density wave orders, ferromagnetic~\cite{naritsuka_compass-like_2023} and anti-ferromagnetic orders~\cite{Armitage_Electronic_2025}, and superconductivity~\cite{bhattacharyya_superconducting_2023,profe_magic_2024}.

The approach thus allows exploring the influence of various terms on the band structure and hence \ac{QPI}. It is particularly useful for strongly correlated electron materials, where \ac{DFT} often does not fully capture the electronic structure and sometimes even not the correct ground state.

The impurity potential is typically the least controlled parameter. It can be estimated from \ac{DFT} calculations by calculating a supercell with a defect and comparing the band structure of the system with and without defect~\cite{berlijn_can_2011,kreisel_towards_2016}, enabling modelling of the \ac{QPI} from a tight-binding model with all parameters fixed from {\it ab-initio} calculations, however this is a comparatively elaborate process. Obtaining the scattering potential from \ac{DFT} calculations enables to explore chemical trends for different impurity species~\cite{kreisel_towards_2016,chi_impact_2016}.

\subsection{Impact of impurity density}
\label{sec:impurity_density}
The scattering theory and $T$-matrix approach outlined in \cref{sec:tmatrix} are typically used to treat a single point like defect, however can rather straightforwardly be extended to multiple impurities. In that case, it includes multiple scattering between the impurities, terms that are, for example, typically included when simulating the standing wave patterns within and around artificial resonator structures~\cite{fiete_colloquium:_2003}. The multi-scattering events typically result only in negligible changes in the density of states, so can safely be discarded. This is likely because typically the scattering potential $V$ is a small perturbation, and so higher order terms in $V$ contribute significantly less. In \ac{QPI} measurements, the random distribution of the defects results in an overall form factor, that can in principle be accounted for a known defect distribution (see, e.g., appendix C of Ref.~\onlinecite{chi_extracting_2017}).
In particular, for initial studies of noble metal (111) surfaces, often \ac{QPI} has been imaged close to extended defects such as step edges because those extended scatterers tend to increase the signal-to-noise ratio of the \ac{QPI}. This approach works well when the dispersion relation can be assumed to be isotropic, so that the anisotropy introduced by an extended scatterer can be ignored, however in general the geometry of the scatterer will affect the intensity distribution of the \ac{QPI} pattern as shown in \cref{eq:LDOS_QPI_Born_scalar}. This means that in general, the interpretation and comparison to theory is more straightforward for point-like scatterers.

\subsection{\textit{Ab-initio} approaches to calculating \ac{QPI}}
\Ac{QPI} can in principle also be calculated directly from Density Functional Theory calculations in a KKR approach or by constructing the Green's function directly from the \ac{DFT}. This approach has been successfully applied, e.g., to the \ac{QPI} of \ce{Bi2Te3}~\cite{rusmann_ab_2021}. The strength of the approach is that it is fully \textit{ab-initio} and allows straightforwardly to include the scattering potential for impurities of different chemical identity. For materials where \ac{DFT} does not capture the electronic structure accurately, for example when electronic correlations are important, it does not offer the same flexibility as an approach via a tight-binding model, where corrections and additional terms can be easily included.




\section{Applications of \ac{QPI}}
In the following, we will provide and overview of applications of \ac{QPI} to a number of systems and materials, starting from simple model systems and the historically first materials where \ac{QPI} has been observed, via a few key results obtained from \ac{QPI} in 2D materials to applications of \ac{QPI} to strongly correlated electron systems and unconventional superconductors, arguably the materials where \ac{QPI} inherently provides the largest advantages compared to other methods for electronic structure determination due to the low energy scales and temperatures required to characterize their ground state.

\subsection{\Ac{QPI} for a nearest neighbour square lattice model system}
\begin{figure}
    \includegraphics[width=\textwidth]{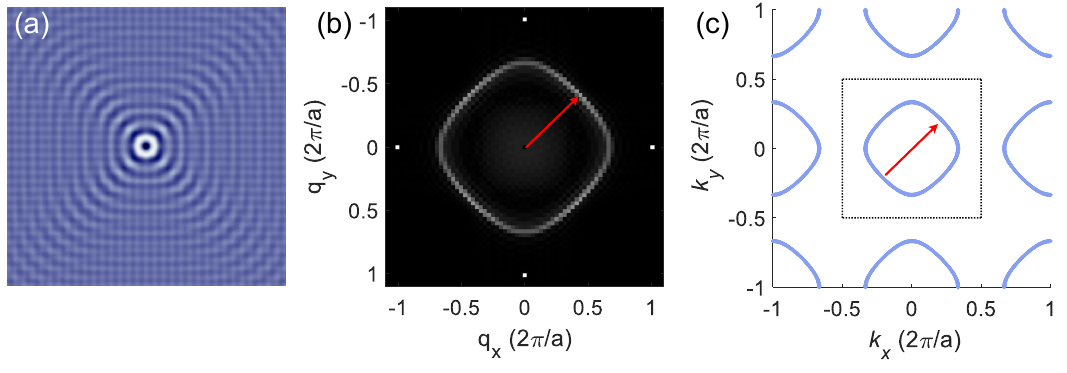}
    \caption{\Ac{QPI} imaging for a simple nearest-neighbour tight-binding model. (a) Real space image of an individual defect in the centre of the image, showing the oscillations in the \ac{LDOS} around the defect due to quantum interference. (b) \ac{FT} of (a), showing a characteristic wave vector. (c) Fermi surface of the same tight-binding model, the dotted square indicates the first Brillouin zone. The red arrow shows the scattering vector as in (b), connecting two parallel sections of the Fermi surface.}
    \label{1nnqpi}
\end{figure}
Before considering experimental \ac{QPI}, to establish the key analysis steps, we discuss the \ac{QPI} of a simple model system, here a two-dimensional nearest-neighbour tight-binding model on a square lattice. \Cref{1nnqpi}(a) shows an \ac{LDOS} map $\rho(\bm{r},eV)$ at a constant height $z_0$ above the surface and calculated within the continuum Green's function approach discussed in \cref{sec:cQPI}. The wave function overlap is assumed to be due to $s$-wave like atomic orbitals centred on the atoms. The \ac{LDOS} map is what would be recorded by the differential conductance in constant height mode in an \ac{STM} experiment.

The typical analysis of \ac{QPI} maps as shown in \cref{1nnqpi}(a) is by analyzing the characteristic scattering vectors in Fourier space, often referred to as $\bm{q}$ space to distinguish it from the $\bm{k}$ space of the wave vectors of the electrons. The \ac{FT} of \cref{1nnqpi}(a) is shown in \cref{1nnqpi}(b) and shows the four atomic peaks due to the square lattice and a rounded square that represents the dominant scattering vector. Experimentally, the ring can be identified as being due to \ac{QPI} from its energy dependence -- the characteristic $\bm{q}$-vector will change as a function of energy. To obtain information about the underlying electronic dispersion, the characteristic $\bm{q}$-vector is compared to a constant energy cut of the band structure, \cref{1nnqpi}(c). For a simple one band model, it is straightforward to identify the dominant nesting vectors that give rise to the \ac{QPI}, and analyzing how they change with energy/bias voltage readily provides information about the dispersion relation. In multi-band systems this typically becomes significantly more complex, and a detailed comparison requires modelling of the scattering vectors. Which vectors are observable depends, e.g., on the orbital overlap between the initial and final state (\cref{sec:selection_rules}), their spin and the spatial decay into the vacuum.

\subsection{Quasi-free two-dimensional electron gases at noble metal surfaces}
\label{sec:noble_metal}
\subsubsection{Surface states}
\begin{figure}
    \includegraphics[width=\textwidth]{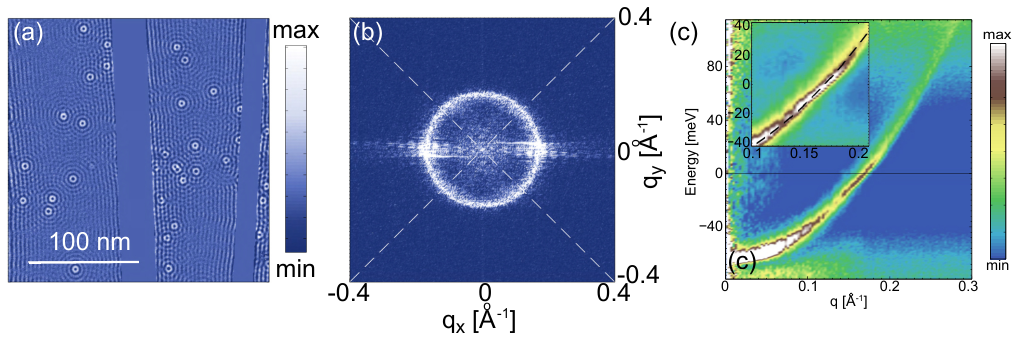}
    \caption{
        Quasi-particle interference imaging of the surface state of Ag(111).
        (a) shows a differential conductance map at the Fermi energy, i.e. $V=\qty{0}{\mV}$, with a lateral size of \qtyproduct[product-units=power]{239 x 239}{\nm}, showing clear quantum interference patterns around point defects as well as along the step edge running vertically through the image.
        (b) \ac{FT} of (a), showing a single dominant ring due to intraband scattering in the surface band.
        (c) Dispersion relation extract from a conductance map, showing a small deviation from the parabolic dispersion close to the Fermi energy which is attributed to self-energy effects due to electron-phonon coupling. Figure adapted from Ref.~\onlinecite{grothe_quantifying_2013}, with permission from American Physical Society.}
    \label{fig:ag111surface}
\end{figure}
Historically, the first observation of \ac{QPI} was for the surface states of the noble metal (111) surfaces~\cite{crommie_imaging_1993,hasegawa_direct_1993,li_local_1997}.  These surface states reside inside a directional band gap of the bulk electronic structure and form a nearly free-electron-like two-dimensional surface band.  Because they lie in a directional band gap, the lifetime of the quasiparticles is comparatively long, resulting in well-resolved quantum interference patterns.  The surface states have mainly  $sp$-derived character, with substantial weight of the wave function extending above the metal surface into the vaccum. As a result, they couple efficiently to the \ac{STM} tip and contribute significantly to the tunneling current and are therefore easily detectable and can even be seen at room temperature~\cite{hasegawa_direct_1993}. The \ac{QPI} can also be detected in point contact measurement, when scanning a contact across the surface~\cite{zhang_experimental_2013}.

\Cref{fig:ag111surface}(a) shows a spatial map of the differential conductance over a large area of a Ag(111) sample, showing clear standing wave patterns near step edges and point defects.  The scattering at step edges has been masked out here to suppress the influence of the form factor of the step edges; nevertheless, even in the far field, the presence of the step edges still affects the \ac{QPI} pattern, as seen in \cref{fig:ag111surface}(b).

For such nearly free-electron surface states, the connection between the Fourier-transformed \ac{QPI} pattern and the band dispersion is particularly direct when the spin splitting remains unresolved in the \ac{QPI} signal.  This is the case for Ag(111), where the Rashba spin splitting is smaller than the resolution of typical \ac{QPI} measurements~\cite{Yaji2018}.  In this effective single-contour picture, the velocity selection rule (\cref{sec:velocity_selection}) selects scattering between states with antiparallel group velocities.  For a nearly circular constant-energy contour, this corresponds to scattering between $\bm{k}$ and $-\bm{k}$, so that the dominant scattering vector is simply $\bm{q}=\bm{k}-(-\bm{k})=2\bm{k}$.  As a result, the measured $\bm{q}$-space dispersion can be directly converted into the band dispersion of the surface state, making the interpretation of the \ac{QPI} particularly straightforward.  Consistent with this picture, the \ac{FT} in \cref{fig:ag111surface}(b) shows a circular pattern whose parabolic dispersion in \cref{fig:ag111surface}(c) agrees well with the band structure expected from \ac{ARPES}.

\subsubsection{Confinement effects in surface states}
The noble metal (111) surfaces not only host a quasi-free 2DEG, but due to their close-packed structure and comparatively low reactivity, adatoms and adsorbates such as CO molecules can be manipulated with the \ac{STM} tip comparatively easily. This is typically achieved by bringing the tip of the \ac{STM} close to the surface, and then using the interaction of the tip apex with the adatom or adsorbate to either push or pull it around. Through this atomic manipulation process, resonator structures can be built which confine the 2D electronic states~\cite{crommie_confinement_1993,braun_engineering_2002}. The scattering in these resonator structures is well captured by scattering theory~\cite{heller_scattering_1994} using the same formalism as introduced in \cref{sec:tmatrix} but including multiple impurities.
In the centre of these resonator structure, the density of states exhibits peaks at the energy of the resonator states~\cite{crommie_confinement_1993}, providing in principle a route to engineer a specific density of states. The resonator states can also be imaged using differential conductance imaging.
Maybe the most spectacular result from these studies is that they allow projecting a spectroscopic feature from one focal point of an elliptical resonator to the other, demonstrated originally for the Kondo resonance of a single cobalt adatom~\cite{manoharan_quantum_2000}. In principle, it can be expected that the modulated density of states in such a resonator also results in a modified many body state, e.g. here in a modified Kondo resonance~\cite{aligia_many-body_2001}. However, the experiments can be explained from scattering theory without the need to invoke such a feedback onto the Kondo state. This suggests that the interaction of the surface state electrons with the many body Kondo state is negligible, and indeed later experiments found that the Kondo state is predominantly due to coupling to the bulk states~\cite{schneider_observing_2002}.

Confinement effects can also be probed between naturally occurring step edges~\cite{burgi_confinement_1998}.

\subsubsection{Probing Fermi liquid effects}
The obvious question is how electron correlations and the formation of Landau quasiparticles affects \ac{QPI}. From Fermi liquid theory, a $1/E^2$ dependence of the lifetime of the quasiparticles is expected, where $E$ is the energy relative to the Fermi energy. One possibility to obtain information about the lifetime is by analyzing the phase coherence length of the \ac{QPI} patterns, for example from scattering at step edges~\cite{burgi_probing_1999, vitali_inter-and_2003} or in resonator structures~\cite{braun_engineering_2002}. In these cases, the spatial decay of the modulation of the density of states is analyzed. This decay typically is dominated by geometric effects, therefore to extract the intrinsic lifetime due to electron-electron or electron-phonon interactions, the lifetime effects need to be separated from the geometric factor.
When performed with sufficiently high resolution in $\bm{q}$-space, which requires the \ac{QPI} to be sampled from comparatively large areas, in principle both the real and imaginary part of the self energy can be assessed, for example to study the change in slope of quasiparticle bands due to interaction with bosonic modes. Such a study was conducted successfully for the surface state of Ag(111), where the renormalization of the dispersion of the surface state in the vicinity of the Fermi energy due to electron-phonon coupling has been detected~\cite{grothe_quantifying_2013} (visible in small deviations of the dispersion close to the Fermi energy (\qty{0}{V}, \cref{fig:ag111surface}(c))).

In principle, also an analysis of the linewidth of the \ac{QPI} features is possible to extract information about the coherence length, and has been successfully carried out in a few cases~\cite{battisti_direct_2020,yim_quasiparticle_2021}. The main challenge is to separate contributions to the width of the \ac{QPI} features from the Fermi surface geometry from self-energy effects, making the interpretation less straightforward than, e.g., for ARPES measurements~\cite{damascelli_angle-resolved_2003}.

\subsubsection{\Ac{QPI} in image potential states}
Among the states observed at metal surfaces are the image potential states, which in the \ac{STM} geometry become states confined between the tip and the sample, forming so-called field-emission resonances. Much like conventional surface states, they show quantum interference patterns near defects which can be imaged by \ac{QPI}~\cite{wahl_quantum_2003,pascual_role_2007}. Different from the surface states, due to the confinement between the tip and the sample, their energies depend on the tip-sample distance, as well as often on the shape of the tip.

\subsubsection{Influence of $k_z$-dispersion}
One of the limitations of \ac{STM} is the restriction of measurements to surfaces and to the two spatial dimensions parallel to the surface. While this restriction is not an issue for the measurement of surface states, monolayer materials or materials with highly anisotropic electronic structure, it poses a challenge for the interpretation of the electronic structure of bulk three-dimensional materials, with non-negligible electronic dispersions in the direction normal to the surface, typically designated the $k_z$-direction.
Nevertheless, the electronic structure of the surface still indirectly encodes the full three-dimensional information about the band structure~\cite{Lounis_Theory_2011}. As a result, with sufficient modelling it has become possible to gain insight into the three-dimensional electronic structures even in the presence of non-negligible out-of-plane dispersion from a \ac{QPI} measurement.

\Ac{QPI} patterns due to subsurface defects were reported for several single crystal metal surfaces, including due to scattering from extended defects at the Al(111) surface~\cite{schmid_quantum_1996} and later from individual Cobalt impurities at different depths near Cu(111) and Cu(100) surfaces \cite{Weismann_Seeing_2009}. The scattering from point defects allows for more straightforward theoretical modelling of the \ac{QPI} patterns because the defect geometry is known. It is found that the real-space propagation, or group velocity, of the \ac{QPI} scattering vectors is directly related to the Fermi velocity of the underlying electronic structure \cite{Lounis_Theory_2011}. Thus, if a state has a finite group velocity in the $k_z$ direction, it will correspond to \ac{QPI} emanating from a defect with a finite $r_z$. This in turn implies that defects below the surface will generate real-space \ac{QPI} that can propagate towards the surface, and produce $k_z$ specific information detectable by \ac{STM}.

For the interpretation of \ac{QPI} of \ce{FeSe}, the \ac{QPI} can be shown to be dominated by scattering vectors connecting parts of the Fermi surface with in-plane group velocity, whereas states with finite group velocity along $z$ quickly lose intensity~\cite{Rhodes_kz_selective_2019}. Thus for interpreting Fourier transformed \ac{LDOS}, for the specific case of \ce{FeSe}, this implies that the \ac{QPI} is dominated by scattering vectors connecting the $k_z=0$ and $k_z=\pi$ planes which agrees with experimental data \cite{Hanaguri_Two_distinct_2018}.

This argument could be generalized in a highly three-dimensional material, galena, which exhibits a rock-salt crystal structure~\cite{Marques_Tomographic_2021}. Experimental measurements revealed \ac{QPI} emanating from defects at energies where no states are present in the $k_z=0$ plane. The \ac{QPI} can be attributed to contributions from states in other $k_z$ planes with zero out-of-plane Fermi velocity. Due to the very three-dimensional nature of the electronic structure, these contributions are highly energy dependent and originate not just from two $k_z$ planes as in the case of \ce{FeSe}, but multiple $k_z$ planes.

Formally, the correct procedure to model \ac{QPI} in three-dimensional systems is to perform calculations that fully take into account the presence of the surface as well as the full three-dimensional electronic structure. This can be achieved from slab calculations as well as {\it ab-initio} schemes~\cite{Weismann_Seeing_2009, rusmann_ab_2021}, or using surface Green's function methods~\cite{Pinon_Surface_2020,rhodes_nature_2023}. By combining these calculations with a continuum calculation that employs Wannier functions to account for the overlap between the \ac{STM} tip and tail of the electronic states into the vacuum~\cite{wahl_calcqpi_2025}, a reasonably realistic modelling of the \ac{QPI} even at surfaces of materials that do not exhibit a quasi-two dimensional electronic structure becomes possible. These calculations reveal that the arguments presented above hold and explicitly how the proximity to the surface influences the electronic states that the defects can scatter with.

\subsection{Probing spin-orbital entanglement}
Over the last twenty years, a multitude of materials have been studied where strong spin-orbit coupling in the bulk results in topologically non-trivial surface states with often complex spin-orbital entanglement. Modelling of these electronic states is often challenging, because calculations need to be done including the spin-orbit coupling, and for surface slabs to capture the surface electronic structure, however, for the classes discussed here -- Rashba surface states, topological insulators and Weyl semimetals -- they are fully captured by \ac{DFT} calculations. \Ac{QPI} has proven to be a powerful tool to confirm the theoretical prediction of these states and connect the high energy electronic structure established from \ac{ARPES} with the low energy electronic structure in the vicinity of the Fermi energy.

\subsubsection{Rashba systems}
\begin{figure}
    \centering
    \includegraphics{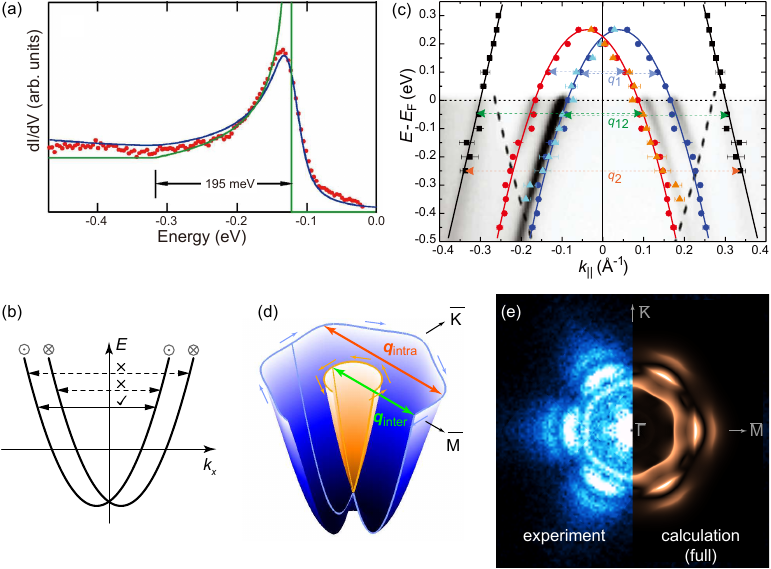}
    \caption{
        Spectroscopic signatures in Rashba systems.
        (a) A $\odv{I}/{V}$ spectrum taken on a Bi/Ag(111) surface alloy, showing a prominent peak originating from a van Hove singularity near the top of the Rashba spin-split band~\cite{ast_local_2007}.
        (b) Schematic of the dispersion of a Rashba system, showing the two branches due to Rashba spin splitting. Three arrows indicate the scattering channels allowed by the velocity selection rule.  The spin selection rule suppresses two of them, leaving only a single allowed scattering channel and the corresponding \ac{QPI} branch, indicated by the solid arrow.
        (c) \Ac{QPI} branches measured on a Bi/Cu(111) surface alloy plotted as points on top of the ARPES measurement, revealing \ac{QPI} signals from Rashba-split bands in the presence of additional spin-degenerate bands~\cite{steinbrecher_rashba-type_2013}.
        (d) Schematic drawing of the Rashba-split electronic structure in BiTeI. The outer branch is hexagonally warped, so that additional scattering channels, $\bm q_\mathrm{intra}$, become available in addition to the interband channels, $\bm q_\mathrm{inter}$.
        (e) Comparison between the measured \ac{QPI} pattern of BiTeI and a numerical simulation including spin-orbit scattering, showing good agreement and confirming the importance of spin-orbit scattering in this system.
        Panels (d) and (e) are adapted from Ref.~\onlinecite{kohsaka_spin-orbit_2017}.
        Panels a, c, d and e reprinted with permission from \cite{ast_local_2007, steinbrecher_rashba-type_2013, kohsaka_spin-orbit_2017}, \copyright (2007, 2013, 2017) American Physical Society.
    }
    \label{fig:qpirashba}
\end{figure}

Rashba-split states are ideal systems for probing spin-orbit-coupled electronic states using \ac{STM} studies. They occur in environments with broken inversion symmetry, as naturally realized at surfaces, and often possess nontrivial spin-orbital textures.  A prototypical example is the Au(111) surface state, where \ac{ARPES} revealed a Rashba splitting on the order of \qty{100}{\mV} near the Fermi energy~\cite{lashell_spin_1996}.  Even larger spin splittings were later found in surface alloys, such as Bi on Ag(111)~\cite{ast_giant_2007}.  These splittings leave clear signatures in the local tunneling spectrum.  In the ordered Bi/Ag(111) surface alloy, tunneling spectroscopy revealed pronounced features in the \ac{LDOS} associated with the van Hove singularity at the band edge of the Rashba-split surface state~\cite{ast_local_2007}.  Fitting this spectroscopic feature provides an estimate of the magnitude of the Rashba spin splitting (\cref{fig:qpirashba}(a)).

In \ac{QPI}, however, the Rashba spin splitting was much less directly visible in early studies. This apparent invisibility provided a prototypical example of how selection rules determine the scattering channels observed in \ac{QPI}.  Although the energy scale of the splitting in Au(111) is large enough that multiple \ac{QPI} branches associated with the two spin-split electronic bands might be expected, \ac{QPI} measurements show only a single ring in the Fourier-transformed \ac{QPI} map~\cite{Petersen1998} and a single parabolic \ac{QPI} dispersion~\cite{hasegawa_direct_1993}.  As shown in \cref{fig:qpirashba}(b), three \ac{QPI} branches are expected from inter- and intra-band scattering processes connecting states with antiparallel group velocities, according to the velocity selection rule (\cref{sec:velocity_selection}).  In a Rashba-split surface state, however, two of these channels are suppressed by the spin selection rule (\cref{sec:spin_selection})~\cite{petersen_simple_2000}.  As a result, only a single spin-allowed scattering channel remains, producing a single ring and a single dispersion.

The visibility of Rashba-split branches in \ac{QPI} is therefore controlled by the compatibility between the velocity- and spin-selection rules.  The suppression of otherwise expected \ac{QPI} signals was demonstrated particularly clearly in \ac{QPI} studies of Bi(110)~\cite{pascual_role_2004}.  Conversely, Rashba-split branches can become visible when additional spin-allowed scattering channels are available.

For example, when an additional spin-degenerate band is present, scattering between that band and the Rashba-split state can give visible signatures of both Rashba-split branches, as shown in \cref{fig:qpirashba}(c)~\cite{steinbrecher_rashba-type_2013}.  This case also illustrates that channels connecting states with parallel group velocities are invisible in the measured \ac{QPI} pattern, consistent with the velocity selection rule (\cref{sec:velocity_selection}).  Another route appears when the Rashba-split contours are warped rather than circular: additional scattering channels can then connect states with antiparallel velocities without connecting opposite spin directions, leading to additional \ac{QPI} branches, as shown in \cref{fig:qpirashba}(d)~\cite{Kohsaka2015}.  Interestingly, the \ac{QPI} signals from the intraband channels in the warped branch ($\bm{q}_\mathrm{intra}$) form the strongest peaks in the $\bar{\Gamma}$--$\bar{\mathrm M}$ direction and are stronger than those from the interband channels ($\bm{q}_\mathrm{inter}$), which appear as the hexagonal ring.  In a scalar-scattering picture, the interband channels would be expected to have larger intensity because they connect states with nearly parallel spin directions.  This reversal of the expected intensity can be explained by spin-orbit scattering (\cref{sec:spin_orbit_selection})~\cite{kohsaka_spin-orbit_2017}.  A numerical simulation including both spin-orbit scattering and the setpoint effect (\cref{sec:setpoint_effect}) quantitatively reproduces the observed \ac{QPI} pattern, including the outermost peaks in the $\bar{\Gamma}$--$\bar{\mathrm M}$ direction that arise from the setpoint effect, as shown in \cref{fig:qpirashba}(e).  This agreement confirms the importance of spin-orbit scattering in this system.

\subsubsection{Topological insulators}
\label{subsectiontopoins}
The interest in topological insulator stems from the realization of topological protection in the surface states. The spin-orbital texture of the surface states means that back scattering is forbidden, similar to the edge states of a quantum Hall system.  The topological surface states are in principle ideal systems for studies by surface spectroscopies since they reside in the surface layer.  The spin-selective scattering which results in the protection against back scattering makes \ac{QPI} sensitive to the topological protection: scattering vectors that would naively be expected to occur become forbidden.  Notably, the protection against backscattering means that for an isotropic state, the scattering vectors spanning the diameter of the circular shape which would typically be the dominant signature are suppressed.
\begin{figure}
   \includegraphics[width=\textwidth]{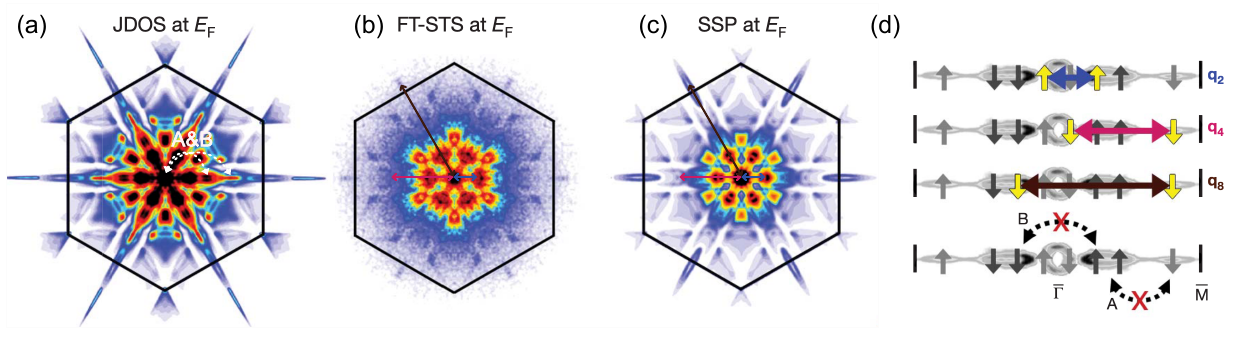}
    \caption{Quasi-particle interference imaging of a topological insulator. (a) \Ac{JDOS}, (b) experimental \ac{QPI} and (c) \Ac{JDOS} accounting for spin selection rules for \ce{Bi_{0.92}Sb_{0.08}}. While the experimental \ac{QPI} pattern (b) is still fairly complex despite the protection against back scattering, comparison with a calculation that does not account for the spin selection rules (a) shows significantly worse agreement compared to a calculation that does account for it (c). (d) shows the dominant scattering vectors allowed by the spin selection rules and visible in (b). The allowed scattering vectors are indicated in (b, c) and the forbidden ones in (a). Figure adapted from Ref.~\onlinecite{roushan_topological_2009}. Reprinted with permission from \onlinecite{roushan_topological_2009}, \copyright (2009) Springer Nature.}
    \label{fig:topoins-bisb}
\end{figure}
The first material in which clear signatures of such spin-selective scattering have been observed is \ce{Bi_{0.92}Sb_{0.08}}~\cite{roushan_topological_2009}, the same material for which topologically protected surface states were reported for the first time from \ac{ARPES}~\cite{hsieh_topological_2008}.  \ce{Bi_{0.92}Sb_{0.08}} does not only exhibit one topological state crossing the Fermi energy but several.  This means that for \ac{QPI}, while the intraband backscattering is forbidden, there are still interband scattering vectors which remain allowed.  As a consequence, to identify the protection against backscattering requires detailed analysis of the scattering vectors (compare \cref{fig:topoins-bisb}).  A \ac{JDOS} calculation based on \cref{eq:JDOS} does not reproduce the experimental pattern well, as shown in Figs.~\ref{fig:topoins-bisb}(a) and \ref{fig:topoins-bisb}(b).  This discrepancy motivated a spin-filtered \ac{JDOS} calculation, \cref{eq:tJDOS}, with the spin selection factor $T(\bm{q},\bm{k})=\bigl|\braket<\bm{S}(\bm{q}+\bm{k}) | \bm{S}(\bm{k})>\bigr|^2$, where $\bm{S}(\bm{k})$ denote the spin state at $\bm{k}$.  The spin-filtered result agrees much better with experiment, as shown in \cref{fig:topoins-bisb}(c).

\ce{Bi2Se3} and \ce{Bi2Te3} are cleaner examples of a topological insulator~\cite{zhang_topological_2009}. In both cases, the bulk exhibits a full band gap in which a single topological surface state with a Dirac cone resides. The corresponding signature in \ac{QPI} is therefore much clearer.

\ce{Bi2Se3} provides an almost ideal case: in \ac{STM} experiments, no clear signature of \ac{QPI} has been reported so far~\cite{hanaguri_momentum-resolved_2010,alpichshev_stm_2012}, consistent with the protection against backscattering.  We note, however, that no \ac{QPI} signal is expected from the topological surface state of \ce{Bi2Se3} even apart from the absence of backscattering.  Since the constant-energy contours of the topological surface state of \ce{Bi2Se3} are nearly circular, the dominant scattering channels expected from the velocity selection rule (\cref{sec:velocity_selection}) are suppressed by the spin selection rule (\cref{sec:spin_selection}).  The very low quasiparticle mass of the topological bands means that they show well-separated Landau levels, and so Landau level spectroscopy can still be used to establish the dispersion of the Dirac cone~\cite{hanaguri_momentum-resolved_2010,cheng_landau_2010}.  While \ce{Bi2Se3} provides a nearly ideal realization of a topological insulator, the closely related compound \ce{Bi2Te3} already shows a more nuanced picture: its topological band is no longer circular but instead exhibits significant warping.  As a consequence, new scattering channels that are not suppressed by the spin selection rule become available and \ac{QPI} signals appear~\cite{Zhang2009,alpichshev_stm_2010}.  The two examples showcase how the information of the matrix elements for scattering encode information about the spin and orbital texture of the band structure and hence ultimately the Berry phase or quantum geometric tensor for more complex systems, making them accessible through \ac{QPI}.

\subsubsection{Weyl semimetals}
\begin{figure}
    \includegraphics{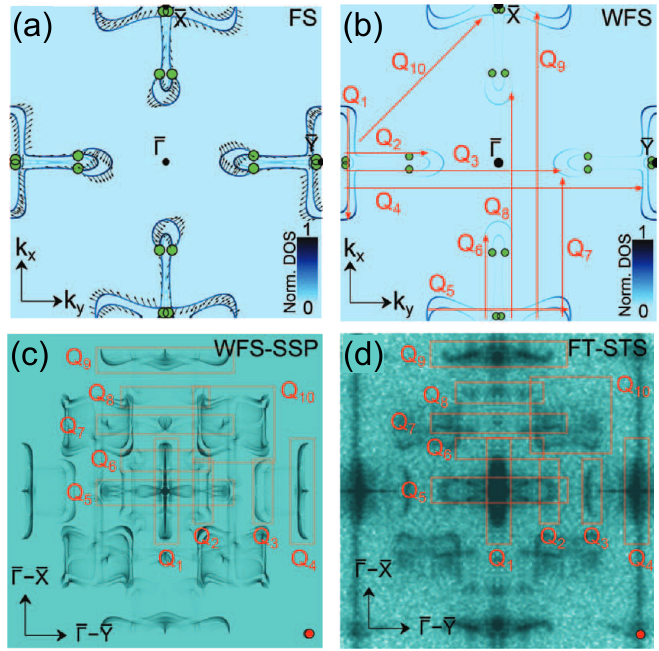}
    \caption{
        Quasi-particle interference imaging of a Weyl semimetal.
        (a) Constant energy contour of the surface electronic structure (projected onto the top-most unit cell of a slab) of TaAs at \qty[print-implicit-plus]{40}{\milli\eV} obtained from \ac{DFT} calculations. Green dots indicate the Weyl points projected onto the surface layer, contours the Fermi arcs and arrows the spin texture.
        (b) Constant energy contour as in (b), now with intensities obtained from projecting onto the top-most As layer to mimic which states contribute most to the tunneling current in an \ac{STM} experiment, orange arrows indicate expected \ac{QPI} wave vectors.
        (c) Weighted \ac{QPI} simulation, accounting for spin selection rules as described by \cref{eq:tJDOS} with the wave vectors from (b) shown.
        (d) \ac{FT} of an experimental \ac{QPI} map, showing good consistency with the calculation in (c) when the spin-selection rules are accounted for.
        Figure adapted from Ref.~\onlinecite{inoue_quasiparticle_2016} with permission, \copyright (2016) American Association for the Advancement of Science.
    }
    \label{fig:weyl}
\end{figure}
The electronic structure of Weyl semimetals exhibits Dirac cones in the bulk electronic structure, the Weyl points.  They occur in pairs and form sources and sinks for the Berry phase.  As a consequence, at the surface, Fermi arcs form which connect these Weyl points.  The Fermi arcs are topologically protected and can be observed in \ac{QPI}~\cite{inoue_quasiparticle_2016,chang_signatures_2016,batabyal_visualizing_2016}.  \Cref{fig:weyl} shows an example of the Fermi surface and \ac{QPI} of a Weyl semimetal.  In \cref{fig:weyl}(a), the surface electronic structure is shown with the Weyl points indicated as green points.  The Weyl points are connected by topologically protected arcs which exhibit a non-trivial spin texture.  For realistic modelling of the \ac{QPI}, the spectral weight on the surface As layer needs to be accounted for, \cref{fig:weyl}(b), which shows that only some parts of the Fermi arcs can be expected to contribute significantly in a \ac{QPI} measurement.  \Cref{fig:weyl}(c) and (d) show the \ac{QPI} calculated from a \ac{JDOS} calculation while accounting for the spin selection rules (c) and the experimental \ac{QPI} (d), showing good agreement.

\subsection{2D Materials}
One of the most powerful applications of \ac{QPI} is its ability to probe local information about the electronic and magnetic properties of monolayers and 2D heterostructures. Systems too thin for many conventional thermodynamic and magnetic measurements, and where surface coverage is not necessarily homogeneous, can challenge conventional spectroscopic methods. With the advancement of techniques such as molecular beam epitaxy, or thin film exfoliation, these materials can be readily synthesized or grown in small islands and flakes, and the characterization of their electronic and magnetic properties via \ac{QPI} has become a routine method. Below we highlight some key examples uncovered by \ac{QPI}.

\subsubsection{Graphene}
\begin{figure}
    \includegraphics{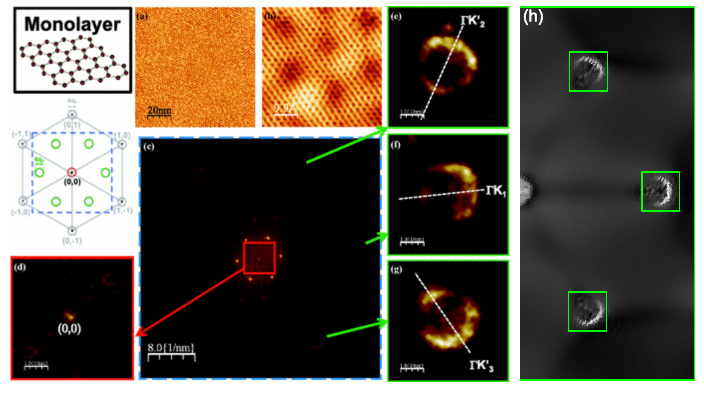}
    \caption{Quasi-particle interference imaging of graphene. (a) Topographic \ac{STM} image of a monolayer of graphene on SiC(0001) with a size of \qtyproduct[product-units=power]{100 x 100}{\nm} taken at $V=\qty{-4}{\mV}$. The lattice visible in the image is due to a moir\'e superstructure between the graphene layer and the SiC, which is clearer visible in a zoom-in (b). (c) shows the \ac{FT} of (a), showing clearly the peaks due to the moir\'e superstructure. No scattering pattern is seen around the centre at $\bm{q}=(0,0)$ (d), while clear half-moon shaped scattering patterns due to the pseudo-spin texture of the electronic structure of graphene are seen in (e-g) around the $(\frac{1}{2},\frac{1}{2})$ points. (h) \ac{cLDOS} calculation of the \ac{QPI} in graphene. Panels a-g adapted from Ref.~\onlinecite{mallet_role_2012} and reprinted with permission, \copyright (2012) American Physical Society.}
    \label{fig:Graphene}
\end{figure}

Graphene is perhaps the most well known and well studied 2D quantum material. Graphene has received significant attention due to its very high purity and conductance, as well as its ability to exhibit the quantum hall effect even at room temperature~\cite{novoselov_room-temperature_2007}. Graphene's low-energy electronic structure is dominated by the Dirac cones around the K-point, bands that have a linear dispersion that exhibit a symmetry protected crossing at half-filling. Due to the two in-equivalent carbon atoms in the unit cell, these Dirac cones have an unusual orbital texture, referred to as a pseudospin \cite{DiVincenzo_psuedospin_1984}. Similar to the protection against backscattering in topological insulators due to the spin texture and spin selection rules~\cite{roushan_topological_2009}, in graphene this pseudospin texture results in protection against backscattering. In \ac{QPI} measurements, this pseudospin is observed as a peculiar half-moon-shaped scattering patterns around the atomic peaks (\cref{fig:Graphene})~\cite{Rutter_Scattering_2007,Brihuega_Quasiparticle_2008,Ugeda_Missing_2010,mallet_role_2012,Sun_Determining_2023}. This half-moon intensity inverts when changing bias across the Dirac point. Importantly, this intensity texture in \ac{QPI} can be reproduced from continuum \ac{QPI} calculations~\cite{Brihuega_Quasiparticle_2008,rhodes_probing_2025}.

\subsubsection{Moir\'e materials}
Measuring free-standing monolayers in \ac{STM} is highly challenging, as the free-standing nature would result in instabilities, as such all 2D materials are measured on top of a substrate, often graphene~\cite{Sun_Determining_2023,saika_flat_2025,Armitage_Electronic_2025}. This almost always results in lattice mismatch which can result in an additional long-range moir\'e superpotential. Recently, moir\'e lattices have received particular attention in twisted bilayer systems due to the opportunity to control bandwidths and interactions based on precise twist angles, inducing a range of strongly correlated phases including unconventional superconductivity and Mott-insulating phases~\cite{cao_correlated_2018,cao_unconventional_2018}.
These moir\'e periodicities can exhibit local variations, structural relaxations and local deformations, which is where \ac{STM} and \ac{QPI} can provide invaluable insight~\cite{Liu_Visualizing_2024}.Furthermore, the small energy scales of the flat bands typically mean that high energy resolution is required to characterize them than is readily available in \ac{ARPES}.

\ac{STM} studies on twisted moiré samples~\cite{li_observation_2010} and in devices~\cite{jiang_charge_2019,choi_electronic_2019,oh_evidence_2021} have been highly successful, for an overview of those work we refer the interested reader to specialized reviews on this topic~\cite{nuckolls_microscopic_2024}. So far, however, the \ac{STM} studies have not addressed the electronic structure via \ac{QPI}.

Due to the large and complex nature of these unit cells, the electronic structure is expected to be complex, yet \ac{QPI} measurements on different twisted systems have revealed patterns often surprisingly similar to the monolayer counterpart. Examples include twisted heterostructures of \ce{NbSe2} on graphene~\cite{naritsuka_superconductivity_2025}, \ce{TiSe2} on graphene~\cite{saika_flat_2025}, and \ce{CrTe2} on graphene~\cite{Armitage_Electronic_2025}. Theoretical calculations on twisted bilayer graphene revealed that as the defect is localised in one layer of the twisted bilayer, the electronic structure in the surrounding environment of the defect is only weakly perturbed by the long-range moir\'e potential~\cite{rhodes_probing_2025}. Nevertheless, the weak perturbation produces distinct signatures in the \ac{QPI} pattern which become most prominent at energies where the scattering vector becomes resonant with the moir\'e potential. Whilst this has not been experimentally verified yet in twisted bilayer graphene, experiments on twisted \ce{NbSe2} and \ce{TiSe2} on graphene confirm this picture~\cite{naritsuka_superconductivity_2025,saika_flat_2025}.

\subsection{Bogoliubov quasiparticle interference}

\ac{BQPI}, originating from quasiparticles in the superconducting state, has played a crucial role in elucidating the momentum-space structure of the often complex superconducting order parameters in unconventional superconductors, such as cuprate and iron-based high-temperature superconductors.
Since the superconducting order parameter directly reflects the nature of Cooper pairs, elucidating its structure is crucial for identifying the mechanisms that mediate the pairing and ultimately stabilize the superconducting state.
While both the amplitude, namely the superconducting gap, and the phase of the order parameter are essentially constant in most phonon-mediated conventional superconductors, unconventional superconductors typically exhibit highly anisotropic gaps, and the phase often changes sign in momentum space resulting in nodes.

\ac{BQPI} imaging is a unique technique that provides both momentum-space resolution and phase sensitivity.
The momentum dependence of the superconducting gap can be inferred from the energy dependence of \ac{BQPI} patterns, in close analogy to normal-state \ac{QPI} that enables the extraction of the underlying band structure.
Phase information can be deduced from coherence effects associated with pairing between states at $\bm{k}$ and $-\bm{k}$ (see also \cref{sec:superconducting_selection}).
Representative applications are discussed in the following subsections.

\subsubsection{Superconducting-gap dispersions}

In the presence of a momentum-dependent order parameter $\Delta(\bm{k})$, the energy dispersion of the Bogoliubov quasiparticles $E(\bm{k})$ is given by
\begin{equation}
	E(\bm{k})=\pm\sqrt{\xi(\bm{k})^2+\Delta(\bm{k})^2},
	\label{eq:Bogolon_spectrum}
\end{equation}
where $\xi(\bm{k})$ denotes the normal-state band dispersion, with the Fermi level at zero.
Since Bogoliubov quasiparticles are coherent superpositions of electron-like and hole-like excitations, $E(\bm{k})$ is particle-hole symmetric about the Fermi level.
This symmetry is generally absent in normal-state single-particle bands and thus helps distinguish \ac{BQPI} from normal-state \ac{QPI}.
At energy $|\omega| \gg |\Delta|$, the electron-hole mixing becomes negligible and the spectral function approaches that of the normal state.

\begin{figure}
\centering
\includegraphics[width=\textwidth]{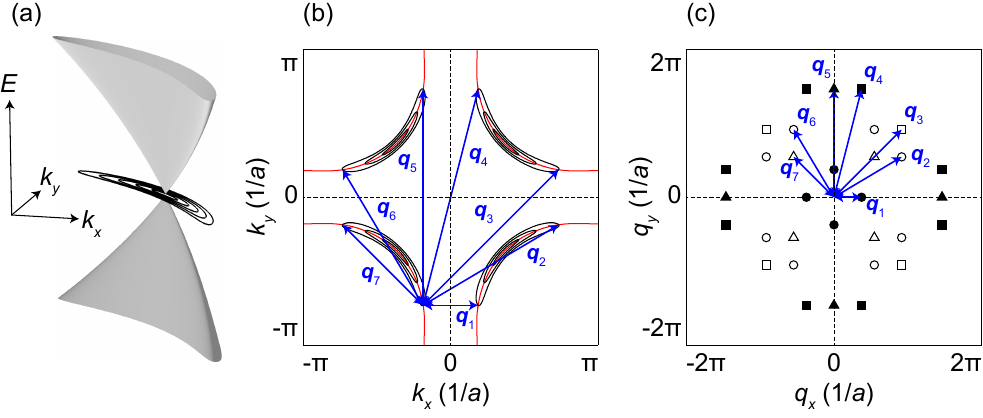}
\caption{
(a)~Schematic illustration of the Bogoliubov quasiparticle dispersion near one of the nodes in a cuprate superconductor.
(b)~Momentum-space electronic structure and \ac{BQPI} scattering vectors (blue arrows) in a cuprate superconductor.
Red and black curves denote the Fermi surface and the constant-energy contour of Bogoliubov quasiparticles, respectively.
(c)~\ac{BQPI} scattering vectors in $\bm{q}$ space accessible via \ac{FT}.
}
\label{fig:octet}
\end{figure}

\Cref{fig:octet}(a) schematically shows $E(\bm{k})$ for a cuprate superconductor where the $d_{x^2-y^2}$-wave symmetry of the order parameter has been established by various experimental techniques~\cite{tsuei_pairing_1996,fong_superconductivity-induced_1997,deutscher_andreevsaint-james_2005}.
As in normal-state \ac{QPI}, $E(\bm{k})$ can be inferred from the \ac{BQPI} patterns, which are associated with scattering between states on the same constant energy contour shown in \cref{fig:octet}(b).
The dominant scattering vectors are expected to connect eight spots at the tips of banana-shaped constant energy contours, where the dispersion becomes shallow and the spectral weight is enhanced~\cite{Hoffman_Imaging_2002,mcelroy_relating_2003,wang_quasiparticle_2003}.

\begin{figure}
\centering
\includegraphics[width=\textwidth]{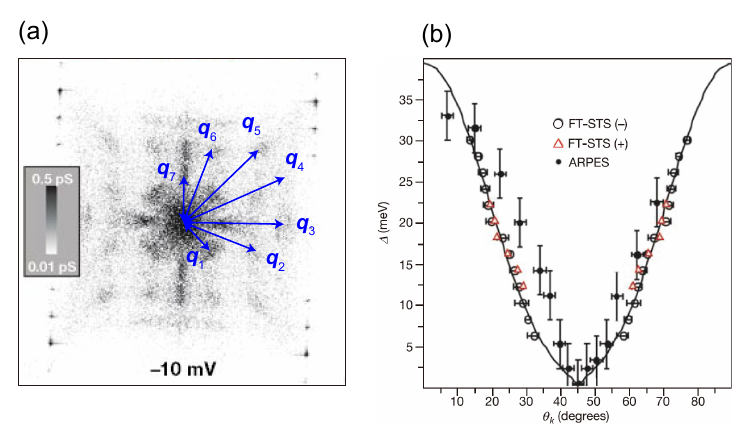}
\caption{
(a)~Experimentally observed \ac{BQPI} pattern in \ce{Bi2Sr2CaCu2O_{8+\delta}} at \qty{-10}{\milli\eV}.
(b)~Superconducting-gap dispersion $|\Delta(\bm{k})|$ deduced using the octet model.
$\theta_{\bm{k}}$ denotes the Fermi surface angle about $(\pi, \pi)$.
Figure adapted from Ref.~\onlinecite{mcelroy_relating_2003}, reprinted with permission from \onlinecite{mcelroy_relating_2003}, \copyright (2003) Springer Nature.
}
\label{fig:Bi2212BQPI}
\end{figure}

This so-called octet model predicts a set of seven scattering vectors $\bm{q}_i$ ($i=1 \sim 7$) in $\bm{q}$ space [\cref{fig:octet}(c)], all of which are observed in the Fourier-transformed \ac{BQPI} patterns [\cref{fig:Bi2212BQPI}(a)].

The octet model enables the extraction of the superconducting-gap dispersion $|\Delta(\bm{k})|$.
From the observed octet scattering vectors at a given energy $\omega = eV$, the positions of the tips of the banana-shaped constant-energy contours in the Brillouin zone can be determined.
This energy-momentum relation directly yields $|\Delta(\bm{k})|$.
Continuum Green's function calculations have shown that, despite its phenomenological joint-density-of-states basis, the octet model correctly captures the dominant scattering vectors~\cite{kreisel_interpretation_2015}.

The gap dispersions obtained from \ac{BQPI} patterns of \ce{Bi2Sr2CaCu2O_{8+\delta}}~\cite{mcelroy_relating_2003} and \ce{Ca_{2-x}Na_xCuO2Cl2}~\cite{Hanaguri2007NatPhys} are consistent with a $d_{x^2-y^2}$-wave superconducting gap and semi-quantitatively reproduce $|\Delta(\bm{k})|$ measured by \ac{ARPES}, except that apparent gapless regions tend to appear near the nodes in the \ac{BQPI} results~\cite{Kohsaka2008Nature}.
This apparent discrepancy can be resolved by recent comprehensive \ac{BQPI} simulations, incorporating self-energy effects~\cite{Sakai2026PRX}.

The concept of the octet model can generally be applied to superconductors with an anisotropic gap.
Since $\Delta(\bm{k})$ varies much more slowly in momentum space than the band dispersion $\epsilon(\bm{k})$, the constant-energy contours of $E(\bm{k})$ that enclose the gap minima are highly elongated, with well-defined tips that evolve along the Fermi surface with energy.
The scattering vectors connecting these tips are expected to dominate the \ac{BQPI} patterns, allowing the extraction of $|\Delta(\bm{k})|$.

While cuprate superconductors constitute a single-band system, the analysis scheme of the octet model can be transferred to multi-band superconductors as well, provided that intra-band scattering vectors are correctly identified.
Superconducting gaps in iron-based superconductors~\cite{allan_anisotropic_2012, sprau_discovery_2017, Nag2025NatPhys} and the heavy-fermion superconductor \ce{CeCoIn5}~\cite{allan_imaging_2013} have been investigated.
In particular, the gap amplitude, anisotropy, and their relative orientations in momentum space have been separately identified for different Fermi surfaces of LiFeAs and FeSe~\cite{allan_anisotropic_2012, sprau_discovery_2017}.
\begin{figure}
\centering
\includegraphics[width=\textwidth]{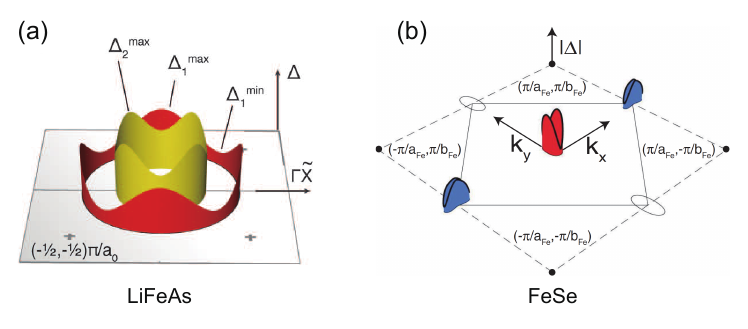}
\caption{
Superconducting-gap dispersions in multi-band iron-based superconductors deduced from \ac{BQPI} experiments.
(a)~LiFeAs, figure adapted from Ref.~\onlinecite{allan_anisotropic_2012}.
(b)~FeSe, figure adapted from Ref.~\onlinecite{sprau_discovery_2017}.
Reprinted with permission from \cite{allan_anisotropic_2012,sprau_discovery_2017} (\copyright (2017,2021) by the American Association for the Advancement of Science).
}
\label{fig:IBS_BQPI}
\end{figure}

Although $|\Delta(\bm{k})|$ can be investigated more directly by \ac{ARPES}, \ac{BQPI} offers several advantages.
For instance, since the attainable temperature is lower and the energy resolution is higher in \ac{STM} than in \ac{ARPES}, \ac{BQPI} imaging can be performed on superconductors with transition temperatures as low as a few kelvin~\cite{allan_imaging_2013}.
In addition, \ac{BQPI} enables the investigation of magnetic-field effects on the superconducting gap~\cite{hanaguri_coherence_2009}.
The magnetic field plays a role in accessing the phase information.

\subsubsection{Phase-sensitive \ac{BQPI}}
\label{sec:phase_sensitive_BQPI}

While the analysis of \ac{BQPI} patterns within the framework of the octet model provides information on the gap amplitude $|\Delta(\bm{k})|$, it remains challenging to determine the phase structure of the superconducting order parameter.
For example, a highly anisotropic $s$-wave order parameter with a constant phase in momentum space may produce \ac{BQPI} patterns similar to those expected for a sign-changing $d$-wave order parameter.
Accessing phase information is essential for elucidating the pairing mechanism, because superconductivity arising from unconventional repulsive interactions, such as those mediated by spin fluctuations, generally results in a sign-changing order parameter.

\begin{figure}
\centering
\includegraphics[width=0.5\textwidth]{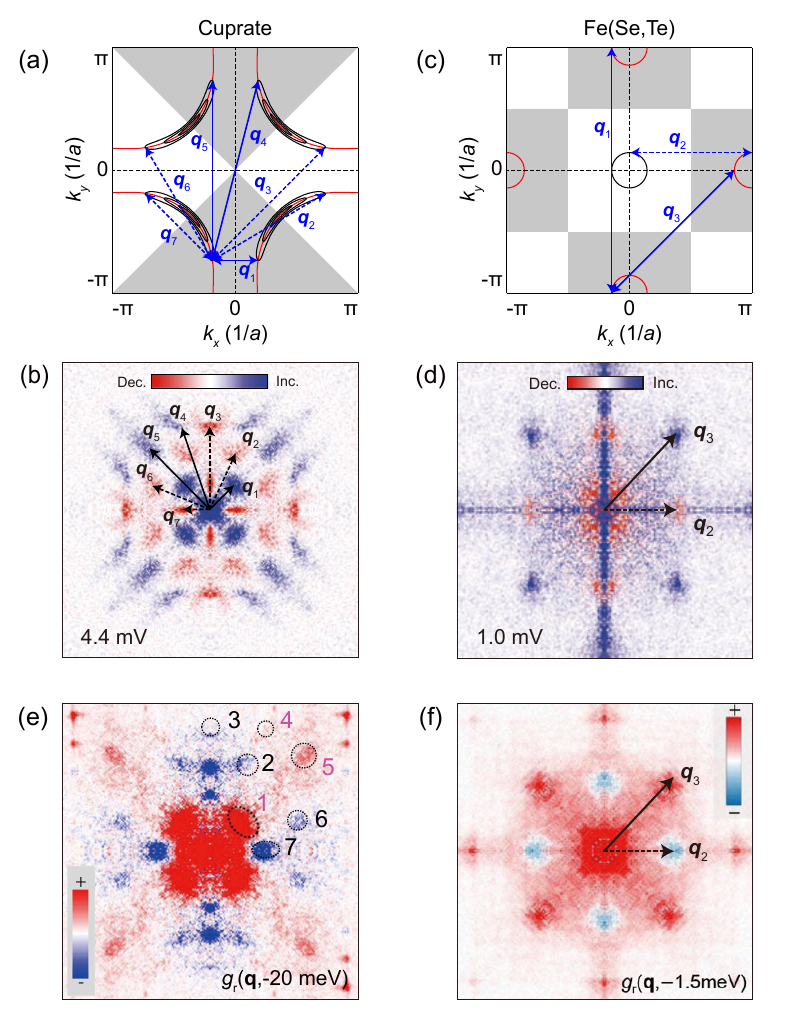}
\caption{
Phase-sensitive \ac{BQPI} experiments.
(a)~The sign structure (gray and white background) of the $d_{x^2-y^2}$-wave order parameter in a cuprate superconductor, together with the momentum-space electronic structure and \ac{BQPI} scattering vectors shown in \cref{fig:octet}(b).
(b)~Magnetic-field induced changes in the \ac{BQPI} intensity in \ce{Ca_{2-x}Na_xCuO2Cl2}.
Blue and red indicate enhancement and suppression by the magnetic field, respectively.
Figure adapted from Ref.~\onlinecite{hanaguri_coherence_2009}.
(c)~The sign structure (gray and white background) of the $s_\pm$-wave order parameter expected in iron-based superconductors.
Black and red circles schematically represent the hole and electron Fermi surfaces, respectively.
(d)~Magnetic-field induced change in the \ac{BQPI} intensity in Fe(Se,Te).
Figure adapted from Ref.~\onlinecite{hanaguri_unconventional_2010}.
(e)~Phase-referenced \ac{QPI} results for \ce{Bi2Sr2CaCu2O_{8+\delta}}.
Figure adapted from Ref.~\onlinecite{gu_directly_2019}.
(f)~Phase-referenced \ac{QPI} results for Fe(Se,Te).
Figure adapted from Ref.~\onlinecite{Chen2019PRB}.
Note that in (b) and (d), blue and red correspond to sign-preserving and sign-reversing scattering signals, respectively, whereas this color assignment is reversed in (e) and (f). Panels a, b reprinted with permission from \cite{hanaguri_coherence_2009}, d with permission from \cite{hanaguri_unconventional_2010}, \copyright (2009, 2010) by the American Association for the Advancement of Science. e reprinted from \cite{gu_directly_2019} under CC-BY 4.0, f reprinted from 
\cite{Chen2019PRB} with permission, \copyright (2019) American Physical Society.}
\label{fig:phase_BQPI}
\end{figure}

\ac{BQPI} becomes phase sensitive through coherence effects in the Bogoliubov quasiparticle scattering processes from $\bm{k}_i$ to $\bm{k}_f$ as discussed in \cref{sec:superconducting_selection}.
There are two types of Bogoliubov quasiparticle scattering processes: sign-preserving scattering, in which the signs of the order parameter at $\bm{k}_i$ and $\bm{k}_f$ are the same, and sign-reversing scattering, in which they are different.
For example, among the seven scattering vectors $\bm{q}_i$ ($i=1 \sim 7$) in the octet model for $d_{x^2-y^2}$-wave cuprates, $\bm{q}_1$, $\bm{q}_4$, and $\bm{q}_5$ correspond to sign-preserving scattering whereas the others are sign-reversing scattering [\cref{fig:phase_BQPI}(a)].

Depending on the nature of the scatterers, the corresponding coherence factors can favor sign-preserving or sign-reversing scattering.
Therefore, when the dominant contribution to the \ac{BQPI} signal can be associated with a particular type of scattering process, the observed \ac{BQPI} pattern itself provides phase information.
However, as shown in \cref{fig:Bi2212BQPI}(a), \ac{BQPI} signals are equally observed at all $\bm{q}_i$ in \ce{Bi2Sr2CaCu2O_{8+\delta}}.
This observation suggests that, in real samples, multiple types of scattering centers activate both sign-preserving and sign-reversing scattering processes, thereby suppressing the phase sensitivity of \ac{BQPI}.

The application of a magnetic field is expected to recover the hidden coherence effects by introducing vortices as scattering centers~\cite{hanaguri_coherence_2009,Pereg2008PRB,Maltseva2009PRB}.
The suppression of the superconducting gap within the vortex core may induce $\tau_1$-type scattering, in addition to the $\tau_0$-type scattering arising from the circulating supercurrent.
While the latter has little impact on QPI, the former selectively enhances sign-preserving scattering, as discussed in \cref{sec:superconducting_selection}.
This idea has indeed been demonstrated in the cuprate \ce{Ca_{2-x}Na_xCuO2Cl2}~\cite{hanaguri_coherence_2009}.
As shown in \cref{fig:phase_BQPI}(b), sign-preserving and sign-reversing scattering signals exhibit contrasting magnetic-field dependence: the former is enhanced whereas the latter is suppressed.
The suppression of sign-reversing scattering may originate from the Doppler energy shift induced by the circulating supercurrent.

This method has been applied to the iron-based superconductor \ce{Fe(Se,Te)}, in which the sign structure of the order parameter on different Fermi surfaces have been debated~\cite{hanaguri_unconventional_2010}.
The leading candidate is $s_\pm$-wave superconductivity, in which the order parameter changes sign between the hole-like Fermi surface at the center of the Brillouin zone and the electron-like Fermi surfaces at the zone boundary [\cref{fig:phase_BQPI}(c)].
The observed \ac{BQPI} signal associated with the scattering between the hole and electron Fermi surfaces is suppressed by a magnetic field, whereas that associated with scattering between the electron Fermi surfaces is enhanced, consistent with $s_\pm$-wave pairing [\cref{fig:phase_BQPI}(d)].

Although \ac{BQPI} imaging under a magnetic field helps to recover hidden coherence effects, several issues remain.
First, since a magnetic field suppresses superconductivity itself, this method is applicable only to superconductors with a sufficiently high upper critical field.
Second, the distinction based on magnetic-field dependence is phenomenological and lacks solid support from microscopic theory.
In particular, this limitation becomes critical in multi-band systems such as iron-based superconductors, where the use of coherence factors as scattering matrix elements is not fully justified~\cite{Hirschfeld2015PRB,Yamakawa2015PRB}.

By analyzing a general two-band model theoretically, Hirschfeld \textit{et al.} proposed a robust method to extract the sign structure of the order parameter from \ac{BQPI} data~\cite{Hirschfeld2015PRB}.
They focused on the relationship between the spatial modulations of \ac{BQPI} patterns at positive and negative energies and pointed out that these modulations are out of phase for sign-reversing scattering, whereas they are in phase for sign-preserving scattering.
This method has been used to demonstrate the sign reversal between the hole and electron Fermi surfaces in FeSe~\cite{sprau_discovery_2017}.

To analyze experimental data systematically, Chi \textit{et al.} introduced a scheme known as phase-referenced \ac{QPI}~\cite{chi_extracting_2017,Chi2017arXiv2}.
To incorporate the phase difference $\theta(\bm{q},V)$ between the \ac{BQPI} modulations at symmetric energies $\pm eV$ with respect to the Fermi level, a quantity $|\tilde{g}(\bm{q},V)|\cos\theta(\bm{q},V)$ is analyzed instead of the simple amplitude of the \ac{FT} from the differential conductance, $|\tilde{g}(\bm{q},V)|$.
In phase-referenced \ac{QPI}, the signal becomes negative when the modulations between the filled and empty states are out of phase, whereas it becomes positive when they are in phase.
These correspond to sign-reversing and sign-preserving scattering processes, respectively.
Phase-referenced \ac{QPI} analyses have been successfully applied to \ce{Bi2Sr2CaCu2O_{8+\delta}}~\cite{gu_directly_2019} [\cref{fig:phase_BQPI}(e)] and Fe(Se,Te)~\cite{Chen2019PRB} [\cref{fig:phase_BQPI}(f)] and can potentially be applied to a wide range of superconductors, as this is a field-free technique.

\subsection{\Ac{QPI} on putative triplet superconductors}
As shown in the previous section, \ac{QPI} is in principle an ideal technique to detect unconventional pairing symmetries from Bogoliubov \ac{QPI}, however in the case of candidate triplet superconductors, the evidence has been ambiguous so far. In this section, we discuss some of the leading candidate materials and describe the current state of research into detecting evidence for triplet pairing.

\subsubsection{\ce{Sr2RuO4}}
The strontium ruthenates adopt a perovskite crystal structure and form a Ruddlesden-Popper series of compounds with composition \ce{Sr_{n+1}Ru_nO_{3n+1}}. The crystal structure is closely related to that of the cuprate high-temperature superconductors. For this reason, the discovery of superconductivity in the $n=1$ member \ce{Sr2RuO4}\cite{maeno_superconductivity_1994} generated significant interest due to the hope that understanding superconductivity in this material might also uncover the origin of high-temperature superconductivity in the cuprate. Shortly after, it was suggested that it is also a triplet superconductor,\cite{ishida_spin-triplet_1998}, a claim refuted more recently~\cite{pustogow_constraints_2019}. The material exhibits a well-defined cleavage plane and high-quality surfaces. In the normal state, \ac{QPI} shows an electronic dispersion consistent with what is seen in \ac{ARPES}, with strong signatures of electron correlations~\cite{wang_quasiparticle_2017}. The clean surface exhibits a surface reconstruction~\cite{matzdorf_ferromagnetism_2000} that appears to suppress superconductivity,\cite{barker_stm_2003,kambara_scanning_2006,marques_magneticfield_2021} with only one report suggesting that a superconducting gap can be detected on a clean surface~\cite{sharma_momentum-resolved_2020}. The photoemission~\cite{morales_hierarchy_2023} and \ac{QPI} data~\cite{marques_magneticfield_2021} obtained from surfaces of \ce{Sr2RuO4} has allowed for a detailed comparison of the electronic structure between the two techniques and with electronic structure calculations, with an agreement between the experimental techniques within a few meV,\cite{chandrasekaran_engineering_2024} providing a validation of modelling of \ac{QPI}. The low energy electronic structure is found to exhibit multiple van Hove singularities in the vicinity of the Fermi energy. However, due to the absence of clear signatures of a superconducting gap, a resolution of the superconudcting order parameter by \ac{BQPI} has not been possible, and likely requires some modification of the surface~\cite{profe_magic_2024,valadkhani_why_2024}.

\subsubsection{\ce{UTe2}}
Most recently, the unconventional superconductor uranium ditelluride has been extensively studied by \ac{STM}. There are reports of \ac{QPI}\cite{wang_odd-parity_2025,aishwarya_visualizing_2026}, yet the interpretation is complicated by the surface cut. Some of the reported \ac{QPI} has been recorded with a superconducting tip~\cite{wang_odd-parity_2025}, which further complicates the interpretation. The \ac{QPI} has been interpreted as evidence for triplet pairing~\cite{wang_odd-parity_2025}, however is so far lacking the phase sensitivity that has enabled determining the symmetry of the order parameter in the cuprate and iron-based superconductors. Comparison to theory does at this point not provide a clear interpretation~\cite{christiansen_quasiparticle_2025}.

\subsection{Heavy fermions in strongly correlated materials}
\subsubsection{Heavy fermion materials}
The electronic structure of heavy fermion materials is particularly challenging to study: the flat bands and large quasiparticle masses up to one thousand times the free electron mass observed in many of these means that a very high energy resolution is required to resolve the band structure, often exceeding the energy resolution achievable in \ac{ARPES}. At the same time, many heavy fermion materials exhibit complex phase diagrams at low temperatures, with Fermi liquid behaviour often only setting at temperatures well below \qty{10}{\K}. \Ac{QPI} is therefore in principle an ideal tool to study the electronic structure of heavy fermion materials, because it allows for measurements at temperatures well below \qty{1}{\K}, with an energy resolution down to \qty{10}{\micro\eV} as well as in magnetic fields.

Studies of the electronic structure at the surface becomes challenging by multiple complicating factors: (1) The band structure of heavy fermion materials is often rather complex with multiple bands crossing the Fermi energy and the heavy bands arise from hybridization with typically the rare earth $f$-states. The localized nature of the $f$ bands can make detection in \ac{QPI} challenging. (2) Heavy fermion materials are intermetallic compounds some of which have rather complex crystal structures which will exhibit multiple surface terminations and often not cleave easily. (3) Some $f$-elements will change their valence dependent on the chemical environment, as a consequence of which some heavy fermion materials will exhibit significant differences in the surface electronic structure compared to the bulk. (4) The electronic structure is often rather three-dimensional, which complicates interpretation of the features seen in \ac{QPI} and often makes them rather broad.

Nevertheless, for a few heavy fermion material, \ac{QPI} has provided valuable insights into the low energy electronic structure and imaging of the \ac{BQPI}.

\subsubsection{Heavy band formation and surface states in \ce{URu2Si2}}
\begin{figure}
    \includegraphics[width=\textwidth]{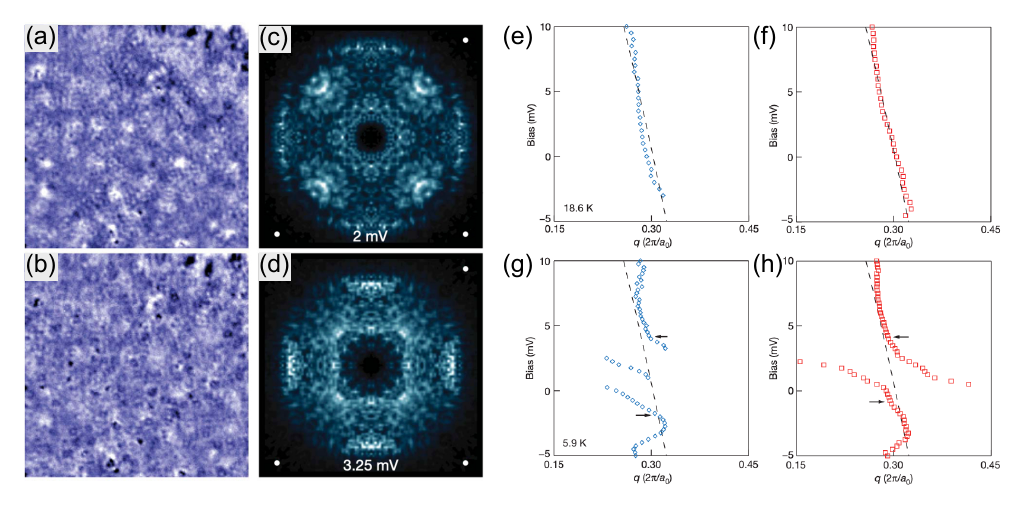}
    \caption{\Ac{QPI} in the heavy fermion material \ce{URu2Si2}. (a, b) Real space differential conductance maps of the \ac{QPI}; (c, d) corresponding \ac{FT}s. (e, f) show the dispersion relation at a temperature above the hidden order transition, and (g, h) below. For the dispersion below the hidden order transition, and avoided crossing can be observed, resulting in the formation of heavy bands. Figure adapted from Ref.~\onlinecite{schmidt_imaging_2010}.}
    \label{fig:uru2si2-qpi}
\end{figure}

The first heavy fermion material for which \ac{QPI} was observed is \ce{URu2Si2}. \ce{URu2Si2} is a heavy fermion superconductor, with a large electronic contribution to the specific heat. Quantum oscillations show effective masses in the bulk between $8m_e$ and $25m_e$\cite{ohkuni_fermi_1999,hassinger_similarity_2010} and a hidden order phase~\cite{palstra_superconducting_1985,mydosh_hidden_2014}, i.e. a phase where specific heat shows a clear phase transition at about \qty{17.5}{\K}, but there is currently no consensus on what the order parameter is. The material becomes superconducting below $\sim\qty{1.5}{\K}$~\cite{palstra_superconducting_1985,schlabitz_superconductivity_1986}. \Ac{QPI} of \ce{URu2Si2} shows rather dramatic changes within a narrow energy range (compare \cref{fig:uru2si2-qpi}(a-d))\cite{schmidt_imaging_2010}, as expected for a large effective mass. Tracing out the dominant scattering vectors at a temperature above and below the hidden order transition shows the formation of a heavy band and formation of a hybridization gap (\cref{fig:uru2si2-qpi}(e-h))~\cite{schmidt_imaging_2010,aynajian_visualizing_2012}. The heavy band reaches an electron mass of $28m_e$\cite{schmidt_imaging_2010}, remarkably close to what has been obtained from quantum oscillations~\cite{ohkuni_fermi_1999,hassinger_similarity_2010}. The \ac{QPI} appears to, however, only show one band crossing the Fermi energy, not capturing the whole band structure. This is not unexpected for a multi-band system and suggests that some of the other bands either have a strong dispersion in the direction normal to the surface, effectively blurring them out, or an orbital character that make their contribution to the tunneling current negligible.

\begin{figure}
    \includegraphics[width=0.5\textwidth]{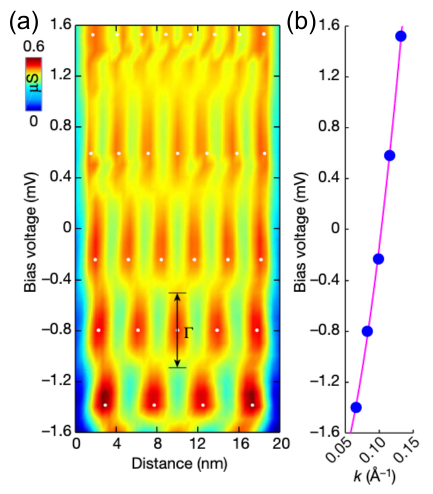}
    \caption{Quantum confinement at the surface of \ce{URu2Si2}. (a) Line cut of spectra across a nano-meter-sized terrace, showing the maxima due to confinement of the heavy electron states on the terrace. (b) From the dimensions of the terrace and the number of maxima and energies of the resonance states, the dispersion relation can be recovered. Figure adapted from Ref.~\onlinecite{herrera_quantum-well_2023} and reused under CC-BY 4.0.}
    \label{fig:uru2si2-confinement}
\end{figure}

Subsequent work showed that the surface of \ce{URu2Si2} also exhibits a surface state~\cite{herrera_quantum-well_2023}, which gives rise to quantum confinement effects (compare \cref{fig:uru2si2-confinement}) as previously seen in the surface states of noble metals. These resonant states show energy differences that are with splittings below \qty{1}{\milli\eV} orders of magnitude smaller than in the case of the surface states of noble metals, reflecting an  effective mass of $m_*\sim 17m_e$, comparable to what is expected from bulk measurements.

\subsubsection{Superconductivity in \ce{CeCoIn5}}
In subsequent studies of the heavy-fermion superconductor \ce{CeCoIn5}, also information about the symmetry of the superconducting order parameter was obtained,\cite{allan_imaging_2013,zhou_visualizing_2013} however without a full mapping of the momentum-space structure of the superconducting gap as was achieved in the cuprate and iron-based superconductors.

Because \ac{QPI} only images a two-dimensional projection of the electronic structure and due to the three-dimensionality of the electronic structure of these materials, features in \ac{QPI} remain typically comparatively broad~\cite{Marques_Tomographic_2021,rhodes_nature_2023}. Still, from \ac{QPI} maps, the dispersion relation of a heavy band can be traced which forms right at the Fermi energy~\cite{allan_imaging_2013}. In the energy range of the superconducting gap, i.e. within \qty{1}{\milli\eV} of the Fermi energy, evidence for the gapping out of states with formation of nodes, suggesting a $d$-wave structure of the order parameter. Similar conclusions about the symmetry of the superconducting order parameter have been obtained by considering the formation of Andreev bound states~\cite{zhou_visualizing_2013}.

\subsubsection{Correlated oxides}
\Ac{QPI} has been applied to a range of transition metal oxides apart from the cuprate superconductors, notably \ce{Sr2RuO4}\cite{kreisel_quasi-particle_2021}, \ce{Sr2RhO4}\cite{battisti_direct_2020}, \ce{Sr3Ru2O7}\cite{lee_heavy_2009} and \ce{Sr4Ru3O10}\cite{marques_spin-orbit_2024} as well as the delafossite oxides~\cite{yim_quasiparticle_2021,mazzola_tuneable_2022,zheng_exchange_2026}. In the Ruddlesden-Popper strontium oxides, a natural cleavage plane exists between the strontium oxide layers which results in a charge neutral surface. The electronic structure in these materials is highly two dimensional even in the bulk, resulting in well-defined \ac{QPI}.

\begin{figure}
    \includegraphics[width=\textwidth]{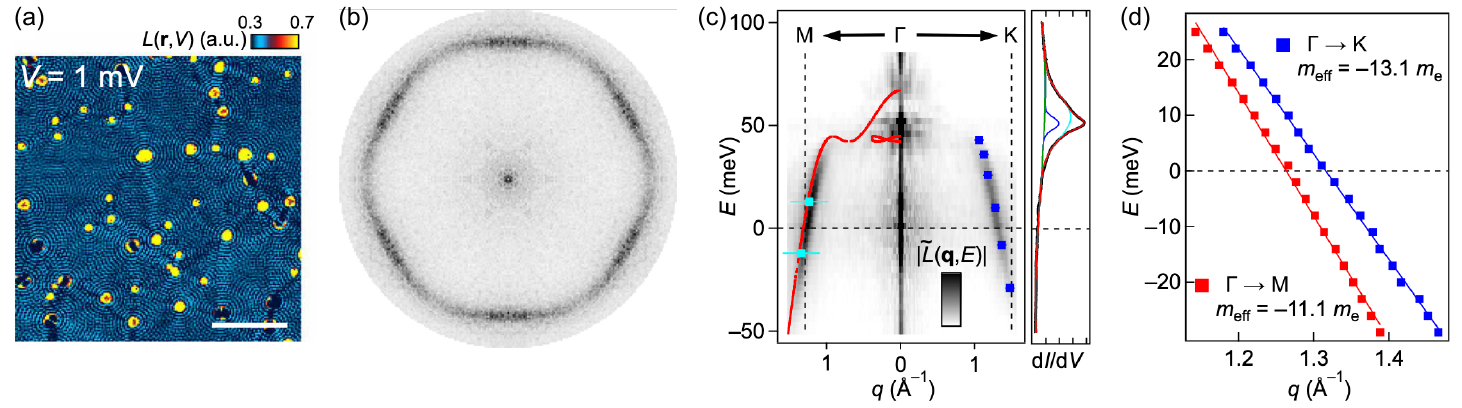}
    \caption{\ac{QPI} at the surface of \ce{PdCoO2}. (a) Spatial map of $L(V)$ of the \ac{QPI} in \ce{PdCoO2}, showing very strong \ac{QPI} signal. (b) \ac{FT} of (a). (c) Cut through the \ac{QPI} map, showing good agreement with a model which takes into account the spin-selection rules in quasiparticle scattering. (d) extracted dispersion relation in the vicinity of the Fermi energy, showing that the bands have rather heavy effective masses of more than ten free electron masses. Figure adapted from Ref.~\onlinecite{yim_quasiparticle_2021} and reused under CC-BY 4.0.}
    \label{fig:pdcoo2-qpi}
\end{figure}

By contrast in the delafossite oxides, cleavage results in two distinct surfaces which are polar and hence show an electronic structure and sometimes even a ground state that differs from that of the bulk. A notable example is the delafossite oxide \ce{PdCoO2}: for the Pd-terminated surface, the electronic structure as determined by \ac{ARPES}\cite{mazzola_itinerant_2018} as well as by \ac{QPI}\cite{mazzola_tuneable_2022,zheng_exchange_2026} suggests that the surface becomes ferromagnetic, while the \ce{CoO2}-terminated surface hosts a metallic two-dimensional electronic state with a huge Rashba spin splitting and comparatively large effective mass in the surface layer~\cite{sunko_maximal_2017}. \Cref{fig:pdcoo2-qpi} shows the \ac{QPI} of this spin-split surface state. The \ac{QPI} shows a heavy electron mass of $m^*\sim 12m_e$ and a dispersion consistent with \ac{ARPES}, however apart from that the \ac{QPI} obeys the usual spin selection rules expected for a Rashba system. The two-dimensional nature of the band and sharp \ac{QPI} allows assessing also the lifetime and phase coherence length of the electronic state from an analysis of the width of the \ac{QPI} signal. The analysis shows a behaviour that is consistent with Fermi liquid theory for a 2D electron gas. A similar consistency between \ac{ARPES}, \ac{QPI} and with Fermi liquid theory has been found for the case of \ce{Sr2RhO4} \onlinecite{battisti_direct_2020}.

\section{Future Directions}
\subsection{\Ac{QPI} as a quantitative tool to study electronic states in quantum materials}
\subsubsection{Quantitative description of \ac{QPI}}
The present review shows that the theoretical understanding of \ac{QPI} has almost reached a quantitative level, where corrections due to, e.g., correlation effects or spin-orbit scattering become relevant and their effect can be characterized. Finer details of the \ac{QPI} can be analysed in terms of their origin from comparison with calculations, and where needed the models can be adjusted until a quantitative match is achieved. For multi-band systems this can still be challenging, but at least in principle has become a possibility. The high energy resolution of \ac{QPI} means that often subtle structural details of the surface can become important to establish a full understanding of the \ac{QPI}.

\subsubsection{Capturing electronic correlation effects}
Modelling of \ac{QPI} so far typically starts from a single particle description, either in the form of a tight-binding model or from \ac{DFT}. In a few cases, Fermi liquid effects~\cite{vitali_inter-and_2003,battisti_direct_2020,yim_quasiparticle_2021} are accounted for, assuming, e.g., an $E^2$ dependence of lifetime broadening.  For correlated materials, typically a phenomenological band renormalization is included (see, e.g., Ref.~\onlinecite{kreisel_quasi-particle_2021}). A more quantitative approach would be to introduce correlation effects, e.g., from Dynamical Mean Field Theory. This approach has frequently been pursued to compare the spectral function from \ac{ARPES} with calculations. Through inclusion of a realistic self energy from DMFT in \ac{QPI} calculations, a fully ab-initio approach to modelling of \ac{QPI} comes within reach and will enable a more quantitative comparison between \ac{QPI} calculations and experiment. Such approaches have recently been demonstrated for cuprates~\cite{Sakai2026PRX} and  \ce{Sr2RuO4}\cite{rhodes_revealing_2026} by some of us, demonstrating simulation of \ac{QPI} from models accounting more accurately for many-body effects and realizing a fully ab-initio approach even in strongly correlated electron materials. However, at the moment, application to materials with more complex band structures remains numerically demanding.

\subsection{Functionalized tips for spin and orbital selectivity}
\subsubsection{Spin and orbital selective tips}
The discussion so far has focused on tips which are point-like metallic tips, and assumed to have a featureless density of states. These assumptions make the interpretation of the \ac{QPI} significantly easier, however one can also use modified tips to make them sensitive for specific states, probing particular aspects of the electronic structure and quantum states of the sample. Such functionalized tips can either project onto particular orbital states or particular spin states.
To achieve orbital selectivity likely requires attaching a molecule to the tip, where the orbital at the apex would then define the projection.

Using spin-polarized tips for \ac{QPI} enables spin-resolved measurements of the band structure, making the \ac{QPI} more sensitive to scattering vectors connecting constant energy contours of a particular spin character. The spin-polarization of the tip affects the matrix element for tunneling between tip and sample, suppressing features with opposite direction of the spin. There is so far relatively little discussion of spin-polarized \ac{QPI} in the literature~\cite{trainer_relating_2022}.

In some cases, significant tip-dependence of the \ac{QPI} has been reported~\cite{sprau_discovery_2017}, however the tip apex has so far not been separately controlled. One of the challenges is that controlled preparation of the tip at the surface of quantum materials tends to be more challenging than, e.g., on noble metal surfaces, lacking typically established procedures to controllably pick up atoms of molecules as is routinely done on metal surfaces.

\subsubsection{Superconducting tips}
In tunneling spectra, for example for measuring Andreev bound states~\cite{franke_competition_2011}, superconducting tips offer superior energy resolution. With a superconducting tip, the resolution is not limited by the Fermi broadening of the tip, but by the sharpness of the coherence peak. Such increased energy resolution is also desirable for \ac{QPI}, however for comparison with calculations requires deconvolution of the tunneling spectrum with that of the tip or, alternatively, accounting for the single particle density of states of the tip, complicating the interpretation of the data. Superconducting tips have notably employed in the study of \ce{UTe2}~\cite{wang_odd-parity_2025}.

\subsection{Magnon \ac{QPI}}
As discussed briefly in \cref{sec:inelastictunneling}, the tunneling current does not only consist of elastically tunneling electrons, but also electrons that tunnel inelastically and lose energy before entering the sample. Usually this contribution is neglected, which is justified in many cases, and the often good agreement between the expected and measured \ac{QPI} confirms this expectation a posteriori. However, in strongly correlated electron materials or also magnetic materials, it is not obvious that this is a reasonable assumption, and early on it was proposed that the overall v-shape of the tunneling spectrum of high temperature cuprate superconductors may be due to contributions from inelastic tunneling~\cite{kirtley_inelastic-tunneling_1990}.
It is indeed the case that vibrational excitations, for example in molecules, result usually in comparatively weak features in inelastic tunnling with step-like increases on order \qty{10}{\percent} (see, e.g., Refs.~\onlinecite{lauhon_single-molecule_1999,heinrich_molecule_2002}). The same is, however, not true for spin-flip excitations, which can result in significant changes in the differential conductance by \qtyrange[range-units=single]{30}{50}{\percent}~\cite{heinrich_single-atom_2004,loth_spin-polarized_2010}.
There is mounting evidence that in strongly correlated electron materials, coupling to inelastic modes, in particular spin excitations or fluctuations, make an important contribution to the overall tunneling current with traces in the spectrum that can be rationalized by coupling to the spin fluctuation spectrum~\cite{chi_imaging_2017,hlobil_tracing_2017,chi_quasiparticle_2025}.

This raises the question what information can be extracted from spatial variations in that inelastic signal, what it reveals about the electronic \ac{QPI}\cite{fransson_surface_2007,chi_quasiparticle_2025} and whether information about the bosonic dispersion can be extracted~\cite{mitra_magnon_2023}. The latter has been proposed theoretically, using similar Green's function techniques to characterize the bosonic \ac{QPI} as are used to model the electronic \ac{QPI}. Detecting such bosonic \ac{QPI} would enable extracting the dispersion relation of the bosonic mode, as well as an evaluation of the electron-boson coupling by providing the electronic and bosonic dispersions and self energies from the same measurement.

While there are some attempts in the literature~\cite{ganguli_visualization_2023} to extract, for example, magnonic dispersion relations, so far demonstration of a magnonic origin of the features, for example from magnetic field dependence, is lacking.

Besides what the inelastic contribution reveals about the modes which are excited, if the contribution is sizeable, it complicates the interpretation of the electronic \ac{QPI} by introducing replica features that are not due to the underlying electronic quasiparticle band structure~\cite{chi_quasiparticle_2025}. It may well be that it is the inelastic contribution which limits the observation of sharp \ac{QPI} features to the vicinity of the Fermi energy in many correlated materials. A full interpretation of the \ac{QPI} will therefore require also accounting for the inelastic sector and its influence on the measured \ac{QPI} signal.

\subsection{Non-equilibrium effects in \ac{QPI}}
Non-equilibrium phases provide new routes to stabilize correlated ground states~\cite{fausti_light-induced_2011,wang_observation_2013} and control them on femtosecond time scales. Tools that enable determining the electronic structure in these ground states are limited -- thermodynamic and transport measurements are practically impossible to perform with sub-ps resolution. Spectroscopically, most work so far has been with optical spectroscopy and angle-resolve photoemission spectroscopy. For \ac{STM}, the methodology to perform time-resolved measurements is becoming established, with promising results in electron spin resonance  (ESR)-type of experiments, manipulating single spins~\cite{loth_measurement_2010}, as well as using THz \ac{STM}~\cite{cocker_ultrafast_2013} of, e.g., vibrational modes of molecules~\cite{cocker_tracking_2016}.

In quantum materials, the goal would be to perform pump-probe \ac{QPI} to establish non-equilibrium electronic structures of transient states. This would enable, for example, full characterization of the transient correlated phases and imaging of their emergent electronic states. \ac{STM} experiments performed on Floquet-Bloch states, for example, not only would provide additional validation for the topological nature of the driven electronic states, but should also enable mapping out protected defect and boundary states that should arise due to the topological nature of the band structure.
The experimental challenges are formidable and may require measurement schemes for fast acquisition of \ac{QPI} maps~\cite{nakanishi-ohno_compressed_2016,oppliger_sparse_2020}.

\subsection{Bayesian fitting of \ac{QPI} -- towards extracting quantitative band structure information}
One of the key drawbacks of the analysis of \ac{QPI} measurements compared to, e.g., the spectral function measured by \ac{ARPES} is the significantly more involved analysis and interpretation: for \ac{ARPES}, calculated spectral functions can be straightforwardly compared with the \ac{ARPES} measurements. For \ac{QPI}, besides the electronic structure, the type of defect and tunneling matrix elements need to be accounted for, increasing the number of parameters and requiring relatively complex modelling. At the moment, despite availability of code to simulate \ac{QPI} calculations from a tight-binding model~\cite{wahl_calcqpi_2025}, the interpretation remains challenging.
The availability of simulation codes and the power of modern high-performance computing clusters does raise the prospect, though, for Bayesian fitting of \ac{QPI}, which would provide parameter distributions, and hence estimates for the parameters of the electronic structure as well as information about the error. In such a scheme, tight-binding models can be numerically matched to the experimental \ac{QPI}. The framework that would enable such Bayesian fitting largely exists, but still requires some gaps to be filled in. A key open aspect is the definition of a suitable cost function that provides a good estimate for the goodness of the calculated \ac{QPI} compared to the measured \ac{QPI}. Often effects not accounted for in the calculation, such as the exact distribution of impurities, their chemical nature, as well as self energy effects and influence of inelastic tunneling, mean that while the calculations can provide qualitatively good results, they lack in quantitative agreement.  While these effects complicate automated fitting of \ac{QPI}, none of them should present an insurmountable obstacle.

\section{Conclusion}
\Ac{QPI} is a technique that, since its discovery three decades ago, has matured from a curiosity found at the surfaces of noble metals into a powerful technique that complements \ac{ARPES} and quantum oscillations in our understanding of the electronic structure of quantum materials. While the interpretation of the spectral function measured by \ac{ARPES} is much more straightforward, \ac{QPI} has the ability to probe the electronic structure in regimes that are not accessible to \ac{ARPES}, including in the unoccupied states, at temperatures below \qty{100}{\milli\K}, and in magnetic fields. Beyond merely complementing other techniques for studying the electronic structure, \ac{QPI} provides access to information not easily accessible by other techniques: for example it is one of only two techniques that can provide momentum-resolved information about the symmetry of a superconducting order parameter.

What, however, the past few decades of research on the interpretation of \ac{QPI} have also shown is that in most cases, a strong theoretical support is required to understand the observed patterns. The development of theoretical modelling of \ac{QPI} has made tremendous progress and in conjunction with broad availability of high performance computing resources, we can now start to read the finer details of \ac{QPI} data. Because \ac{QPI} contains so many contributions from a range of physical phenomena, disentangling these requires realistic modelling of the band structure, matrix element effects, and in some cases even the tunneling process itself.

On the experimental side, the development of reliable lower temperature instruments that operate at temperatures below \qty{100}{\milli\K} has made a significant difference to the quality of \ac{QPI} that can be obtained from strongly correlated electron materials. Even if the relevant phases of a material are accessible at higher temperatures, broadening to the temperature of the tip and electronic interactions often means that the \ac{QPI} often only becomes sharp and clearly discernible at lower temperatures. The increased sharpness of features at lower temperatures has in many cases proven crucial to fully constrain the \ac{QPI} and unlock new physical insight.

In particular for strongly correlated electron materials, the ability to access ultra-low temperatures and high magnetic fields has been crucial to study these materials  in the parameter regimes where the electronic correlation effects are strongest, and which are typically not accessible by \ac{ARPES}. We therefore fully expect that the next thirty years will uncover similar surprises for our understanding of electronic correlation effects and quantum criticality, and help uncover some of the remaining mysteries of condensed matter physics.

\begin{acknowledgements}
We acknowledge insightful discussions with Saeed Bahramy, Seamus Davis, Peter Hirschfeld, Andreas Kreisel, Tadashi Machida and Matthew J. Neat.
PW and TH acknowledge support through a JSPS Invitational fellowship, TH from KAKENHI Grants (No. 24H00198 and No. 25H01249) and the RIKEN TRIP initiative (Many-body Electron Systems), YK from KAKENHI Grants (No. 25K22009), LCR from UKRI3345, and CDAM and PW from Leverhulme Trust through Research Project Grant RPG-2022-315 and UKRI through UKRI1107 and PW from UKRI3584.
\end{acknowledgements}

\bibliography{references}

\end{document}